\documentclass[twocolumn, twocolappendix]{aastex631}

\newcommand{\jwst}{\textit{JWST}}
\newcommand{\hst}{\textit{HST}}
\usepackage{array}
\usepackage{multirow} 

\usepackage{color}
\usepackage{graphicx}
\usepackage{amsmath}
\def\apjl{\rmfamily{ApJ~}}        
\def\apjs{\rmfamily{ApJS~}}       

\def\aap{\rmfamily{A\&A~}}        

\def\ao{\rmfamily{Appl.~Opt.~}}   

\usepackage{multirow}
\begin{document}

\title{SN\,2022riv in RX\,J2129: Discovery, Spectroscopic Classification, and Microlensing of a Strongly Lensed Type Ia Supernova from JWST and HST Observations}

\author[0000-0003-0203-3853]{Birendra Dhanasingham}
\affiliation{Minnesota Institute for Astrophysics, University of Minnesota, 116 Church Street Southeast, Minneapolis, MN 55455, USA}

\author[0000-0003-3142-997X]{Patrick L. Kelly} 
\affiliation{Minnesota Institute for Astrophysics, University of Minnesota, 116 Church Street Southeast, Minneapolis, MN 55455, USA}

\author[0000-0003-1060-0723]{Wenlei Chen}
\affiliation{Department of Physics, Oklahoma State University, 145 Physical Sciences Building, Stillwater, OK 74078, USA}

\author[0000-0002-2361-7201]{Justin Pierel}
\affiliation{Space Telescope Science Institute, 3700 San Martin Drive, Baltimore, MD 21218, USA}
\affiliation{NASA Einstein Fellow}

\author[0000-0003-3484-399X]{Masamune Oguri}
\affiliation{Center for Frontier Science, Chiba University, 1-33 Yayoi-cho, Inage-ku, Chiba 263-8522, Japan}
\affiliation{Department of Physics, Graduate School of Science, Chiba University, 1-33 Yayoi-Cho, Inage-Ku, Chiba 263-8522, Japan}

\author[0000-0002-4693-0700]{Derek Perera}
\affiliation{Minnesota Institute for Astrophysics, University of Minnesota, 116 Church Street Southeast, Minneapolis, MN 55455, USA}

\author[0000-0001-9065-3926]{Jose M. Diego}
\affiliation{Instituto de F\'isica de Cantabria, Edificio Juan Jord\'a, Avenida de los Castros, 39005 Santander, Spain}

\author[0000-0002-0350-4488]{Adi Zitrin}
\affiliation{Department of Physics, Ben-Gurion University of the Negev, P.O. Box 653, Be'er-Sheva 84105, Israel}

\author[0000-0002-7876-4321]{Ashish K. Meena}
\affiliation{Department of Physics, Indian Institute of Science, Bengaluru 560012, India}

\author[0000-0003-1974-8732]{Mathilde Jauzac}
\affiliation{Department of Physics, Centre for Extragalactic Astronomy, Durham University, South Road, Durham, DH1 3LE, UK}
\affiliation{Department of Physics, Institute for Computational Cosmology, Durham University, South Road, Durham, DH1 3LE, UK}
\affiliation{Astrophysics Research Centre, University of KwaZulu-Natal, Westville Campus, Durban 4041, South Africa}
\affiliation{School of Mathematics, Statistics \& Computer Science, University of KwaZulu-Natal, Westville Campus, Durban 4041, South Africa}
\affiliation{Institut de Recherche en Astrophysique et Plan\'etologie (IRAP), Universit\'e de Toulouse, CNRS, CNES, 9 avenue du colonel Roche, 31028 Toulouse, France}

\author[0000-0003-3266-2001]{Guillaume Mahler}
\affiliation{Department of Physics, Centre for Extragalactic Astronomy, Durham University, South Road, Durham, DH1 3LE, UK}
\affiliation{Department of Physics, Institute for Computational Cosmology, Durham University, South Road, Durham, DH1 3LE, UK}
\affiliation{STAR Institute, University of Li{\`e}ge, Quartier Agora, All\'ee du six Ao\^ut 19c, 4000 Li\`ege, Belgium}

\author[0009-0009-7962-656X]{Elias Mamuzic}
\affiliation{Technical University of Munich, TUM School of Natural Sciences, Physics Department, James-Franck-Straße 1, 85748 Garching, Germany}
\affiliation{Max Planck Institute for Astrophysics, Karl-Schwarzschild Straße 1, 85748 Garching, Germany}

\author[0000-0002-6039-8706]{Liliya L.R. Williams}
\affiliation{Minnesota Institute for Astrophysics, University of Minnesota, 116 Church Street Southeast, Minneapolis, MN 55455, USA}

\author[0000-0002-0992-5742]{Yoon Chan Taak}
\affiliation{Department of Physics, Oklahoma State University, 145 Physical Sciences Bldg, Stillwater, OK 74078, USA}

\author[0000-0002-6610-2048]{Anton M. Koekemoer}
\affiliation{Space Telescope Science Institute, 3700 San Martin Drive, Baltimore, MD 21218, USA}

\author[0000-0002-8785-8979]{Thomas J. Broadhurst}
\affiliation{Department of Theoretical Physics, University of Basque Country UPV/EHU, Bilbao, Spain}
\affiliation{Ikerbasque, Basque Foundation for Science, Bilbao, Spain}
\affiliation{Donostia International Physics Center, Paseo Manuel de Lardizabal, 4, San Sebastián, 20018, Spain}

\author[0000-0001-6278-032X]{Lukas J. Furtak}
\affiliation{Department of Astronomy, The University of Texas at Austin, Austin, TX 78712, USA}
\affiliation{Cosmic Frontier Center, The University of Texas at Austin, Austin, TX 78712, USA}

\author[0000-0002-7633-2883]{David Lagattuta}
\affiliation{Centre for Astrophysics Research, Department of Physics, Astronomy and Mathematics, University of Hertfordshire, Hatfield, AL10 9AB, UK}

\author[0000-0002-1681-0767]{Hayley Williams}
\affiliation{School of Earth and Space Exploration, Arizona State University, Tempe, AZ 85287-6004, USA}

\author[0000-0001-8737-9700]{Kyle Dalrymple}
\affiliation{William H. Miller III Department of
Physics \& Astronomy, Johns Hopkins University, 3400 N Charles St, Baltimore, MD 21218, USA}

\author[0000-0003-3460-0103]{Alexei V. Filippenko}
\affiliation{Department of Astronomy, University of California, Berkeley, CA, 94720-3211, USA}

\author[0000-0002-8526-3963]{Christa Gall}
\affiliation{DARK, Niels Bohr Institute, University of Copenhagen, 2100 Copenhagen, Denmark}

\author[0000-0002-5116-7287]{Daniel Gilman}
\affiliation{Department of Astronomy and Astrophysics, University of Chicago, 5640 S Ellis Ave, Chicago, IL 60637, USA}
\affiliation{Brinson Prize Fellow}

\author[0000-0002-4571-2306]{Jens Hjorth}
\affiliation{DARK, Niels Bohr Institute, University of Copenhagen, 2100 Copenhagen, Denmark}

\author[0000-0001-8738-6011]{Saurabh W. Jha}
\affiliation{Department of Physics and Astronomy, Rutgers, the State University of New Jersey, Piscataway, NJ 08854, USA}

\author[0000-0003-2037-4619]{Conor Larison}
\affiliation{Space Telescope Science Institute, 3700 San Martin Drive, Baltimore, MD 21218, USA}

\author[0000-0003-1700-5740]{Chien-Hsiu Lee}
\affiliation{Hobby-Eberly Telescope, McDonald Observatory, The University of Texas at Austin, Fort Davis, TX 79734, USA}

\author[0000-0001-6876-8284]{Paolo A. Mazzali}
\affiliation{Astrophysics Research Institute, Liverpool John Moores University, 146 Brownlow Hill, Liverpool L3 5RF, UK}
\affiliation{Max Planck Institute for Astrophysics, Karl-Schwarzschild Str. 1, 85748 Garching, Germany}

\author[0000-0002-7559-0864]{Keren Sharon}
\affiliation{Department of Astronomy, University of Michigan, 500 Church Street, Ann Arbor, MI 48109, USA}

\author[0000-0001-5568-6052]{Sherry H.~Suyu}
\affiliation{Technical University of Munich, TUM School of Natural Sciences, Physics Department, James-Franck-Straße 1, 85748 Garching, Germany}
\affiliation{Max Planck Institute for Astrophysics, Karl-Schwarzschild Str. 1, 85748 Garching, Germany}



\begin{abstract}
The multiply imaged SN\,2022riv was discovered through a search of galaxy cluster fields as part of a {\it Hubble Space Telescope (HST)} SNAP program to find highly magnified stars. The supernova (SN) was detected in the image corresponding to the longest time delay of a galaxy at redshift $z=1.522$, strongly lensed by the foreground galaxy cluster RX\,J2129.7+0005. Follow up {\it James Webb Space Telescope (JWST)} NIRSpec G140M and PRISM spectroscopy yields an SN Ia classification. Using the SALT3-NIR light-curve fitter, we obtain a cosmology-independent measurement of the magnification of $5.35\pm1.01$ for the last-to-arrive image of the SN, with multiple SALT SN spectral time-series models yielding consistent constraints. The last-to-arrive image of SN\,2022riv we detect appeared adjacent to the brightest cluster galaxy (BCG) at a location with an exceptionally high stellar mass density ($\sim$ 1--2 dex higher than that of SN\,Refsdal), where microlensing is expected to introduce a 20--50\% modulation of the magnification. Analyzing six independent lens models of the cluster, we find that four predict the magnification with much greater precision ($p < 0.05$) than would be expected by random chance, given the large effect anticipated from microlensing. Five models yield magnifications of roughly 4--7 (within  1$\sigma$) prior to accounting for microlensing, whereas \texttt{HoliGRALE} favors a significantly higher value of $15.39 \pm 0.85$. After incorporating nominal microlensing, the \texttt{HoliGRALE} prediction is within 1$\sigma$ tension with our measurement. A companion paper will present constraints on the relative time delay of the image that arrived earlier.
\end{abstract}



\section{Introduction} 
\label{sec:intro}

Galaxy clusters are the largest gravitationally bound objects in the Universe and consist predominantly of dark matter. Therefore, constructing accurate cluster mass models requires assumptions about the nature of dark matter and how its distribution is connected to that of a luminous baryonic component i.e., the hot intracluster medium, the luminous components of the cluster galaxies, and the intracluster light (ICL) \citep[see, e.g.,][]{2005ApJ...621...53B, 2005MNRAS.360..477D, 2009MNRAS.396.1985Z, oguri10, 2011ascl.soft02004K, 2014MNRAS.443.1549J, 2015MNRAS.446.4132J, 2014MNRAS.444..268R, 2014MNRAS.437.2642S, 2015ApJ...801...44Z, 2019RPPh...82l6901O, 2020RNAAS...4..215C, 2021PASP..133g4504O, 2022ApJ...931..127C, 2023MNRAS.523.4568F, 2024ApJ...961..186C, 2024ApJ...973...25K, perera25}. Gravitational lensing by sufficiently massive foreground objects---such as galaxies and galaxy clusters---deflects photon trajectories and increases light-travel times, producing multiple magnified images of background sources and introducing relative time delays between the images. Consequently, galaxy clusters serve as powerful tools for probing the high-redshift Universe by magnifying faint background sources, offering a unique opportunity to detect intrinsically faint galaxy populations beyond the unaided limits of current instrumentation \citep[see, e.g.,][]{2013ApJ...762...32C, 2014ApJ...795..126B, 2014ApJ...793L..12Z, 2018ApJ...864L..22S, 2022ApJ...940...55B, 2023ApJ...948L..14C, williamskellychen23}. However, their effective use requires accurate lens models. The expanding sample of resolved lensed images---made possible by advances in high-resolution instrumentation---has substantially enhanced the precision of lens models, largely independent of the specific modeling framework employed.

Strongly lensed supernovae (SNe) provide a comparatively new, precise tool for evaluating the accuracy of lens models, including the assumptions used to construct them. Lens-model-independent measurements of the magnification and time delays of strongly lensed SNe---particularly SNe Ia, which function as standardizable candles---can then be directly compared to lens model predictions at the corresponding image positions. Such comparisons offer a robust validation of the lens model and can be used to refine it or to identify discrepancies arising from the substructure in the main lens plane or from structures along the line of sight. Examples where we can measure a time delay---the difference in arrival time between multiple images of the same source arising from different geometric paths and gravitational potentials---also offer a new opportunity to measure the Hubble constant, $H_0$ (since the time delay is inversely proportional to $H_0$) \citep[e.g.,][]{Refsdal_1964, 2023Sci...380.1322K, 2025ApJ...979...13P, 2026A&A...708A.291S, pierel+26}, complementing the well-established approach based on strongly lensed quasars \citep{2020A&A...642A.193M, 2020A&A...640A.105M, 2024SSRv..220...48B, 2024A&A...689A.168W, 2025A&A...697A.139D, 2026A&A...706A.270P}. Although the probability of discovering an SN in a multiply imaged galaxy is relatively low compared to lensed quasars, strongly lensed SNe also offer several characteristics that are advantageous for time-delay cosmography. Because SNe fade with time, the underlying host-galaxy light can eventually be observed with reduced contamination from the transient, enabling more accurate modeling of the host galaxy’s light distribution and, consequently, more precise photometry \citep{2021MNRAS.504.5621D}. This also benefits the modeling of the spatially resolved stellar kinematics of the lensed host galaxy, which can help to break the mass-sheet degeneracy \citep{1996ApJ...464...92G, 1999ApJ...516...18R, 2002MNRAS.337L...6T, 2011MNRAS.415.2215B, 2012MNRAS.423.1073B}. In addition, SN light curves exhibit substantial variations on timescales of weeks to months, allowing for precise time-delay measurements with relatively short observing campaigns \citep{
2019ApJ...876..107P, 2022A&A...658A.157H, 2024A&A...692A.132H, 2024MNRAS.534.1077C}. Furthermore, time-delay measurements from SNe Ia are less susceptible to chromatic microlensing effects, provided that observations are conducted during the early phase when microlensing is approximately achromatic \citep[typically within the first three weeks after explosion;][]{2018ApJ...855...22G, 2018MNRAS.478.5081F, huber_chromatic_microlensing} and multiband light curves are employed \citep{2019A&A...621A..55B}.

The luminosities of SNe Ia, the thermonuclear explosions of white dwarf stars \citep[e.g.,][]{web84}, can be estimated using the shape and color of their light curves. Therefore, they can be used to measure the magnification due to a foreground gravitational lens.  \citet{patelmccullyjha14} directly measured the magnifications of three SNe Ia in the CLASH \citep[Cluster Lensing And Supernova survey with \textit{Hubble};][]{postmancoebenitez12} cluster observations with lens model magnifications of $2.1 \pm 0.15$, $1.29 \pm 0.08$, and $1.48 \pm 0.11$, with similar results presented by \citet{Nordin:2014}. 
An SN Ia discovered in Abell 2744 during the Hubble Frontier Fields campaign was magnified by a factor of $\mu = 2.03 \pm 0.29$, and the majority of models were within $<$1\,$\sigma$ agreement with lens model predictions \citep{rodneypatelscolnic15}. To date, several lensed SNe Ia---spanning both galaxy-scale \citep[see, e.g.,][]{2017ApJ...835L..25M, 2020MNRAS.496.3270M, 2022A&A...662A..34D, 2023ApJ...948..115P, 2023NatAs...7.1098G, 2025ApJ...980..172L, 2025OJAp....8E.166A} and cluster-scale lens systems \citep[see, e.g.,][]{fryepascalepierel24, 2025A&A...702A.157E, 2025MNRAS.544..708X, 2026A&A...708A.291S} --- have been used to test the accuracy of lens model predictions.

The relative magnifications and ratios of time delays of five SN images --- independent of $H_0$ --- were also used in the case of SN\,Refsdal \citep{2023Sci...380.1322K, delaykellyrodneytreu23} to distinguish among a large set of lens models of the galaxy cluster MACS\,J1149.5+2223 constructed before its reappearance, although several required updates following the reappearance. The measurements found that simply parameterized models consisting of cluster-scale dark-matter halos, and halos associated with individual galaxies, were strongly favored over more flexible methods.  SN\,H0pe  was an SN Ia strongly lensed by the galaxy cluster PLCK\,G165.7+67.0 \citep[redshift $z = 0.35$;][]{fryepascalepierel24}. \citet{2025ApJ...979...13P} measured the magnification and time delays of three images from their light curves, and \citet{chenkellyfrye24} determined the relative time delays from the phase of their spectra. Furthermore, \cite{2025A&A...702A.157E}, \citet{acebron+25}, and \citet{2026A&A...708A.291S} used the lensed SN\,Encore \citep{2024ApJ...967L..37P}, discovered in a lensed galaxy that had previously hosted SN\,Requiem \citep[the first photometrically classified SN Ia strongly lensed by a galaxy cluster;][]{2021NatAs...5.1118R}, for the cluster lens modeling of MACS\,J0138.0-2155.

\begin{figure*}[th!]
\centering
\includegraphics[angle=0,width=0.99\textwidth]{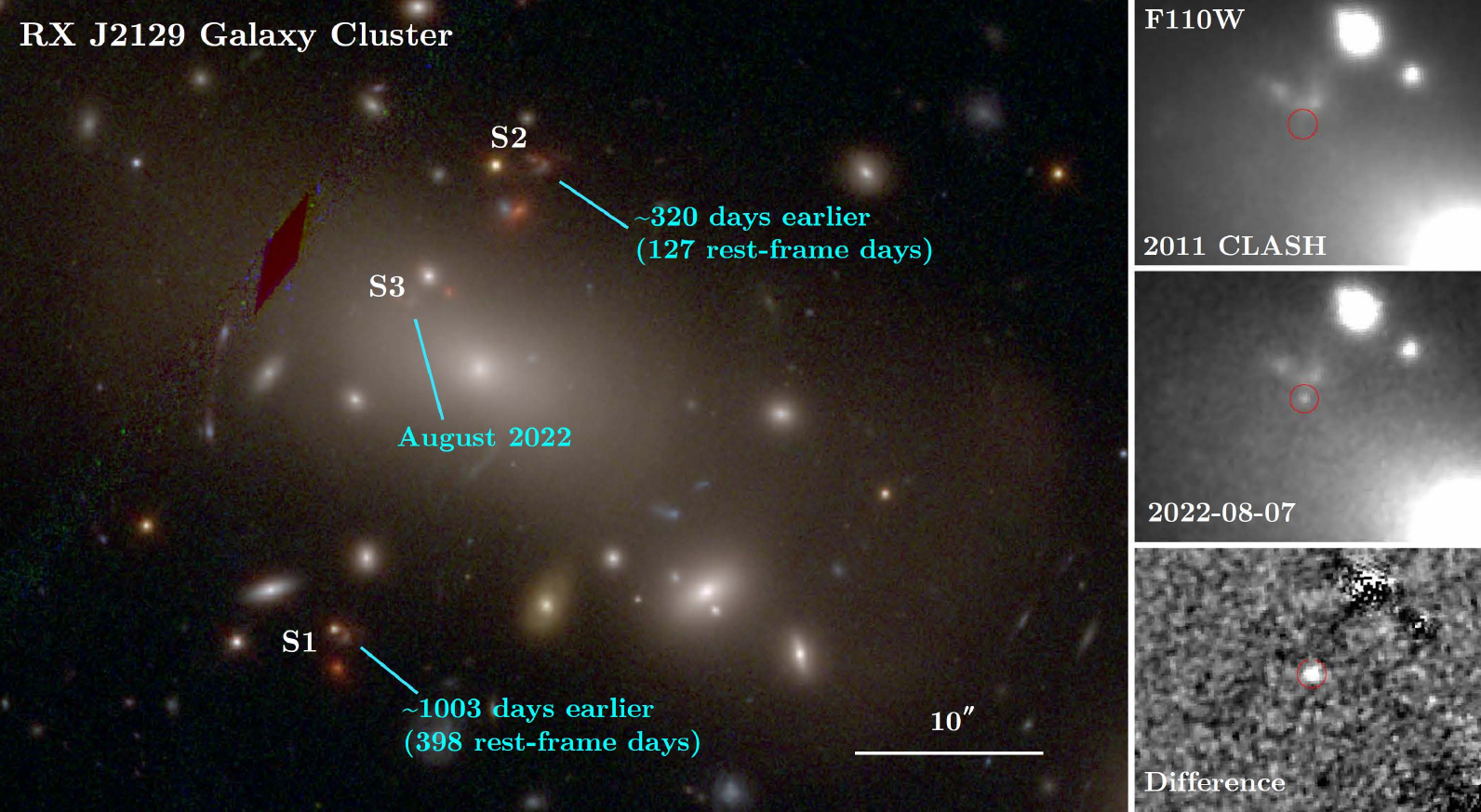}
\caption{The \textit{HST} WFC3/IR image of the RX\,J2129 galaxy cluster with three marked images (S1, S2, and S3) of the SN\,2022riv host galaxy with the positions of the SN indicated. From top to bottom in the right-hand panels we show close-up views of the observation centered on the SN location taken by the CLASH survey in 2011 using the WFC3/IR F110W filter,  the observation under the SNAP program on 2022 August 7 (S3, also using the WFC3/IR F110W filter), and the difference image.
\label{fig:detection}}
\end{figure*}

The strongly lensed SN\,2022riv \citep{kellyzitrinoguri22} in the RX\,J2129.7+0005 galaxy cluster field (hereafter RX\,J2129; $z=0.234$) was discovered as part of the \textit{Hubble Space Telescope} Snapshot (\textit{HST}-SNAP) program GO-16729 (Cycle 29, PI P. L. Kelly). WFC3/IR F110W (369\,s) and WFC3/UVIS F606W (880\,s) imaging, acquired on 2022 August 7 UTC (MJD 59798), was used to identify the SN by subtracting deeper archival images obtained by the CLASH survey in 2011 \citep{postmancoebenitez12} from coadded data. As shown in Figure~\ref{fig:detection}, the transient labeled S3 exhibits a significance of $\sim 10\sigma$ with a 0\farcs3 aperture in the {\it HST} WFC3/IR F110W difference image. The transient was coincident with a multiply imaged galaxy that has a published spectroscopic redshift of 1.52 \citep{caminharosatigrillo19} and is offset from the host nucleus. 

The RX\,J2129 galaxy cluster lens forms three images (S1, S2, and S3) of the SN's host galaxy. The image detected as S3 is one of the ``trailing'' images, i.e., last-to-arrive. At the time of discovery, a simply parameterized {\tt GLAFIC} model \citep{oguri10,2021PASP..133g4504O}, which had accurately reproduced the time delays of SN\,Refsdal in the MACS\,J1149 galaxy cluster, was used to ``postdict'' that S2 arrived $\sim$320 days before S3, and S1 arrived $\sim$1003 days before S3. Simply parameterized models by \cite{2015ApJ...801...44Z} yielded nearly identical values, with a $\sim$320-day delay between S3 and S2, and a $\sim$1200-day delay between S3 and S1.

In this paper, we present the discovery, spectroscopic and photometric classification, and  inference of the absolute magnification of SN\,2022riv. Additionally, we present six independent lens model analyses of the RX\,J2129 galaxy cluster using astrometry of the lensed images based on high-resolution {\it James Webb Space Telescope (JWST)} Near-Infrared Camera (NIRCam) imaging. These models are used to compute predictions for the magnifications, time delays, and image positions of SN\,2022riv, which we then compare with the absolute magnification inferred from photometry. We also constrain the possible effects of stellar microlensing and dark matter substructure millilensing on the magnification.

This paper is organized as follows. Section~\ref{sec:observations} presents the discovery and the \textit{HST} and {\it JWST} follow-up observations of SN\,2022riv. In Section~\ref{sec:classification}, we classify the object as an SN Ia using both spectroscopic and photometric data and infer its absolute magnification using SALT SN Ia spectral time-series models. In Section~\ref{lens_models}, we present and discuss our double-blind analysis of the magnifications, time delays, and image positions of SN\,2022riv, using updated RX\,J2129 cluster lens models, and constrain the effects of millilensing and microlensing. The results of our analysis are discussed in Section~\ref{sec:discussion}.

\section{Observations} 
\label{sec:observations}

SN\,2022riv was first detected in observations of the galaxy cluster RX\,J2129 obtained as part of the \textit{HST}-SNAP program GO-16729 (Cycle 29, PI P. L. Kelly). The transient was detected in Wide Field Camera 3 infrared (WFC3/IR) F110W imaging acquired on 2022 August 7 (UTC dates are used throughout this paper), but was not detected in the contemporaneous WFC3 ultraviolet and visible light (UVIS) F606W imaging. 

The apparent F110W AB magnitude of $24.780\pm0.104$ and the nondetection in F606W at the discovery epoch implied an 80\% probability that the transient was an SN Ia. A {\it JWST} Director’s Discretionary Time (DDT) program was approved to obtain a spectroscopic classification of the SN and detect the fading next-to-last to arrive image (Cycle 1, \textit{JWST}-DD-2767, PI P. L. Kelly). \textit{JWST} NIRCam imaging (Cycle 1, \textit{JWST}-DD-2767, PI P. L. Kelly) was acquired on October 6 using the short-wavelength filters F115W, F150W, and F200W, and the long-wavelength filters F277W, F356W, and F444W. The exposure times in F150W and F356W  were $\sim 2.5$ times longer than those in the other bands to maximize sensitivity.

The \textit{JWST} Near Infrared Spectrograph (NIRSpec) in MultiObject Spectroscopy (MOS) mode was used with the microshutter assembly (MSA). Spectra of both the SN and the host galaxy were acquired using the medium-resolution G140M/F070LP (0.70--1.27\, $\mu$m) and G140M/F100LP (0.97--1.89\,$\mu$m) disperser-filter combinations, as well as the low-resolution PRISM/CLEAR (0.6--5.3\, $\mu$m) disperser-filter combination on 2022 October 22 under program DD-2767. The G140M spectra were expected to probe the rest-frame wavelengths from roughly 2700 to 7500\,\AA, thus providing coverage of key spectroscopic features used to classify SNe (e.g., the Balmer series for SNe~II and the 6150\,\AA\ Si~II absorption feature for SNe Ia). The PRISM spectrum was expected to cover wavelengths from 2400 to 21000\,\AA\ in the rest frame. A summary of all imaging and spectroscopic observations conducted for the classification and analysis of SN\,2022riv is provided in Table~\ref{observations}. 

Following the discovery, a nondisruptive \textit{HST} Target-of-Opportunity (ToO) program (LensWatch; Cycle 28, GO-16264, PI J. Pierel) was triggered to monitor the transient and measure its light curve, with follow-up observations obtained between 2022 September 20 and October 23. A complementary \textit{HST} DDT imaging (Cycle 29, \textit{HST}-GO/DD-17253, PI P. L. Kelly) program acquired an additional two epochs of imaging.  Consequently, the follow-up imaging was conducted using the WFC3/IR F110W and F160W filters over four successful epochs, with no parallel instruments.

\begin{deluxetable*}{ccccccc}
\tablecaption{Summary of observations of SN\,2022riv in the RX\,J2129 galaxy cluster. \label{observations}}
\tablehead{
\colhead{Proposal ID} & \colhead{Obs. Type} & \colhead{Telescope} & \colhead{Instrument} & \colhead{MJD}& \colhead{Grating/Filter} & \colhead{Exp. [s]}\\ [-0.4cm]}
\startdata
16729 & Imaging & \hst & WFC3/IR & 59798 & F110W & 369\\
16729 & Imaging & \hst & WFC3/UVIS & 59798 & F606W & 880\\
\hline
2767 & Imaging & \jwst & NIRCam/SW & 59858 & F115W & 2061\\
2767 & Imaging & \jwst & NIRCam/SW & 59858 & F150W & 4982\\
2767 & Imaging & \jwst & NIRCam/SW & 59858 & F200W & 2061\\
2767 & Imaging & \jwst & NIRCam/LW & 59858 & F277W & 2061\\
2767 & Imaging & \jwst & NIRCam/LW & 59858 & F356W & 4982\\
2767 & Imaging & \jwst & NIRCam/LW & 59858 & F444W & 2061\\
2767 &  Spectroscopy & \jwst & NIRSpec MSA & 59874 & G140M/F070LP & 3939\\
2767 &  Spectroscopy & \jwst & NIRSpec MSA & 59874 & G140M/F100LP & 3939\\
2767 &  Spectroscopy & \jwst & NIRSpec MSA & 59874 & PRISM/CLEAR & 4377\\
\hline
16264 & Imaging & \hst & WFC3/IR & 59842 & F110W & 1098\\
16264 & Imaging & \hst & WFC3/IR & 59842 & F160W & 1098\\
16264 & Imaging & \hst & WFC3/IR & 59853 & F110W & 1098\\
16264 & Imaging & \hst & WFC3/IR & 59853 & F160W & 1098\\
16264 & Imaging & \hst & WFC3/IR & 59865 & F110W & 1098\\
16264 & Imaging & \hst & WFC3/IR & 59865 & F160W & 1098\\
16264 & Imaging & \hst & WFC3/IR & 59875 & F110W & 1098\\
16264 & Imaging & \hst & WFC3/IR & 59875 & F160W & 1098\\
\hline
17253 & Imaging & \hst & WFC3/IR & 59916 & F110W & 1098\\
17253 & Imaging & \hst & WFC3/IR & 59916 & F160W & 1048\\
17253 & Imaging & \hst & WFC3/IR & 59931 & F110W & 2195\\
17253 & Imaging & \hst & WFC3/IR & 59931 & F160W & 2095\\
\enddata
\tablecomments{Columns: \textit{JWST/HST} Proposal ID, Observation Type, Telescope Name, Instrument Name, Modified Julian Date, Grating/ Filter, Exposure Time (in seconds).}
\end{deluxetable*}


\section{Classification}
\label{sec:classification}
In this section, we present a classification of SN\,2022riv as an SN Ia using two approaches: spectroscopic analysis and photometry.

\subsection{Spectroscopy}

We spectroscopically classify SN\,2022riv from our \textit{JWST} NIRSpec MOS observations. We processed and analyzed data products retrieved from the Mikulski Archive for Space Telescopes (MAST) portal, which includes both low-resolution PRISM spectroscopy and higher-resolution G140M grating spectroscopy.

\subsubsection{NIRSpec MOS Observations and Data Reduction}

\begin{figure*}
\centering
\includegraphics[angle=0,width=5.5in]{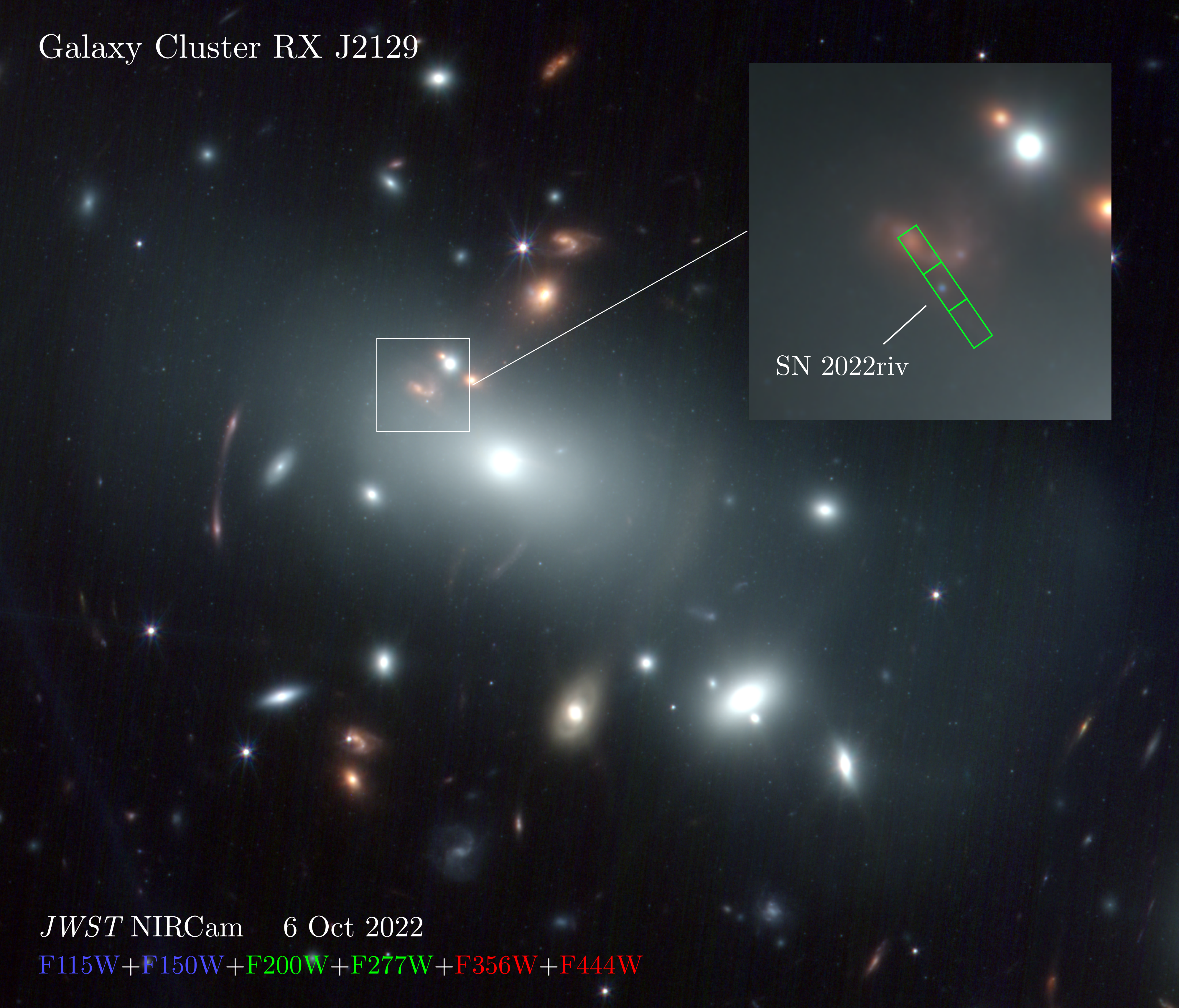}
\caption{The \textit{JWST} NIRCam image of the RX\,J2129 galaxy cluster. The stamp image superposed on the main image shows the detected SN and its host galaxy, where green rectangles mark the open MSA shutters used in the NIRSpec MOS observations.}
\label{fig:rxj2129}
\end{figure*}

During the SN spectral reduction process, we first designed the MSA to take the G140M and PRISM spectra of the bright SN image. The green rectangles in Figure~\ref{fig:rxj2129} mark the open MSA shutters used in our NIRSpec MOS observations. During the observation, the telescope executed nods to place the SN image at the center of each of the three shutters in consecutive exposures.

As shown in Figure~\ref{fig:rxj2129}, the shutter occupied by the SN was not significantly contaminated by its host galaxy. This allowed us to use the standard \textit{JWST} science calibration pipeline \citep[\texttt{jwst};][]{bushouse_2022_7229890} from the Space Telescope Science Institute (STScI) to process the background subtraction in the SN region. This pixel-to-pixel method first combines one or more consecutive nodded exposures to be treated as backgrounds into a mean background image and then subtracted the background directly from the source exposure. In our study, we excluded nodded background exposures in which the SN shutter of the source exposure was contaminated by the host galaxy's bright core, ensuring it did not affect the background subtraction in the SN region. We then used the standard pipeline for data reduction and source extraction. The \texttt{jwst} pipeline was executed with the context file \texttt{jwst\_1293.pmap}.

\begin{figure*}[th!]
\centering
\includegraphics[clip, trim=0cm 0cm 0cm 0cm, width=0.96\textwidth]{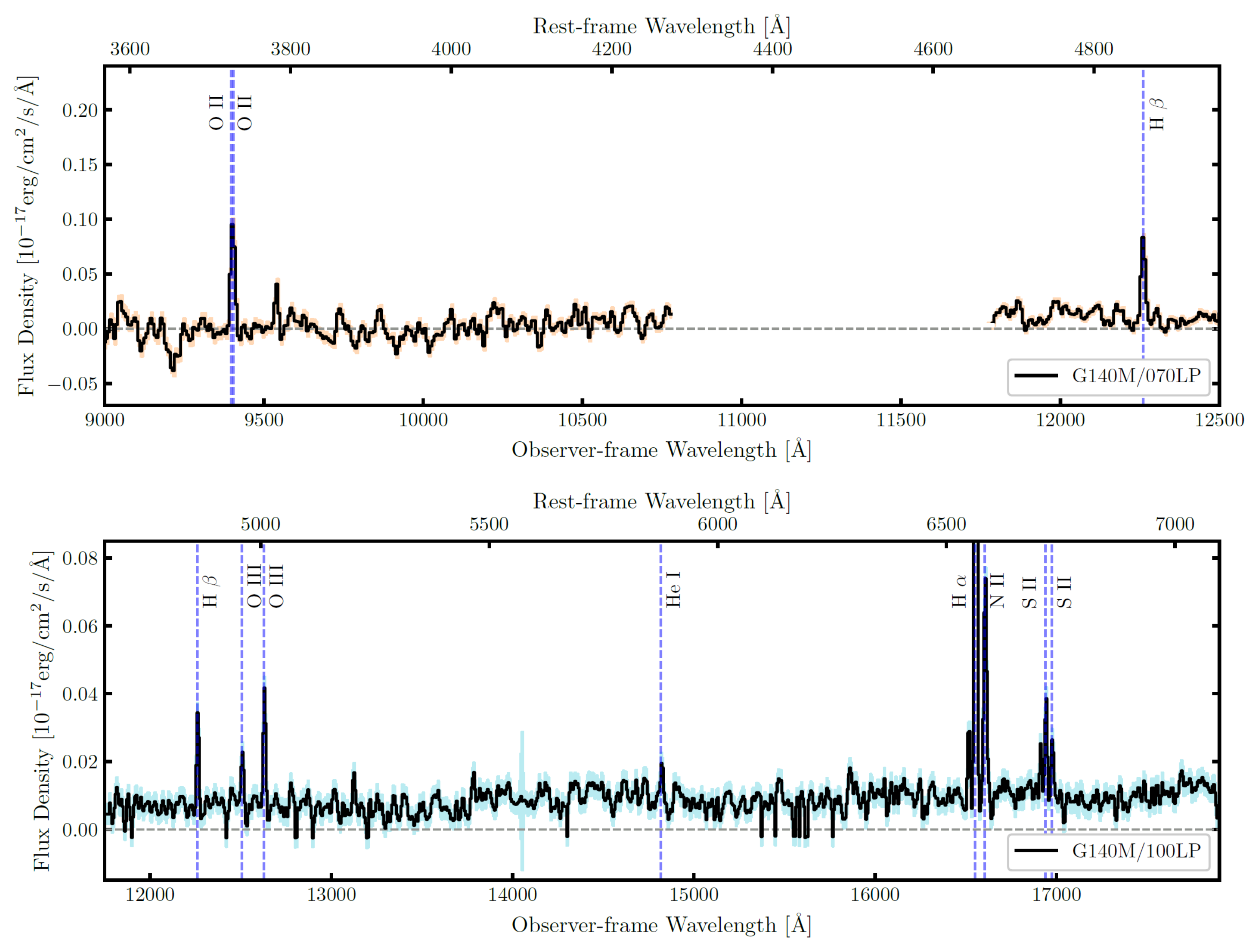}
\caption{NIRSpec G140M spectra of the host galaxy (top, G140M/070LP; bottom, G140M/100LP). Blue vertical dashed lines indicate the positions of the detected emission lines. Wavelengths are shown in both the observed frame (bottom) and the rest frame at $z = 1.522$ (top). The shaded colored regions in both panels represent the uncertainty in the flux-density measurements. The gap in the G140M/070LP spectrum (top) near the observed wavelength of 11000\,\AA{} is due to the physical separation between the two NIRSpec detector chips.
\label{galaxy_spectra}}
\end{figure*}

After extracting the observed G140M and PRISM spectra, the first step in our spectroscopic analysis was to verify that the redshift of the transient's host galaxy is consistent with that of the coincident multiply imaged galaxy. We accomplished this by identifying the emission lines of the host galaxy within the slitlet. As illustrated in Figure~\ref{galaxy_spectra}, the emission lines in the \textit{JWST} NIRSpec G140M spectra of the host galaxy yield a robust spectroscopic determination of $z = 1.522$ for the host galaxy. This value is in excellent agreement with the spectroscopic redshift reported by \cite{caminharosatigrillo19} and it is adopted for all subsequent analyses presented in this study.

\subsubsection{Classification with \texttt{SNID}}

We next classified the extracted G140M and PRISM spectra of SN\,2022riv using the Supernova Identification \citep[\texttt{SNID};][]{blondin07} package, which divides each spectral energy distribution (SED) by a smooth cubic spline fit to mitigate potential distortions in the SN's pseudocontinuum caused by dust extinction, slit losses, or inaccuracy of the spectroscopic calibration. This method removes the continuum shape of the SN spectrum, leaving a flattened SED with only the spectral absorption and emission features. These features were then cross-correlated with the \texttt{SNID} template library (``template 2.0'') using the correlation techniques of \cite{1979AJ.....84.1511T}. The template library includes all major SN spectroscopic types, including the subtypes as defined by \cite{blondin07}.

For cross-matching, we focused on portions within the rest wavelengths of 3500 to 10000\,$\rm \AA$ for the CLEAR/PRISM spectrum and 4700 to 7400\,$\rm \AA$ for the G140M/100LP spectrum. The goodness of the cross match is primarily evaluated using the $r\times lap$ parameter, where $r$ represents the correlation height-to-noise ratio and {\it lap} indicates the degree of wavelength overlap between the SN spectrum and each template spectrum \citep{1979AJ.....84.1511T, blondin07}. To be considered an acceptable match in this color-independent approach, which focuses on spectral features, an $r\times lap$ value greater than 5 is required.

Using the default template-matching thresholds of \texttt{SNID}, the PRISM spectrum yielded 1115 matching templates, of which 1067 were classified as SNe Ia, with 795 of these matching SN~Ia-norm templates. Similarly, the G140M/100LP spectrum produced 729 matches, 676 of which were identified as SNe Ia, with 518 corresponding to SN~Ia-norm templates. Table~\ref{SNID_results} presents the 10 best-fitting SN templates from the SNID analysis for the G140M/100LP and CLEAR/PRISM spectra of SN\,2022riv. The key features identified were Si II ($\sim6150$\,\AA), S II (near 5700\,\AA), and the Ca II near-IR triplet (near 8300\,\AA) P Cygni absorption features. The prominent absorption trough near 6150\,\AA, caused by blueshifted Si II $\lambda$6355\,\AA\ emission, is a distinguishing characteristic of SN Ia spectra, persisting for several weeks past maximum light and absent from the spectra of other SN types \citep[e.g.,][]{Filippenko_1997_SN_spectra}. 

\begin{figure*}[th!]
\centering
\includegraphics[clip, trim=0cm 0cm -0.45cm 0cm, width=0.9\textwidth]{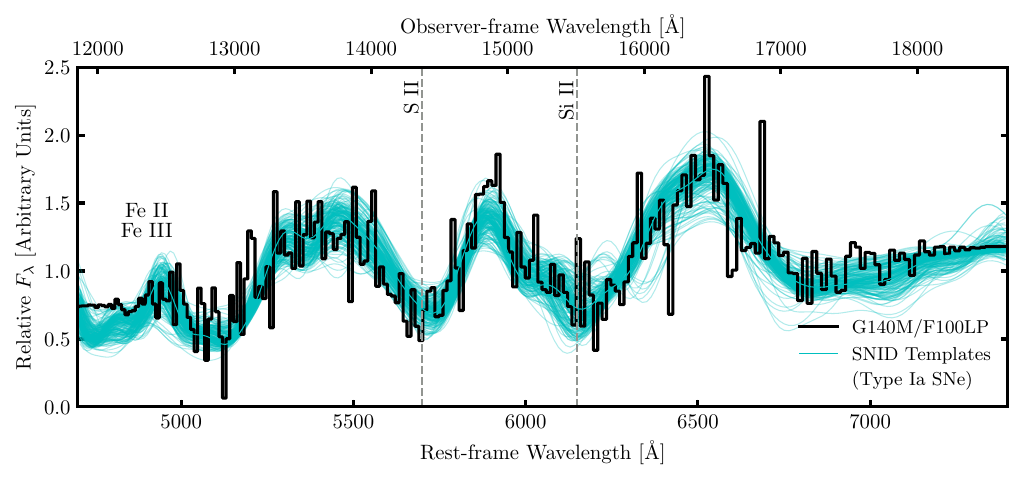}
\includegraphics[clip, trim=0cm 0cm 0cm 0cm, width=0.9\textwidth]{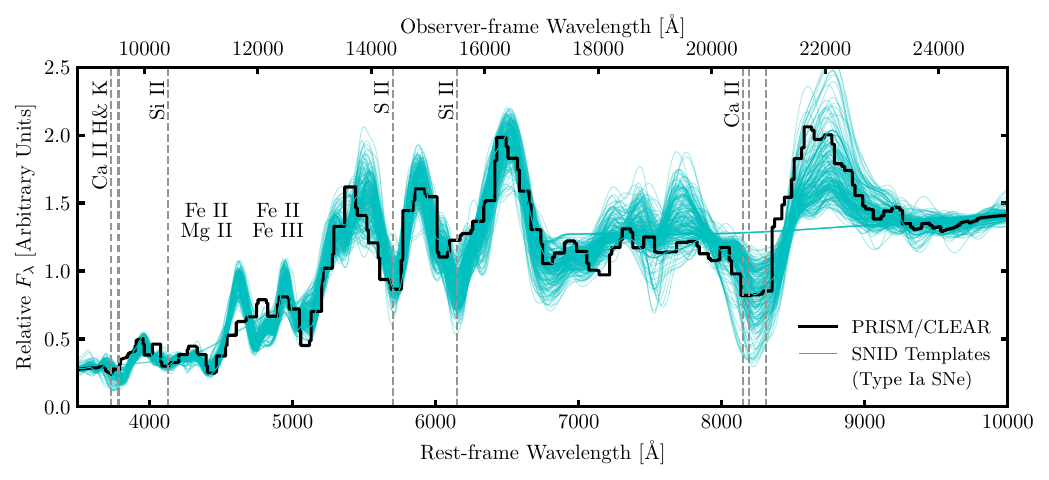}
\caption{
NIRSpec G140M/100LP (top) and CLEAR/PRISM (bottom) spectra of SN\,2022riv used for the \texttt{SNID} analysis (black line). The 200 best-matching SN~Ia templates from \texttt{SNID} are overlaid in cyan, with key spectral features labeled, confirming SN\,2022riv as an SN Ia.
\label{fig:PRISM_100LP_Ia}}
\end{figure*}
Figure~\ref{fig:PRISM_100LP_Ia} presents the 
G140M/100LP (top) and PRISM (bottom) spectra of SN\,2022riv, overlaid with the 200 best-fitting SN Ia spectra with $r \times lap >5$. This color-independent classification confirms SN\,2022riv as an SN~Ia with $>90\%$ confidence. Figures~\ref{fig:100lp_Ib/c} and~\ref{fig:PRISM_Ib/c_II} in Appendix~\ref{App: A} show the continuum-removed G140M/100LP and PRISM spectra, respectively, overlaid with the best-fitting SN Ib, SN Ic, and SN II template spectra. The best matches are SN\,2001V (age = 22.17 rest-frame days after peak) for the G140M/100LP spectrum and SN\,2005cf (age = 16.50 rest-frame days after peak) for the PRISM spectrum. The best match for the G140M/100LP spectrum belongs to the SN\,Ia-91T subclass, known for its spectroscopic similarity to normal SNe Ia at late stages. 

Finally, the \texttt{SNID} classification package constrains the phase to $21 \pm 20$ rest-frame days for the G140M/100LP spectrum and $26 \pm 22$ rest-frame days for the PRISM spectrum.

\begin{deluxetable}{lclcr}
\tablecaption{Top ten SN templates matching the observed G140M/100LP and CLEAR/PRISM spectra based on the \texttt{SNID} analysis.} \label{SNID_results}
\tablehead{
\colhead{Template} & \colhead{Type} & \colhead{Subtype} & \colhead{$r \times lap$} & \colhead{Phase [days]}\\ [-0.4cm]}
\startdata
\multicolumn{5}{c}{G140M/100LP}\\
SN\,2001V  & Ia & Ia-91T  & 29.56 & 22.17 \\
SN\,2001el & Ia & Ia-norm & 28.31 & 18.03 \\
SN\,2001V  & Ia & Ia-91T  & 27.84 & 21.34 \\
SN\,1990N  & Ia & Ia-norm & 27.69 & 20.69 \\
SN\,2001V  & Ia & Ia-91T  & 27.58 & 21.11 \\
SN\,2000fa & Ia & Ia-norm & 25.73 & 20.85 \\
SN\,2007fs & Ia & Ia-norm & 25.63 & 22.67 \\
SN\,1995al & Ia & Ia-norm & 24.68 & 22.26 \\
SN\,2001bg & Ia & Ia-norm & 22.70 & 18.47 \\
SN\,2001bg & Ia & Ia-norm & 22.22 & 18.56 \\
\hline
\multicolumn{5}{c}{PRISM/CLEAR}\\
SN\,2005cf & Ia & Ia-norm & 12.31 & 16.50 \\
SN\,2005cf & Ia & Ia-norm & 11.87 & 16.10 \\
SN\,2005cf & Ia & Ia-norm & 11.77 & 18.38 \\
SN\,2001cp & Ia & Ia-norm & 11.55 & 26.06 \\
SN\,2007cq & Ia & Ia-91T  & 11.47 & 20.22 \\
SN\,2007cq & Ia & Ia-91T  & 11.26 & 15.40 \\
SN\,2001cp & Ia & Ia-norm & 11.18 & 28.01 \\
SN\,2007cq & Ia & Ia-91T  & 11.12 & 37.77 \\
SN\,2001cp & Ia & Ia-norm & 11.03 & 22.09 \\
SN\,1999aa & Ia & Ia-91T  & 10.84 & 20.15 \\
\enddata 
\tablecomments{
Columns include SN template names, SN types, SN subtypes, $r \times lap$ values (which assess the fit quality), and phases in rest-frame days after peak brightness.}
\end{deluxetable}

\subsubsection{Classification with Si II Absorption}
For SNe Ia, the Si II line at 6355\,\AA{} forms a  blueshifted absorption feature near $\sim6150$\,\AA{}, and its blueshifted velocity as a function of the SN phase can be utilized as a classification method. At phases of 20--30\,days, typical SN Ia Si II velocity curves exhibit velocities of 7000--12000\,km\,s$^{-1}$ \citep{2005ApJ...623.1011B,2013ApJ...770...29C, 2015ApJS..220...20Z}, while super-Chandra SNe Ia have smaller velocities (4000--7000\,km\,s$^{-1}$) at a given phase \citep[see, e.g.,][]{2011MNRAS.412.2735T, 2012ApJ...757...12S, 2014MNRAS.443.1663C}. 

The Si II blueshift velocity was measured as follows. First, we fitted a pseudocontinuum using the two intervals straddling the 6150\,\AA{} region that are the least absorbed. The intervals were determined by eye to be 5870--5950\,\AA{} and 6480--6560\,\AA{}. To account for uncertainty arising from the selection of the background intervals, we varied the fitting region by introducing a random offset to each of the four boundary wavelengths by up to 20\,\AA{}. Second, the Si II absorption was fitted by a single Gaussian function in the region 6030--6400\,\AA{}. This fitting interval was chosen to be as small as possible to minimize the effect of adjacent absorption features. Finally, we repeated this fitting procedure 10$^4$ times, while varying the continuum-fitting region and introducing random offsets to the flux density consistent with the uncertainties. 

The resulting constraint on the Si II velocity is $(7.7\pm2.2)\times10^{3}$\,km\,s$^{-1}$. Compared with the velocity curves of other SNe Ia at phases of 20--30\,days, this measurement is consistent with the Si II 6355 velocities of low-redshift, SNe Ia sample \citep{2013ApJ...773...53F}, as well as other lensed normal SNe Ia such as SN\,H0pe \citep[$z=1.78$;][]{chenkellyfrye24}, SN\,Encore \citep[$z=1.95$;][]{2024MNRAS.535.2939D}, and SN\,iPTF16geu \citep[$z=0.409$;][]{2021MNRAS.502..510J}.

\subsection{Photometric Classification}

In this section, we present the results of an independent photometric classification of SN\,2022riv using \textit{JWST}'s NIRCam and \textit{HST}'s WFC3/IR observations. We find that the light curves are consistent with our spectroscopic classification of SN\,2022riv obtained through \texttt{SNID}, as discussed in the previous section.

\subsubsection{\textit{HST} Photometry}

To perform photometry of SN\,2022riv using \textit{HST} WFC3/IR imaging in the F110W and F160W bands across six epochs of follow-up observations along with the discovery epoch, we first retrieved the calibrated ``FLT'' files from the MAST archive for the imaging we acquired and the CLASH survey \citep{postmancoebenitez12}. These individual exposures are bias subtracted, dark-current corrected, and flat fielded. We began by aligning all exposures  using the \texttt{DrizzlePac}\footnote{https://github.com/spacetelescope/drizzlepac} task, \texttt{TweakReg}. Using the \texttt{DrizzlePac} task, \texttt{AstroDrizzle}, we projected the data onto a common pixel grid with 0\farcs06 pixel$^{-1}$, and combined the dithered observations into mosaics for each epoch.

To perform differential photometry using the F110W and F160W  observations, we created difference images by subtracting coadded archival ``template'' images from the CLASH survey \citep{postmancoebenitez12} from coadditions of our imaging. For aperture photometry, we use {\tt PythonPhot3} \citep{2015ascl.soft01010J} and apply a circular aperture with a radius of $0\farcs2$ at the SN position. The photometry, along with the corresponding uncertainties and details, is provided in Table~\ref{Photometry_hst_jwst}.

\subsubsection{\textit{JWST} Photometry}

\begin{figure}
\centering
\includegraphics[width=0.465\textwidth]{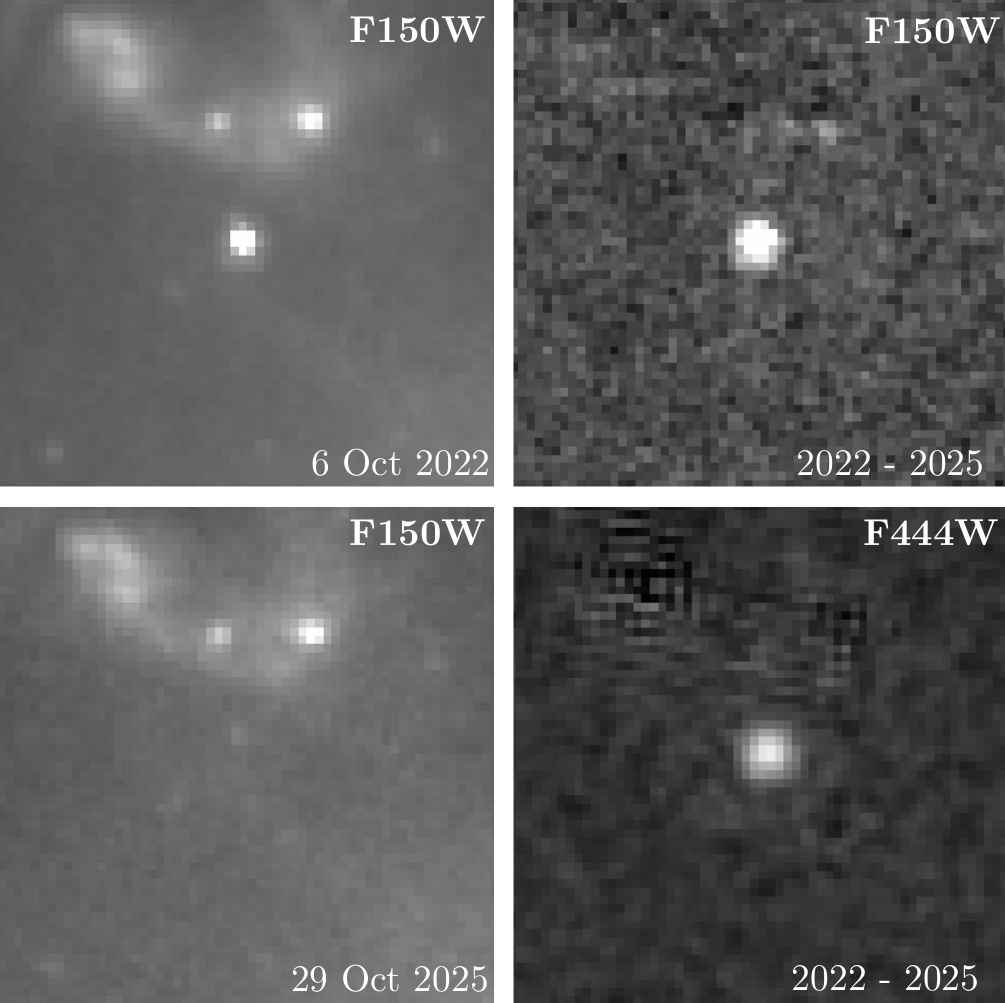}
\caption{\textit{Left:} \textit{JWST}/NIRCam images in the F150W filter, showing SN\,2022riv from the 2022 observation (top) and the 2025 visit to the cluster under the VENUS program (2025 October). \textit{Right:} Difference images created from F150W and F444W observations taken during the 2022 October visit and the 2025 October visit to the cluster field. All images are drizzled to a pixel scale of 0\farcs03 pixel$^{-1}$ and cover the same spatial extent.
\label{fig:new_obs}}
\end{figure}

In the initial stage of \textit{JWST} NIR photometry of SN\,2022riv, we used version 1.15.0 of the \texttt{jwst} pipeline to generate final, science-ready Level 3 coadds from Level 2b NIRCam data products obtained from the MAST portal. These Level 2 products (denoted as ``CAL" files) are individual exposures that have undergone bias subtraction, dark subtraction, and flat-field correction to account for pixel-to-pixel variations and large-scale instrument and telescope responsivity using the pipeline. During the generation of science-ready Level 3 data products, these calibrated CAL files are combined into a single integrated product, utilizing reference files from \texttt{jwst\_1253.pmap} and all default parameters. This process included aligning individual dithered exposures, matching the background, rejecting outliers, and then coadding the exposures onto a common grid with a 0\farcs03 pixel$^{-1}$ scale.

Following the methodology described by \cite{2000PASP..112.1360A} and subsequent improvements outlined by \cite{2016wfc..rept...12A}, we used the \texttt{photutils} \citep{larry_bradley_2024_13989456} package to construct the effective point-spread function (ePSF) for the \textit{JWST} observation of the RX\,J2129 galaxy cluster, using stars in the field. To guarantee the precise construction of the ePSF through a least-squares fitting routine for each NIRCam observation, we chose eight high signal-to-noise ratio (S/N) stars in the cluster field, chosen for their absence of detector artifacts, unsaturation, and adequate isolation to minimize contamination from nearby objects.

After constructing the ePSF for each NIRCam filter, we used \texttt{photutils} to fit the PSF to the SN using a nonlinear least-squares routine, within which a standard median-based background estimator was applied for uniform local background subtraction, thereby constraining its flux and centroid for each band. To estimate the background uncertainty, synthetic sources with flux equal to the measured flux were injected into 150 nearby positions, and the flux was recovered at each location using PSF photometry. The standard deviation of these recovered fluxes was then taken as the uncertainty. To conduct more accurate differential photometry for the F150W and F444W bands, we incorporated NIRCam observations of the RX\,J2129 cluster field, which were obtained through the \textit{JWST} Treasury Survey, ``Vast Exploration for Nascent, Unexplored Sources'' (VENUS; Cycle 4, GO-6882, PIs S. Fujimoto \& D. Coe). These observations served as the ``templates'' for the F150W and F444W  imaging data. We created difference images by subtracting these template images from the NIRCam observations that we acquired in 2022 October. Figure~\ref{fig:new_obs} presents the difference images for the F150W and F444W bands, as well as the 2022 October and 2025 October observations of the SN\,2022riv location using the NIRCam F150W filter.
None of these difference images showed any evidence of late-time flux from the two earlier-arriving images of SN\,2022riv. The final photometric measurements, along with their associated uncertainties and details, are presented in Table~\ref{Photometry_hst_jwst}.
\begin{deluxetable}{lllc}
\setlength{\tabcolsep}{6.5pt}
\tablecaption{Photometry of SN\,2022riv} \label{Photometry_hst_jwst}
\tablehead{
\colhead{MJD} & \colhead{Instrument} & \colhead{Filter} & \colhead{$m_{\rm AB}$} \\ [-0.4cm]}
\startdata
59798.958	&	\textit{HST}-WFC3/IR	&	F110W	&	$	24.78	\pm	0.10	$  \\
59842.224	&	\textit{HST}-WFC3/IR	&	F160W	&	$	24.44	\pm	0.06	$  \\
59842.226	&	\textit{HST}-WFC3/IR	&	F110W	&	$	25.02	\pm	0.06	$  \\
59853.536	&	\textit{HST}-WFC3/IR	&	F110W	&	$	25.20	\pm	0.06	$  \\
59853.536	&	\textit{HST}-WFC3/IR	&	F160W	&	$	24.70	\pm	0.12	$  \\
59858.412	&	\textit{JWST}-NIRCam	&	F444W	&	$	25.01	\pm	0.04	$  \\
59858.412	&	\textit{JWST}-NIRCam	&	F115W	&	$	25.16	\pm	0.03	$  \\
59858.461	&	\textit{JWST}-NIRCam	&	F150W	&	$	24.47	\pm	0.01	$  \\
59858.461	&	\textit{JWST}-NIRCam	&	F356W	&	$	25.96	\pm	0.11	$  \\
59858.511	&	\textit{JWST}-NIRCam	&	F200W	&	$	24.74	\pm	0.03	$  \\
59858.511	&	\textit{JWST}-NIRCam	&	F277W	&	$	25.31	\pm	0.05	$  \\
59865.370	&	\textit{HST}-WFC3/IR	&	F160W	&	$	24.77	\pm	0.12	$  \\
59865.372	&	\textit{HST}-WFC3/IR	&	F110W	&	$	25.65	\pm	0.13	$  \\
59875.622	&	\textit{HST}-WFC3/IR	&	F160W	&	$	25.26	\pm	0.14	$  \\
59875.624	&	\textit{HST}-WFC3/IR	&	F110W	&	$	26.35	\pm	0.20	$  \\
59916.482	&	\textit{HST}-WFC3/IR	&	F160W	&	$	25.46	\pm	0.19	$  \\
59916.485	&	\textit{HST}-WFC3/IR	&	F110W	&	$	26.53	\pm	0.26	$  \\
59931.392	&	\textit{HST}-WFC3/IR	&	F160W	&	$	25.49	\pm	0.10	$  \\
59931.394	&	\textit{HST}-WFC3/IR	&	F110W	&	$	26.78	\pm	0.26	$  \\
\enddata 
\end{deluxetable}

\subsubsection{Classification with \texttt{sncosmo} and \texttt{StarDust2}}

After performing photometry on the observations of SN\,2022riv as outlined above, we independently analyzed the photometric data using a Bayesian photometric classifier. This analysis provides independent confirmation of the classification of SN\,2022riv obtained with \texttt{SNID}. We set the redshift to $z=1.522 \pm 0.0001$ and utilize the \texttt{sncosmo} software package \citep{barbary_2024_14025775} to simulate SN light curves. The classification probability was evaluated with the \texttt{StarDust2} software \citep{satrdust_2014}, which employs traditional Bayesian model selection \citep[see, e.g.,][]{2013ApJ...768..166J, satrdust_2014, 2014ApJ...783...28G}.

For this analysis, we represented SNe Ia with 15 templates from the SuperNova Analysis software \citep[\texttt{SNANA;}][]{SNANA}. The likelihoods were calculated by comparing observed fluxes with model predictions across all defined passbands. The photometric classification yielded a probability of $p(\text{Ia}|\text{\textbf{D}}) = 1.0$, with the SALT3-NIR model providing the best fit for an SN~Ia light curve. 
We note that $p(\text{Ia}|\text{\textbf{D}})$, as reported by \texttt{StarDust2}, is a Bayesian classification probability derived from the relative evidence of the competing SN classes included in the built-in template library available in \texttt{sncosmo}. Therefore, it should not be interpreted as an absolute goodness-of-fit statistic. The reported probability is conditional on the set of templates considered and quantifies the relative preference of the data for the Type Ia model compared to all core-collapse SN subclasses.

\begin{figure*}
\centering
\includegraphics[width=0.75\textwidth]{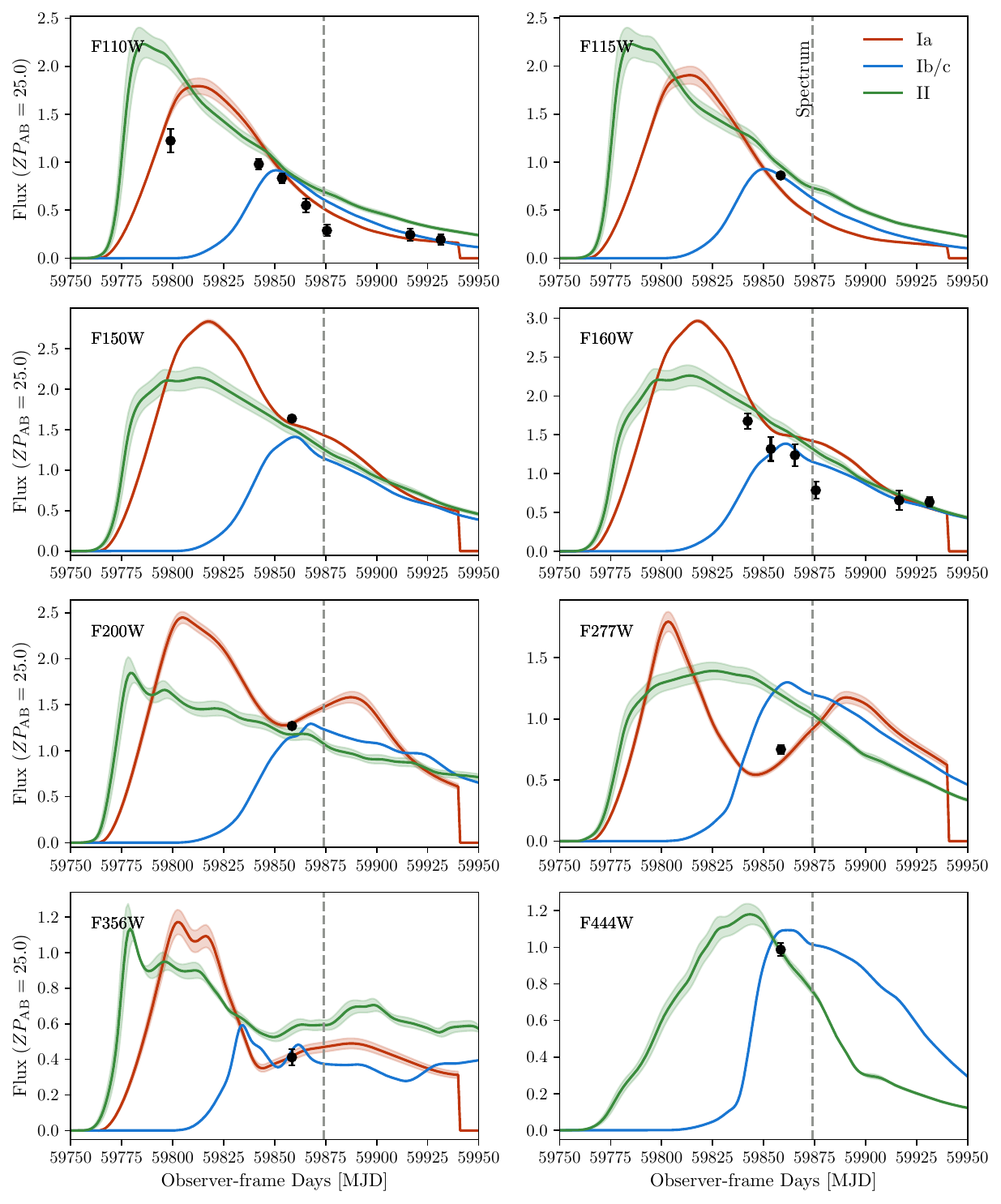}
\caption{Maximum-likelihood light curves for each SN subclass obtained using Bayesian model selection based on the photometric data from \textit{HST} and \textit{JWST}. The black points with error bars indicate the observed photometry of SN\,2022riv. The SN~Ia model, shown by a red solid line, is based on the SALT3-NIR template at $z = 1.522$, with the F444W observation excluded from the plot since it falls outside the model’s wavelength range. The best fit for SNe~Ib/c is SN\,2004ib (blue), while the best fit for SNe~II is SN\,2007kn (green). The date of the \textit{JWST} spectral observations is indicated by a dashed gray vertical line. The color bands represent the $1\sigma$ uncertainty. SN~Ia is the most probable type based on light-curve fitting, while the light curves for the SN~Ib/c and SN~II models poorly match the data.
\label{fig:photometry_SALT3}}
\end{figure*}

Figure~\ref{fig:photometry_SALT3} presents the best-fit SALT3-NIR model \citep{SALT3-NIR} for an SN~Ia, alongside the best-fit models for SNe Ib, Ic, II. F444W observations of SN\,2022riv were excluded from this analysis because they were too red for the models at $z=1.522$, falling outside the SALT3-NIR model's wavelength range. Although the best-fitting SALT3-NIR model does not perfectly reproduce every photometric measurement, the Bayesian evidence strongly favors the SN Ia model ($\log Z_{\rm Ia}=-90.6$) over the SN II ($\log Z_{\rm II}=-221.7$) and SN Ib/c ($\log Z_{\rm Ib/c}=-348.5$) models, with differences in log-evidence of $\sim131$ and $\sim258$, respectively, resulting in $p(\text{Ia}|\text{\textbf{D}}) = 1.0$. The time of peak brightness for the SALT3-NIR model is constrained to $t_{0} = 59813.0^{+1.7}_{-1.8} \,\rm{MJD}$, indicating that the \textit{JWST} G140M/100LP and PRISM spectroscopic data were collected about 24.2 rest-frame days after the peak, in excellent agreement with the results of our spectroscopic analysis.

\subsubsection{Distance Modulus and Magnification of SN\,2022riv}

\begin{figure*}
\centering
\includegraphics[width=0.85\textwidth]{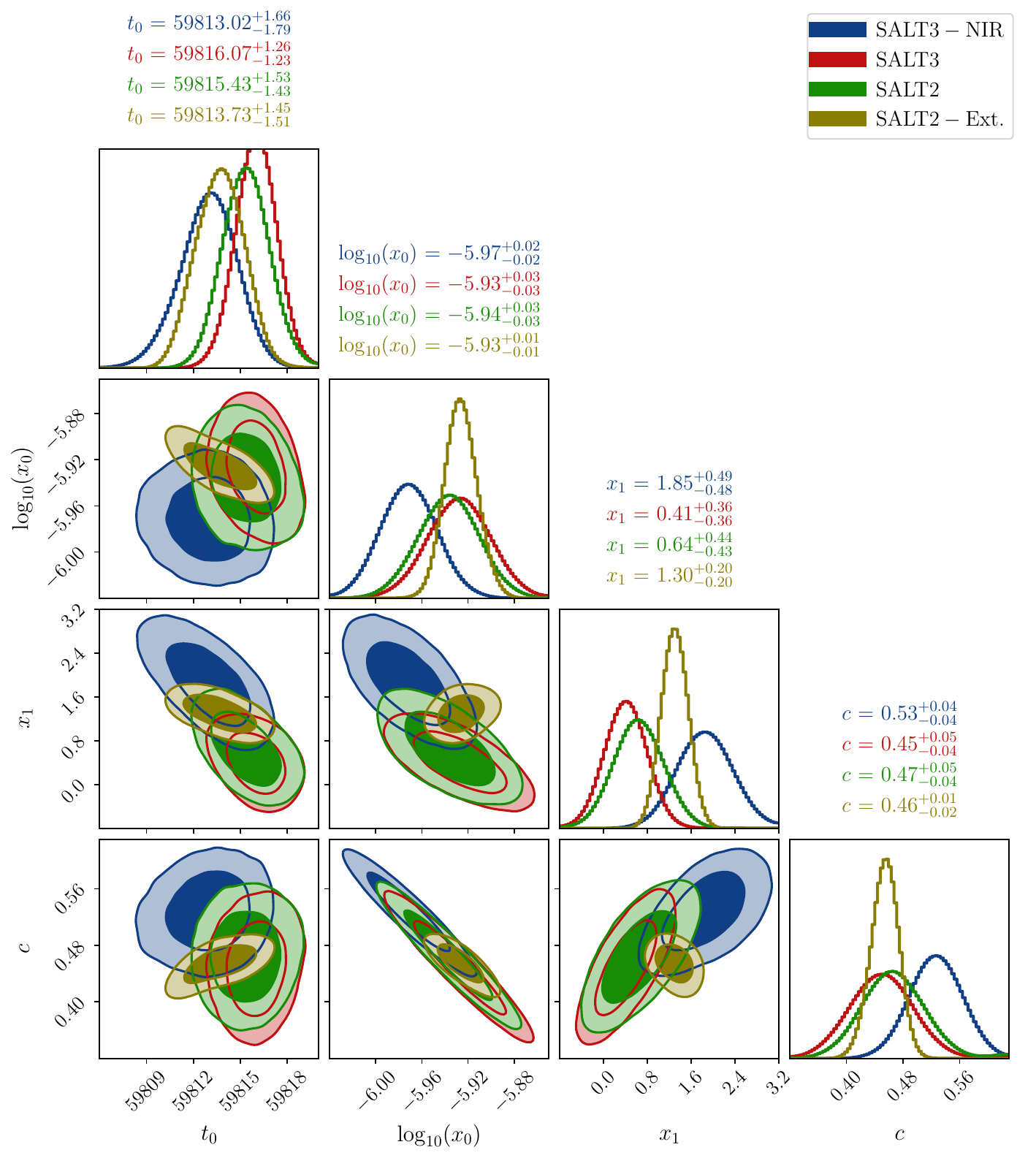}
\caption{Joint posterior distribution of the SALT light-curve parameters $t_0$, $x_0$, $x_1$, and $c$ obtained by fitting the light curves of SN\,2022riv using four different SALT models in \texttt{sncosmo}, with key parameters and Bayesian evidence estimated via a nested sampling algorithm. The contours represent the 68\% and 95\% confidence intervals.
\label{fig:corner_SALT}}
\end{figure*}

With the object securely confirmed as an SN Ia at $z = 1.522$, we then used the SALT models \citep{SALT2, SAT2_Ext, SALT3, SALT3-NIR} to fit the light curves of SN\,2022riv utilizing time-series photometry. We used \texttt{sncosmo} to constrain the light-curve parameters and compute Bayesian evidence via a nested sampling algorithm. These include $x_0$, which represents the overall flux normalization and captures the peak flux in the $B$ band; $x_1$, a time-scaling factor indicating the stretch of the light curve (with smaller values corresponding to faster-evolving light curves); and $c$, which characterizes the SN's color, where larger values indicate a red rest-frame color. During the SALT model fitting, SALT3-NIR does not use F444W photometry, since it falls outside of the model's wavelength range. SALT3-NIR models the F277W and F356W photometry, while SALT3 does not use those filters. SALT2-Extended fits photometry acquired in all filters, while SALT2 does not include the F277W, F356W, or F444W long-wavelength observations. Table~\ref{SALT_bounds} summarizes the bounds and values of the free and fixed SALT light-curve parameters used in the fitting process, while Figure~\ref{fig:corner_SALT} displays the joint posterior distributions obtained from the fitting with a nested sampling algorithm. Table~\ref{SALT_results} provides the best-fit values with associated uncertainties for each SALT model.

\begin{deluxetable}{ccc}
\setlength{\tabcolsep}{9.5pt}
\tablecaption{Summary of the SALT light-curve model parameters used in fitting SN\,2022riv.} \label{SALT_bounds}
\tablehead{
Parameter & Free/ Fixed & Bounds/ Values
}
\startdata
$z$ & Fixed & 1.522 \\
$t_0$ & Free & $[59778.96, \,\, 59951.39]$ \\
$x_0$ & Free & $[0, \,\, 1]$ \\
$x_1$ & Free & $[-5.0, \,\, 5.0]$ \\
$c$ & Free & $[-0.5, \,\, 5.0]$
\enddata
\end{deluxetable}

These fitted parameters enable us to derive the distance modulus (Dist. Mod.) using the \citet{Tripp_1998} equation, expressed as 
\begin{equation}
    {\rm Dist. \,\, Mod.}= -2.5\log_{10}(x_0) + \alpha x_1 -\beta c - M_{\rm 0}\, ,
\end{equation} where $M_0$ is the absolute magnitude of an SN Ia with $c=x_1=0$, and the global nuisance parameters $\alpha$ and $\beta$ describe the stretch–luminosity and color–luminosity relationships, respectively. 
In this study, we adopt $\alpha = 0.154 \pm 0.006$ and $\beta = 3.02 \pm 0.06$ from \cite{2018ApJ...859..101S}, derived for the Pantheon sample, with an intrinsic scatter of $\sigma_{\rm int}=0.09$ based on the Gaussian intrinsic scatter model of SNe Ia from \cite{2010A&A...523A...7G}. The final uncertainty is calculated by adding $\sigma_{\rm int}$ in quadrature to the quadrature of the statistical uncertainty, which accounts for uncertainties from both the observational data and the modeling process.
As discussed below, we consider the relative difference between the distance moduli of lensed and unlensed SNe at the same redshift; therefore, we ignore corrections for selection bias \citep{2017ApJ...836...56K}. 
For the same reason, we also do not include a correction for the dependence of corrected SN Ia luminosities on host-galaxy stellar mass \citep[e.g.,][]{2010ApJ...715..743K, 2010MNRAS.406..782S, 2010ApJ...722..566L}. 

\begin{figure*}[th!]
\centering
\includegraphics[clip, trim=0cm 0cm 0cm 0cm, width=1.0\textwidth]{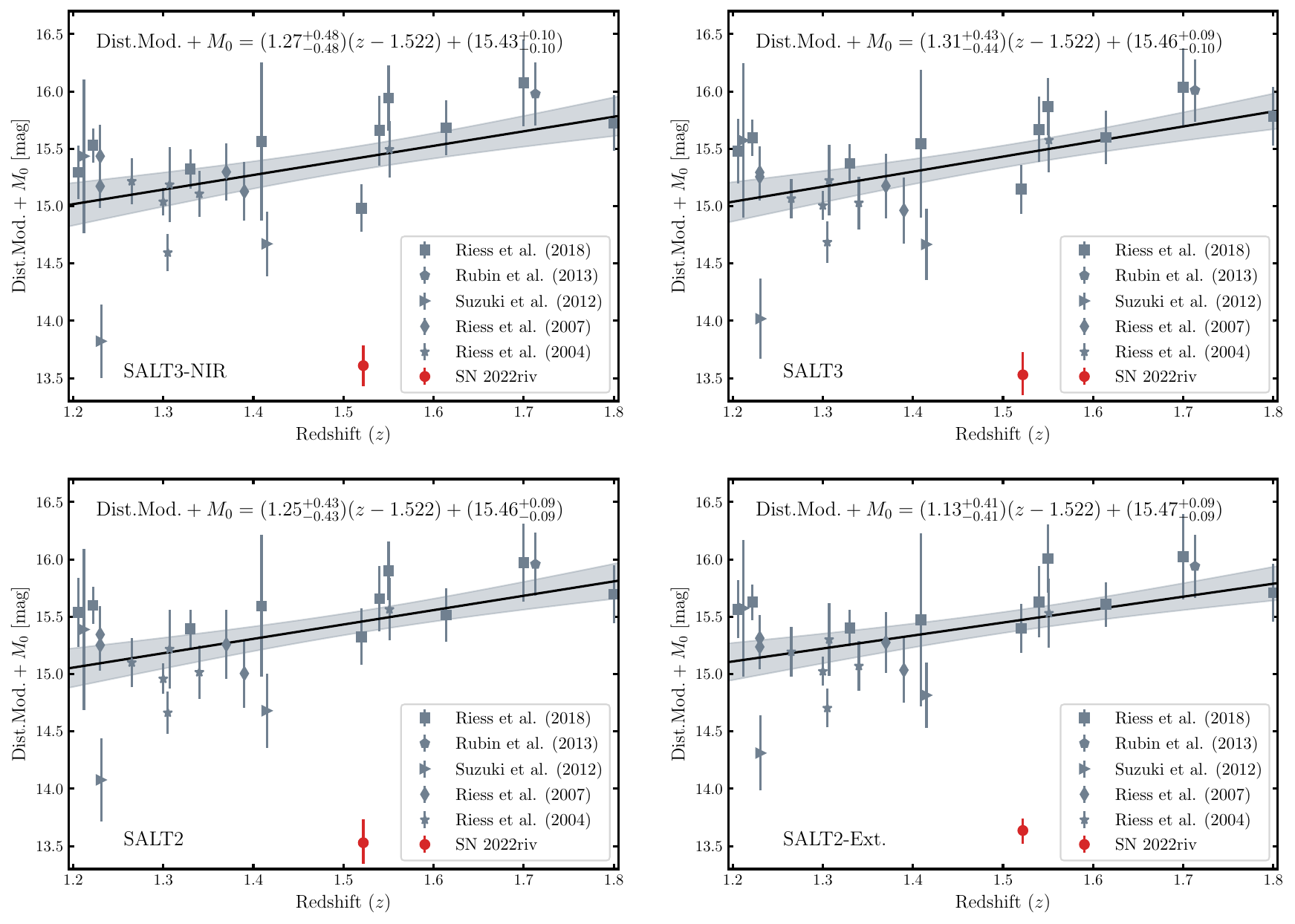}
\caption{Comparison of the inferred distance modulus of SN\,2022riv (red data points) with the distance moduli for our sample of unlensed field SNe (gray data points) for each SALT light curve fitter. The black line, anchored at the redshift of SN\,2022riv, represents the best linear fit to the unlensed control sample, with the gray-shaded region indicating the associated $1\sigma$ error. The best-fit linear relationships for each SALT model are provided at the top of each panel. 
\label{salt_dm_plots}}
\end{figure*}

To conduct a cosmology-independent inference about the magnification ($\mu$) for each SALT fitter, we selected a sample of 24 spectroscopically confirmed, unlensed SNe~Ia within the redshift range $1.2 < z < 1.8$. This sample includes 10 SNe from \cite{2018ApJ...853..126R}, one SN from \cite{2013ApJ...763...35R}, three SNe from \cite{2012ApJ...746...85S}, four SNe from \cite{2007ApJ...659...98R}, and six SNe from \cite{2004ApJ...607..665R}. The distance moduli for each SN in this control sample, computed using SALT models, were fitted to a linear relationship using the weighted least squares fitting method, as illustrated in Figure~\ref{salt_dm_plots}, where the relationship between distance modulus and redshift is highlighted at the top of each panel. The difference in distance modulus between a lensed and an unlensed SN at the same redshift is 
\begin{equation}
    {\rm Dist. \,\, Mod.}_{\rm unlensed} - {\rm Dist. \,\, Mod.}_{\rm lensed} = 2.5\log_{10}(\mu).
\end{equation} 

The magnifications derived from the SALT light-curve models are summarized in Table~\ref{SALT_results}. For a concordance cosmology described by the Lambda cold dark matter ($\Lambda$CDM) model with $\Omega_{\rm m}=0.3$, $\Omega_{\Lambda}=0.7$, and $H_0 = 70 \,{\rm km\,s^{-1}\, Mpc^{-1}}$, we found 
${\rm Dist. \,\, Mod.}+M_0=15.48$\,mag at $z=1.522$ (a $\leq0.5\sigma$ difference for all four SALT models). The cosmology-dependent magnification measurements computed using this value are also listed in Table~\ref{SALT_results}. The absolute magnifications inferred by considering a specific cosmology and those obtained without assuming any cosmological model are in excellent agreement, with a statistical tension of $\sim0.2\sigma$ or smaller for all SALT SN~Ia spectral time-series models. In addition, we note that although the model wavelength ranges differ among the various SALT SN spectral time-series models considered here, they yield consistent constraints on the magnification.

In the following sections, we present the magnifications predicted by six independent lens models of the galaxy cluster RX\,J2129 and discuss potential effects on the magnifications of SN\,2022riv due to microlensing and millilensing. Generally, stellar microlenses have Einstein radii on the same order as the size of the SN photosphere ($\sim 10^{14}$--$10^{15}$\,cm). As the photosphere expands and crosses a microlensing caustic, different regions of the photosphere experience different magnification. Because the intensity profile and effective size of the SN photosphere vary with wavelength, the microlensing magnification is chromatic, producing wavelength-dependent variations in the SN spectrum \citep{2018ApJ...855...22G, 2018MNRAS.478.5081F, 2019A&A...631A.161H}. These variations impact both the magnification and the measured time delays of the lensed SN images because microlensing induces time-dependent distortions in the observed SN light curves. In contrast, dark matter substructure millilenses primarily influence the magnification of the SN, since the photosphere is much smaller than the size of a millilensing caustic. A companion paper (Dalrymple et al.; in preparation) will provide constraints on the relative time delay of the image that arrived earlier, incorporating the effects of microlensing on the time delays.

\renewcommand{\arraystretch}{1.1}
\begin{deluxetable*}{c|c|c|c|c}
\setlength{\tabcolsep}{0.8pt}
\tablecaption{Best-fit parameters and magnifications derived from the SALT light curve models for SN\,2022riv. } \label{SALT_results}
\tablehead{
\colhead{Parameter} & \colhead{SALT3-NIR} & \colhead{SALT3} & \colhead{SALT2} & \colhead{SALT2-Extended}\\ [-0.4cm]
}
\startdata
\multirow{3}{*}{Bands used}  & F110W, F115W, F150W,  & F110W, F115W, F150W,  & F110W, F115W, F150W, & F110W, F115W, F150W,  \\
& F160W, F200W, F277W, & F160W, F200W & F160W, F200W &  F160W, F200W, F277W, \\
& F356W & & & F356W, F444W\\
\hline
$t_{0}$ [MJD] & $59813.0^{+1.7}_{-1.8}$ & $59816.1^{+1.3}_{-1.2}$ & $59815.4^{+1.5}_{-1.4}$ & $59813.7^{+1.5}_{-1.5}$\\
$x_1$ & $1.85^{+0.49}_{-0.48}$ & $0.41^{+0.36}_{-0.36}$ & $0.64^{+0.44}_{-0.43}$ & $1.30^{+0.20}_{-0.20}$\\
$c$ & $0.53^{+0.04}_{-0.04}$ & $0.45^{+0.05}_{-0.04}$ & $0.47^{+0.05}_{-0.04}$ & $0.46^{+0.01}_{-0.02}$\\
$\log_{10}(x_0)$ & $-5.97^{+0.02}_{-0.02}$ & $-5.93^{+0.03}_{-0.03}$ & $-5.94^{+0.03}_{-0.03}$ & $-5.93^{+0.01}_{-0.01}$  \\
\hline 
Dist. Mod.$+ M_0$ [mag] & $13.61^{+0.18}_{-0.18}$ & $13.53^{+0.20}_{-0.18}$ & $13.53^{+0.20}_{-0.18}$ & $13.64^{+0.11}_{-0.12}$ \\
\hline
& \multicolumn{4}{c}{Control Sample Comparison}\\
\cline{2-5}\\[-0.45cm]
$\mu$ & $5.35^{+1.01}_{-1.01}$ & $5.92^{+1.22}_{-1.09}$ & $5.92^{+1.22}_{-1.11}$ & $5.42^{+0.70}_{-0.74}$ \\
\hline
& \multicolumn{4}{c}{Comparison with $\Lambda$CDM Model}\\
\cline{2-5}\\[-0.45cm]
$\mu$ & $5.60^{+0.92}_{-0.92}$ & $6.03^{+1.12}_{-1.00}$ & $6.03^{+1.14}_{-1.02}$ & $5.47^{+0.54}_{-0.60}$ \\
\enddata
\end{deluxetable*}

\vspace{-0.5cm}

\section{Lens Model Predictions}\label{lens_models}

In this section, we report the magnification predicted by six independent lens models for the RX\,J2129 galaxy cluster and explore how microlensing and millilensing could impact the magnification of SN\,2022riv.

\subsection{Lens Models}

Six independent lens modeling teams analyzed the RX\,J2129 galaxy cluster. They derived model predictions for the magnifications, time delays, and image positions of SN\,2022riv. To ensure consistency across all models, we placed all lensed images on a common astrometric grid, defined by our coadditions of the {\it JWST}/NIRCam imaging, which were aligned to Gaia Data Release 3 \citep{GAIA_DR3}. \cite{caminharosatigrillo19} and \cite{jauzac21} reported seven spectroscopically confirmed strongly lensed multiple-image systems in the RX\,J2129 cluster, using data from the CLASH survey. Expanding on this, \cite{williamskellychen23} identified a strongly lensed, compact galaxy at $z=9.51$ using imaging data obtained with NIRCam.

For the seven systems (designated Systems 1--7) originally reported by \cite{caminharosatigrillo19}, we revised the image positions based on high-resolution \textit{JWST}'s NIRCam observation with F150W filter. We adopted spectroscopic redshifts obtained from the Multi Unit Spectroscopic Explorer \citep[MUSE;][]{2014Msngr.157...13B} observations, as summarized in Table~\ref{Systems_RA_and_dec}. Notably, Systems 6 and 7 were not clearly detected in the NIRCam observations. Therefore, to refine their positions, we employed \texttt{SExtractor} \citep{1996A&AS..117..393B} to detect neighboring sources in both \textit{HST}-CLASH survey WFC3/IR F140W and NIRCam F150W observations. We then calculated median offsets in R.A. and decl. based on the matched positions of these nearby sources. The revised positions of Systems 6 and 7 were estimated by applying the median astrometric offsets to the original coordinates reported by \cite{caminharosatigrillo19}. 

Similarly, we revised the position of the high-redshift lensed galaxy (System G) using updated astrometry from our NIRCam coadditions of the galaxy cluster field and adopted the redshift derived from \textit{JWST}’s NIRSpec MOS observations reported by \cite{williamskellychen23}. In addition to these previously known eight systems, we identified three new candidate multiple-image systems, clearly detected in the NIRCam imaging. These are labeled as S3-(1--2)a, S3-(1--2)b, and S3-(1--2)c in Figure~\ref{cutouts_11_S}, with their positions summarized in Table~\ref{Systems_RA_and_dec} in Appendix~\ref{lensedImages}. We interpreted these as components of System 3 and adopt the spectroscopic redshift of $z=1.5194$, as measured by MUSE for this system. With these updated image positions and precise spectroscopic redshifts, all six lens modeling teams agreed on a consistent set of 11 systems, as listed in Table~\ref{Systems_RA_and_dec}.

Additionally, we report three prospective image systems which we did not include as lens model constraints given their uncertain identification. These systems are listed in Table~\ref{fuzzy_systems}, and their cutouts are shown in Figure~\ref{new_systems_invisible}, with dashed circles indicating their inferred positions. Among these three, System 11 is well localized in the NIRCam data, though no spectroscopic redshift is currently available, so it is listed as a prospective candidate. Systems 10 and 12 have faint, extended morphologies that complicate secure identification in the NIRCam images. While these systems were not used by our lens modeling teams, they could be included in future modeling efforts once precise astrometry and spectroscopic redshifts are available.

All six modeling teams assumed a fiducial cosmology with $\Omega_{\rm m}=0.3$, $\Omega_{\Lambda}=0.7$, and $H_0 = 70 \,{\rm km\,s^{-1}\, Mpc^{-1}}$, and conducted a double-blind analysis to ensure that no team had prior knowledge of the absolute magnifications from the analysis of photometry or magnifications, relative time delays, or image positions predicted by the other teams. A brief summary of the six independent lens modeling approaches is provided below.

\subsubsection{{\tt GLAFIC}}
This mass model is a simply parametrized mass model constructed with the {\tt GLAFIC} software \citep{oguri10,2021PASP..133g4504O}. The lens potential was described by a superposition of a halo component modeled by an elliptical Navarro-Frenk-White \citep[NFW;][]{NFW} density profile, cluster member galaxies modeled by elliptical pseudo-Jaffe profiles with their velocity dispersions and the truncation radii scaled with their luminosities, and the external shear. The centroid position of the halo component was kept as a free parameter. Assuming the positional error of $0\farcs4$, the best-fitting mass model had a $\chi^2$ value of $19$ for 29 degrees of freedom. The small reduced $\chi^2$ results from our conservative assumption of a positional error of $0\farcs4$, which is much larger than the measurement errors in order to account for, e.g., substructures in the dark matter halo. The rms of differences between observed and model-predicted multiple image positions was $0\farcs39$. Uncertainties in model parameters were derived by the Markov Chain Monte Carlo (MCMC) method.

\subsubsection{\tt Chen2020}
This lens model algorithm was originally developed for blind prediction of time delays between SN\,Refsdal images \citep{2020RNAAS...4..215C}. As a nonparametric lens model, it assumes that the cluster mass is smoothly distributed across the field, with a characteristic scale much larger than that of an individual galaxy halo, and that the cluster mass vanishes at sufficiently large distances. A perturbation function was introduced to deform the two-dimensional gravitational potential generated by the cluster. Using this function, a global numerical perturbation was applied with respect to the positions of the multiple images, such that the inferred source-plane positions from each set of observed multiple images converged to a single location in the source plane. A large number of perturbations were performed, driven by an iterative procedure that began from an initial condition in which each cluster member was assigned a spherically symmetric NFW halo. The iteration converged on a solution for the gravitational potential of the cluster that successfully reproduced all the strongly lensed images. The initial condition no longer held after a number of iterations of the global perturbation function. Therefore, the final solution of the cluster-scale mass distribution was only weakly dependent on the galactic halo properties.

\subsubsection{\tt HoliGRALE}
This is the first official lens model of a real system using \texttt{HoliGRALE} (Hybrid, only light included, GRAvitational LEnsing). \texttt{HoliGRALE} is a hybrid modeling paradigm and extension of \texttt{GRALE} \citep{liesenborgs06,liesenborgs07,liesenborgs20} that incorporates parametric lens components within the traditional free-form genetic algorithm \citep[see][for further details]{liesenborgs06}. Prototypical versions of \texttt{HoliGRALE} have previously been used to model cluster lenses \citep{perera24, perera25}, and a forthcoming paper (Perera et al., in preparation) will rigorously describe the methodology of \texttt{HoliGRALE} and its systematics. Here, we summarize the basic approach. 

First, following the standard free-form approach of \texttt{GRALE}, we define an initial mass basis on a grid consisting of projected Plummer spheres, which have projected surface mass densities of
\begin{equation}\label{eq:plummerdens}
    \Sigma(\boldsymbol{\theta}) = \frac{M}{\pi D_{\mathrm{d}}^2}\frac{\theta_\mathrm{P}^2}{(\boldsymbol{\theta}^2 + \theta_\mathrm{P}^2)^2}\, ,
\end{equation}
where $\theta_\mathrm{P}$ is the characteristic angular width of the Plummer sphere, $D_{\rm d}$ is the angular diameter distance between the lens and the observer, and $M$ is the total mass. Singular isothermal ellipsoids (SIEs) are then added to this initial basis at the locations of cluster member galaxies. For the SIEs, we first estimate their effective radii using the Kormendy relation \citep{tortorelli18}, thus establishing the aperture at which the velocity dispersion is calculated. The initial velocity dispersion is then determined by the fundamental plane of elliptical galaxies \citep{djorgovski87}. We emphasize that this initial value is not so important because \texttt{HoliGRALE} will accordingly rescale each basis function's weight with the genetic algorithm. The addition of observationally motivated SIE basis functions renders this modeling paradigm a hybrid union of parametric and free-form methods.

\texttt{HoliGRALE} uses a genetic algorithm to optimize the weights of each basis function component based on two fitness measures \citep[see][for further details]{zitrin10}: (1) disfavoring iterations which produce images that map to the null space (source-plane area where no images form), and (2) images of the same source which are projected to the source plane and more heavily weighted based on their degree of overlap. Once optimized, the grid of Plummers is subdivided further based on regions exhibiting greater density, while the SIEs are held constant with weight preserved. This entire process continued for 10 subdivision steps, each consisting of 100--200 additional subdivisions (depending on the fitness at that stage), yielding $\sim 1000$ Plummer spheres in the final model. We ran this 40 times and took the average lens potential to be the complete model. Source positions were optimized a posteriori according to the statistical methodology outlined by \cite{perera24}. Time delay, magnification, and image-position predictions were calculated from the complete model with optimized source positions, with uncertainties calculated from bootstrapping the sample of 40 runs \citep[see Appendix B of ][]{perera25}. The rms of differences between observed and model-predicted multiple image positions was $0\farcs09$. 


\subsubsection{\tt WSLAP+}
This is considered a hybrid-class method where a free-form parameterization is used to describe the smooth, or large-scale distribution of mass, and a compact component is used to describe the small-scale contribution from member galaxies. For the small-scale component, we directly used the distribution of light of member galaxies as observed by {\it JWST}. The member galaxies were distributed in layers, with each layer having a constant mass-to-light ratio. For this particular lens model, we used four layers. Layers 1 and 2 contained the central brightest cluster galaxy (BCG) and the second-largest galaxy, respectively. Layer 2 contained a member galaxy near image S3. Layer 4 consisted of all remaining member galaxies. For the smooth component, we used an adaptive grid, in which a Gaussian was placed at each grid. The full width at half-maximum (FWHM) intensity of the Gaussian functions was smaller in regions with the highest mass at the center of the lens. Conversely, the FWHM of these Gaussians was larger in the outskirts of the lens model.

The adaptive grid contained 340 grid points with their distribution determined after a first model (with a regular grid) was obtained.  The model was optimized by minimizing a quadratic scalar function that contains the square of the residual between the observed and predicted positions. The optimal model was obtained under the constraint that all masses in the Gaussian functions and in the four layers must be positive. Owing to the quadratic nature of the function being optimized, a solution can be obtained in $\sim 1$ minute when the algorithm is run on a commercial laptop. This method is described in detail by \citet{2005MNRAS.360..477D,WSLAP2,WSLAP3}.
    
\subsubsection{\tt Zitrin-Analytic}
This lens model was reconstructed using a revised version of the parametric approach described in \cite{2015ApJ...801...44Z}, often dubbed \texttt{Zitrin-Analytic}. This method is similar in nature to other common parametric techniques~\citep[e.g.,][]{jullokneiblimousin07, oguri10} and has been used to model a large number of clusters in \textit{HST} and \textit{JWST} data~\citep[e.g.,][]{2022ApJ...938L...6P, 2023MNRAS.523.4568F, 2024MNRAS.533.2242F, 2023ApJ...944L...6M, 2026ApJ..1005..215A}. The model employs two main mass components: pseudoisothermal elliptical mass distribution (PIEMD) halos to model dark matter, and dual pseudoisothermal ellipsoid (dPIE) halos to model cluster galaxies, scaled by common scaling relations. The various lensing quantities are directly calculated at the observed image positions, followed by a~$\chi^2$ MCMC minimization in the source plane to find the best-fit lens model. All resulting quantities are then cited in the image plane.

For RX\,J2129, we placed a single dark-matter halo roughly centered on the BCG, with its position allowed to vary within $(-1\farcs3, 5\farcs2)$ and $(-3\farcs9, 1\farcs3)$ relative to the BCG center in the $x$ (EW) and $y$ (NS) directions, respectively. We also left the BCG parameters (i.e., weight, ellipticity, position angle, and core radius) free. In addition, we left the weights of four cluster galaxies free. The (R.A., decl.) coordinates (in degrees) of these galaxies were (322.4195283, 0.0891626), (322.4063092, 0.0829271), (322.4132154, 0.0860124), and (322.4172344, 0.0905544). We also allowed the power-law indices of the scaling relations to vary. No external shear was included in the model. We assumed positional uncertainties of $0\farcs5$ for all lensed images except those belonging to systems 3, 4, and G in Table~\ref{tab:images}. For these systems, we assumed a positional uncertainty of $0\farcs1$. With these choices, the lens model had 17 free parameters.

To optimize the lens model, we first performed a ``burn-in'' stage in which 100 independent chains were run to explore the parameter space. The final model and corresponding uncertainties are then obtained from an MCMC run with annealing, of approximately~$5\times 10^5$ steps. The final lens model has an image-reproduction rms, in the image plane, of 0\farcs2. The corresponding $\chi^2$ is 6.9. Since we have 31 lensed images from 11 sources and 17 free parameters in the lens model, we have 23 degrees of freedom in total. Hence, the resulting reduced$-\chi^2$ is 0.3. This overfitting stems from our relatively large error bars of $0\farcs5$ associated with image positions. If we were to use positional errors of $0\farcs2$ for all lensed images, we would obtain a reduced$-\chi^2$ of $\sim 1$.

\subsubsection{\tt LENSTOOL}

This lens model is based on the model presented by \cite{jauzac21} which used the parametric approach in \texttt{LENSTOOL} algorithm \citep{jullokneiblimousin07}. \texttt{LENSTOOL} uses an MCMC process to sample the posterior density of the assumed model, expressed as a function of the likelihood of the model as defined in \cite{jullokneiblimousin07}. \texttt{LENSTOOL} minimizes the $\chi^{2}$ defined as: 
\begin{eqnarray}
\chi^{2} = \sum\limits_{i} \chi_{i}^{2}\ ,
\end{eqnarray}
where
\begin{eqnarray}
\chi_{i}^{2} = \sum_{j=1}^{n_{i}} \frac{(\theta^{j}_{\rm obs} - \theta^{j}(\textbf{p}))^{2}}{\sigma_{ij}^{2}}\ .
\end{eqnarray}
$\theta^{j}_{\rm obs}$ is the vector position of the observed multiple image $j$, $\theta^{j}$ is the predicted vector position of image $j$, $n_{i}$ is the number of images in System $i$, and $\sigma_{ij}$ is the error on the position of image $j$ \citep[fixed at $\sim 0\farcs5$ for multiple images to account for errors on image positions, unaccounted dark matter overdensities within the cluster mass distribution, and line-of-sight effects as described in][]{jullokneiblimousin07, jullo10}. The most likely model minimizes the distance between the observed and predicted multiple images, i.e., the rms. \texttt{LENSTOOL} is publicly available and has been used broadly within the community to model cluster strong lenses using both \emph{HST} and \emph{JWST} observations \citep[see e.g.,][]{richard10, sharonjohnson15, jauzacrichardlimousin16, 2023A&A...670A..60B, mahler23, sharon23, lagattuta23, beauchesne24, acebron+25}.

The updated \texttt{LENSTOOL} model of RX\,J2129 included one cluster-scale halo to model the overall cluster potential modeled as a PIEMD. In addition, to account for secondary lensing effects due to the presence of cluster member galaxies, we included 70 galaxy-scale halos also modeled as PIEMD. However, in order to avoid an underconstrained model, 68 galaxy-scale halos were optimized using the \cite{FJ76} scaling relation. With this, we assumed that the luminosity of a galaxy is a good tracer of its mass. Their positions and ellipticities were thus fixed to those of their luminous counterparts and their velocity dispersions, core radii, and cut radii were then measured through rescaling to those of a reference galaxy with luminosity $L^{\ast}$. The BCG and an isolated cluster galaxy, which acts as the lens for one of our multiple-image systems, were modeled as independent PIEMDs.

The mass model was then constrained by 31 spectroscopically confirmed multiple images belonging to the 11 systems mentioned earlier.
We note that 10 multiple images (corresponding to 4 systems) mapped the structure of the SN host galaxy. While lying in the same redshift plane, fitting multiple emission knots of one galaxy constrains the gradient of the lensing potential, and therefore, the shear and magnification of the host galaxy. The RMS of our best-fit mass model was $0\farcs32$.

\begin{deluxetable*}{l|
    >{\raggedleft\arraybackslash}p{2.5cm}|
    >{\raggedleft\arraybackslash}p{2.5cm}|
    >{\raggedleft\arraybackslash}p{3.1cm}|
    >{\raggedleft\arraybackslash}p{2.5cm}|
    >{\raggedleft\arraybackslash}p{2.5cm}
}
\setlength{\tabcolsep}{5.4pt}
\tablecaption{Lens modeling constraints on the magnifications ($\mu_{\rm model}$) of the lensed SN\,2022riv images.} \label{tab:mu_model}
\tablehead{
\multirow{2}{*}{Model} & \multicolumn{5}{c}{$\mu_{\rm model}$}\\
\cline{2-6}\\[-0.45cm]
& \colhead{Image S1} & \colhead{Image S2} & \colhead{Image S3} & \colhead{Image S4} & \colhead{Image S5} \\[-0.55cm]}
\startdata
\texttt{GLAFIC} 	 & $4.01^{+0.33}_{-0.31}$ &	 $4.97^{+0.52}_{-0.50}$ &	 $-4.83^{+0.50}_{-0.45}$ ($0.5\sigma$) &	 -- & --\\
\texttt{Chen2020} 	 & $2.60\pm0.13$ 	& $4.46\pm0.18$ &	 $-6.62\pm0.76$ ($1.0\sigma$)	& $-0.85\pm0.03$& --\\
\texttt{HoliGRALE} 	 & $4.60\pm0.05$ 	& $6.04\pm0.08$ 	& $-15.39\pm0.85$ ($7.6\sigma$) &	 -- & --\\
\texttt{WSLAP+} 	 & $4.35\pm0.17$ 	& $5.85\pm0.32$ 	& $-5.32\pm0.16$ ($0.03\sigma$) &	 --& --\\
\texttt{Zitrin-Analytic} 	 & $3.49^{+0.05}_{-0.07}$ &	 $4.71_{-0.07}^{+0.10}$ 	& $-6.16_{-0.27}^{+0.45}$ ($0.8\sigma$) &	 --& --\\
\texttt{LENSTOOL} 	 & $3.80\pm0.30$ 	& $5.10\pm0.40$ &	 $-4.20\pm0.50$ ($1.0\sigma$) &	 $-1.90\pm0.3$& $0.010\pm0.001$\\
\enddata
\tablecomments{The values in parentheses in the Image S3 column indicate the statistical tension of the absolute magnification derived from fitting the light curve of SN\,2022riv using the SALT3-NIR model with $\mu_{\rm model}$. Note that $\mu_{\rm model}$ represents macromodel predictions that do not account for the effects of stellar microlensing or dark matter substructure.
}
\end{deluxetable*}

\begin{deluxetable*}{l|
>{\raggedleft\arraybackslash}p{1.7cm}
>{\raggedleft\arraybackslash}p{1.7cm}|
>{\raggedleft\arraybackslash}p{1.7cm}
>{\raggedleft\arraybackslash}p{1.7cm}|
>{\raggedleft\arraybackslash}p{1.7cm}
>{\raggedleft\arraybackslash}p{1.7cm}|
>{\raggedleft\arraybackslash}p{1.7cm}
>{\raggedleft\arraybackslash}p{1.7cm}
}
\setlength{\tabcolsep}{2.8pt}
\tablecaption{Predicted relative time delays and positions of SN\,2022riv for each image, based on the cluster lens models.} \label{tab:microlensing}
\tablehead{
\multirow{2}{*}{Model} & \multicolumn{8}{c}{$dt_{\rm model}$ [days]}\\
\cline{2-9}\\[-0.45cm]
& \multicolumn{2}{c|}{Image S1} & \multicolumn{2}{c|}{Image S2} & \multicolumn{2}{c|}{Image S4} & \multicolumn{2}{c}{Image S5} \\[-0.55cm]}
\startdata
\texttt{GLAFIC} &	 \multicolumn{2}{r|}{$-976^{+69}_{-74}$} &	 \multicolumn{2}{r|}{$-361^{+41}_{-46}$} &	 \multicolumn{2}{r|}{--} & \multicolumn{2}{r}{--} \\
\texttt{Chen2020} 	&  \multicolumn{2}{r|}{$-1000\pm7$} 	&  \multicolumn{2}{r|}{$-383\pm12$} 	& \multicolumn{2}{r|}{$223\pm11$} & \multicolumn{2}{r}{--}\\
\texttt{HoliGRALE} &	 \multicolumn{2}{r|}{$-871.4\pm8.7$} &	 \multicolumn{2}{r|}{$-299.3\pm2.4$} 	& \multicolumn{2}{r|}{--} & \multicolumn{2}{r}{--}\\
\texttt{WSLAP+} 	& \multicolumn{2}{r|}{ $-957.4\pm2.3$} 	&  \multicolumn{2}{r|}{$-395.9\pm3.6$ }&	\multicolumn{2}{r|}{--} & \multicolumn{2}{r}{--}\\
\texttt{Zitrin-Analytic} &	 \multicolumn{2}{r|}{$-1015^{+38}_{-47}$} &	 \multicolumn{2}{r|}{$-378_{-18}^{+25}$}&	\multicolumn{2}{r|}{ --} & \multicolumn{2}{r}{--} \\
\texttt{LENSTOOL} 	& \multicolumn{2}{r|}{$-960\pm360$} &	 \multicolumn{2}{r|}{$-80\pm400$} & \multicolumn{2}{r|}{$463\pm219$} & \multicolumn{2}{r}{$438\pm254$} \\
\hline
& \multicolumn{8}{c}{Predicted Positions}\\
\cline{2-9}\\[-0.45cm]
 & \colhead{$x\, ['']$} & \colhead{$y\, ['']$} & \colhead{$x\, ['']$} & \colhead{$y\, ['']$} & \colhead{$x\, ['']$} & \colhead{$y\, ['']$} & \colhead{$x\, ['']$} & \colhead{$y\, ['']$}\\
\hline
\texttt{GLAFIC} & $-6.70^{+0.26}_{-0.27}$ & $-13.90^{+0.33}_{-0.33}$ & $3.67^{+0.30}_{-0.29}$ & $10.52^{+0.25}_{-0.25}$ & -- & -- & -- & --\\
\texttt{Chen2020} & $-6.86$ & $-13.72$ & $3.55$ & $10.74$ & $2.38$ & $-0.63$ & -- & --\\
\texttt{HoliGRALE} & $-6.89^{+0.01}_{-0.01}$ & $-13.80^{+0.01}_{-0.01}$ & $3.46^{+0.01}_{-0.01}$ & $10.63^{+0.01}_{-0.01}$ & -- & -- & -- & --\\
\texttt{WSLAP+} & $-6.86$ & $-13.54$ &  $3.61$ & $10.89$  & -- & -- & -- & --\\
\texttt{Zitrin-Analytic} & $-6.83$ & $-13.67$ & $3.53$ & $10.83$ & -- & -- & -- & --\\
\texttt{LENSTOOL} & $-6.62$ & $-13.55$ & $4.26$ & $10.91$&$2.42$ & $-0.88$ & $-0.035$ & $0.005$
\enddata
\tablecomments{ $dt_{\rm model}$ refers to the relative time delay between the $i^{\rm th}$ image and image S3 (i.e., $t_{\rm Si}-t_{\rm S3}$); $x$ and $y$ (in arcsec) are positions relative to (R.A., decl.) = (322\fdg4164769, 0\fdg0892336).}
\end{deluxetable*}

\vspace{-0.5cm}

Table~\ref{tab:mu_model} summarizes the constraints on the magnifications of the lensed SN\,2022riv images from lens modeling. These values represent macromodel predictions and do not account for the effects of microlensing or millilensing. The numbers in parentheses in the Image S3 column indicate the statistical discrepancy between the absolute magnification derived from fitting the light curve of SN\,2022riv using the SALT3-NIR model and the lens model predictions. Five out of six lens models we considered here show strong statistical agreement even without considering microlensing and millilensing, although significant microlensing effects are expected owing to Image S3 being relatively close to the center of the BCG. The galaxy cluster mass reconstruction performed with the lens inversion method \texttt{HoliGRALE} yields a predicted magnification that is in significant statistical tension with the absolute magnification inferred from light-curve fitting, at the level of $7.6\sigma$. In this reconstruction, \texttt{HoliGRALE} includes the cluster member galaxy located near Image S3 but, unlike other lens modeling approaches, does not explicitly model the BCG. Consequently, the mass component associated with the BCG in the \texttt{HoliGRALE} reconstruction is not centered on the observed BCG position, which results in a larger magnification at the location of Image S3. Additional tests identifying a systematic effect in \texttt{HoliGRALE} that accounts for this discrepancy will be presented in detail in a forthcoming paper (Perera et al., in preparation), along with updated results that explicitly include the BCG. Table~\ref{tab:microlensing} presents the predicted time delays relative to Image S3 and the predicted positions of the other SN images based on the cluster lens modeling. These lens models can be downloaded from doi:\dataset[10.5281/zenodo.21460864]{https://doi.org/10.5281/zenodo.21460864}.

\begin{deluxetable}{lrrcc}
\setlength{\tabcolsep}{1pt}
\tablecaption{ Lensing parameter estimates ($\kappa$ and $\gamma$) at the position of Image S3 from different lens modeling efforts, used in microlensing and millilensing simulations with stellar mass densities derived using the {\tt Prospector} algorithm.} \label{tab:glaficpred_gamma_kappa}
\tablehead{
\colhead{Model} & \colhead{$\kappa$} & \colhead{$\gamma$} & \colhead{$\log_{10}\left(\frac{\Sigma_*}{M_\odot\, \rm  kpc^{-2}}\right)$} & \colhead{$\kappa_*/\kappa$}
}
\startdata
\texttt{GLAFIC} & $1.058$ & $0.466$ & \multirow{6}{*}{$8.729^{+0.014}_{-0.037}$} & $0.1836^{+0.0060}_{-0.0150}$\\
\texttt{Chen2020} & $0.869$ & $0.395$ &  & $0.2235^{+0.0073}_{-0.0183}$\\
\texttt{HoliGRALE} & $0.875$ & $0.289$ &  & $0.2219^{+0.0073}_{-0.0181}$\\
\texttt{WSLAP+} & $0.996$ & $0.438$ &  & $0.1949^{+0.0064}_{-0.0159}$\\
\texttt{Zitrin-Analytic} & $1.031$ & $0.384$ &  & $0.1884^{+0.0062}_{-0.0154}$\\
\texttt{LENSTOOL} & $1.013$ & $0.478$ &  & $0.1917^{+0.0063}_{-0.0157}$\\
\enddata 
\end{deluxetable}

\subsection{The Effect of Millilensing}

After estimating the absolute magnifications using multiple SALT models for SN\,2022riv, we now explore the impact of millilensing caused by dark matter subhalos within the galaxy cluster and dark matter halos along the line of sight on the flux estimations of SN\,2022riv. 
\cite{delaykellyrodneytreu23} found that the millilensing's impact on magnification ranges from approximately 4\% to 18\% across the five images of SN\,Refsdal for dark matter substructure mass fractions of 0.5\% and 2\%. Similarly, \cite{pierelfryepascale24} found that this effect ranges from $\sim 4$\% to 8\% for SN\,H0pe with a 5\% dark matter mass fraction.

In practice, we used {\tt pyHalo}\footnote{https://github.com/dangilman/pyHalo} \citep{Gilman_pyHalo}, an open-source {\tt Python} software package that generates realistic simulations of dark matter subhalos and line-of-sight halos, to populate these substructure uniformly around the Image S3. We then employed {\tt lenstronomy}\footnote{https://github.com/lenstronomy} \citep{Birrer:2018xgm, Birrer2021}, an open-source {\tt Python}-based lens modeling package, to perform ray-tracing and generate millilensing simulations for lens systems with convergence ($\kappa$) and shear ($\gamma$) at the Image S3 position, as estimated using six different cluster lens models, which are listed in Table~\ref{tab:glaficpred_gamma_kappa}. We considered subhalo projected mass fractions ($f_{\rm sub}$) of 1\%, 5\%, and 10\% in the main lens plane, assuming a power-law subhalo mass function with a pivot mass of $10^8\,M_\odot$ and a slope of $-1.9$. These choices are motivated by and consistent with both theoretical and observational results \citep{2002ApJ...572...25D, 2008MNRAS.391.1685S, 2014ApJ...784...90O, 2017MNRAS.472..657J, 2018MNRAS.478.2086D, 2020MNRAS.493.1268B, Gilman_pyHalo, 2026MNRAS.547ag486P}. Additionally, we populated line-of-sight halos using the rescaled version \citep{2019MNRAS.487.5721G} of the Sheth--Tormen mass function \citep{2001MNRAS.323....1S}. The dark matter substructure masses were drawn from the range $10^6$--$10^{10}\,M_\odot$, where halos with masses below $10^6\,M_\odot$ are too small to significantly affect lensing magnifications, while those above $10^{10}\,M_\odot$ are massive enough to host a visible counterpart. 
\begin{deluxetable}{l|r|r|r}
\setlength{\tabcolsep}{4.5pt}
\tablecaption{Effect of dark matter substructure millilensing on the magnification of SN\,2022riv Image S3.} \label{tab:millilensing}
\tablehead{
\multirow{2}{*}{Model} & \multicolumn{3}{c}{$\Delta \mu_{\rm milli}$}\\ [0.1cm]
\cline{2-4}\\[-0.4cm]
& $f_{\rm sub}=1\%$ & $f_{\rm sub}=5\%$ & $f_{\rm sub}=10\%$}
\startdata
\texttt{GLAFIC} & $0.00^{+0.02}_{-0.02}$  & $-0.01^{+0.02}_{-0.02}$ & $-0.01^{+0.03}_{-0.02}$\\
\texttt{Chen2020} & $0.01^{+0.03}_{-0.05}$  & $0.01^{+0.03}_{-0.06}$ & $0.01^{+0.04}_{-0.07}$\\
\texttt{HoliGRALE} & $0.03^{+0.07}_{-0.14}$  & $0.03^{+0.10}_{-0.20}$ & $0.04^{+0.12}_{-0.21}$\\
\texttt{WSLAP+} & $0.00^{+0.02}_{-0.02}$  & $0.00^{+0.02}_{-0.03}$ & $0.00^{+0.03}_{-0.03}$\\
\texttt{Zitrin-Analytic} & $-0.01^{+0.03}_{-0.03}$ & $-0.01^{+0.04}_{-0.04}$ & $-0.01^{+0.04}_{-0.04}$\\
\texttt{LENSTOOL} & $0.00^{+0.02}_{-0.01}$  & $0.00^{+0.02}_{-0.02}$ & $0.00^{+0.02}_{-0.02}$\\
\enddata 
\tablecomments{Here, $\Delta \mu_{\rm milli}$ denotes the change in the macromodel magnification due to millilensing, computed as the difference between the total magnification with and without millilensing. Uncertainties are determined from the 16th and 84th percentiles. The results are provided for subhalo mass fractions ($f_{\rm sub}$) of 1\%, 5\%, and 10\% in the main lens plane.}
\end{deluxetable}

Table~\ref{tab:millilensing} presents the additional magnification component introduced by the dark matter substructure millilensing ($\Delta\mu_{\rm milli}$) for Image S3, computed by comparing the total magnification in the presence of millilensing with the magnification from the galaxy cluster and its members alone, assuming no millilensing. The results indicate that millilensing introduces a fractional change in the magnification of SN\,2022riv at the level of $\sim 0.1$\% to 2\%. This effect of dark matter substructure millilensing is relatively small compared to that seen in SN\,Refsdal ($\sim 10$ times) and SN\,H0pe ($\sim 6$ times), which is not surprising given that the last-to-arrive image of SN\,2022riv is close to the BCG. As a result, the dark matter subhalos in this region are expected to be largely tidally stripped, leading to significant mass loss.

\subsection{The Effect of Microlensing}

In addition to accounting for the effect of dark matter substructure millilensing on the absolute magnification of SN\,2022riv, as discussed in the previous section, we now simulate the effects of microlensing on our photometry of Image S3. To account for microlensing effects, we adopted the same estimates of $\kappa$ and $\gamma$ from each cluster lens model that were used for the millilensing simulations, as detailed in Table~\ref{tab:glaficpred_gamma_kappa}. Additionally, we required estimates of the stellar mass density at the Image S3 position to calculate the smooth matter fraction.

The stars along the line of sight to the SN\,2022riv images that affect its photometry primarily consist of two components: the ICL and the BCG of RX\,J2129 cluster field. We modeled the BCG and ICL together as a single underlying stellar population with uniform properties, such as age or initial mass function (IMF). This approximation is reasonable because, at the projected radial distance of $\sim16$\,kpc from the BCG center relevant for our study, the stellar mass is strongly dominated by the BCG. Distinguishing the BCG from the ICL is not straightforward: their surface brightness profiles transition smoothly into one another without a well-defined physical boundary \citep{2014MNRAS.437..816C, 2022NatAs...6..308M}, and although the ICL can, in principle, be distinguished kinematically \citep{2010MNRAS.405.1544D, 2013A&A...558A..42L, 2017Galax...5...49R, 2022A&A...663A..12H}, the BCG and ICL are spatially intermixed, preventing a clean separation using photometry alone \citep{2025A&A...698A..14E, 2026MNRAS.548ag590F}. From an analysis of six \textit{Hubble} Frontier Fields clusters, \cite{2017ApJ...846..139M} found that the BCG’s stellar mass density typically exceeds that of the ICL by about two orders of magnitude at these radial distances, contributing $\gtrsim90\%$ of the total stellar mass components (ICL+BCG+cluster galaxies). To avoid additional uncertainties from photometric decomposition methods \citep{2024MNRAS.528..771B}, we therefore treat the combined BCG+ICL component as a single stellar population, which is expected to have only a minor impact on our microlensing predictions. We selected photometric regions in the 2022 October 6 \textit{JWST}/NIRCam observations to sample the ICL+BCG near SN\,2022riv Image S3, providing a representative estimate of the local underlying stellar population. To avoid contamination from the SN or its host, the apertures were placed at a similar projected distance from the BCG as the SN image, excluding the SN position and sufficiently separated such that neither the SN nor the host galaxy contaminated the measured fluxes. We first measured the fluxes in these regions, accounting for uncertainties, using \textit{JWST}/NIRCam imaging as listed in Table~\ref{observations}, from which we then estimated the stellar mass near the SN images by fitting a six-band SED. The uncertainties in the flux measurements arise from multiple sources, with the dominant contributions coming from errors in the background estimation and the aperture sum. To perform SED fitting and estimate the stellar mass contributions from the ICL and BCG in these regions, we utilized {\tt Prospector} \citep{2021ApJS..254...22J}, which incorporates nonparametric star-formation histories (SFHs) and dust models, while also employing a Bayesian approach to determine the output physical parameters and their uncertainties. We adopted the \cite{Chabrie_2003_IMF} IMF, applied dust attenuation following the \cite{Calzetti_2000_dust} attenuation curve, and employed the continuity SFH \citep{2019ApJ...876....3L, 2021ApJS..254...22J}. 

\begin{figure}
\begin{center} 
\includegraphics[clip, width=0.475\textwidth]{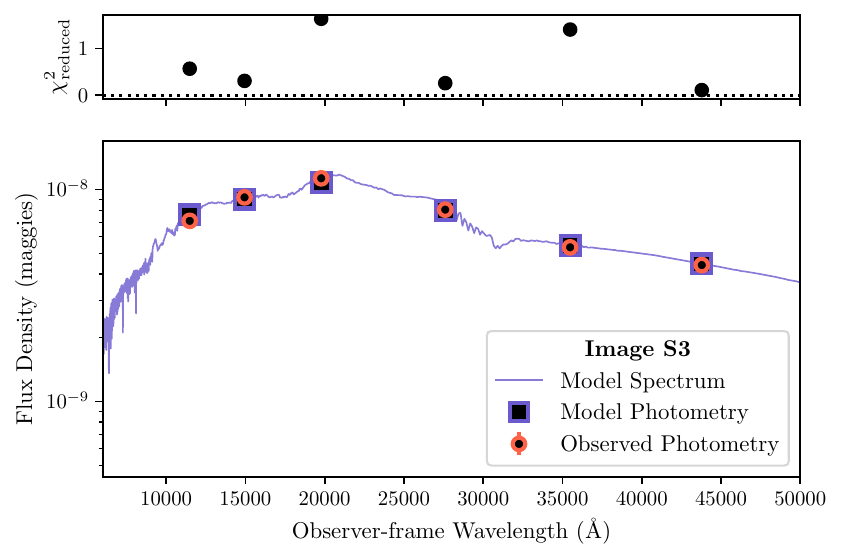}
\end{center}
\caption{\label{fig:SED} SED fitting results for the combined ICL+BCG region at the positions of SN\,2022riv Image S3, obtained using {\tt Prospector}. The filled orange circles with error bars indicate the {\it JWST} photometric data. The best-fit model is shown in blue, with a solid line representing the SED fit and squares denoting the predicted photometry from the model.}
\end{figure}

{\tt Prospector} uses MCMC sampling to fit the observed SED of the combined ICL and BCG region and jointly infer stellar population parameters, including total stellar mass, age, dust attenuation, stellar metallicity, and gas-phase metallicity (when nebular emission is included). The inferred total stellar mass is then used to compute the projected stellar mass density and its uncertainty. Figure~\ref{fig:SED} shows the SED fit for the combined ICL+BCG region at Image S3, obtained using {\tt Prospector}, while Table~\ref{tab:glaficpred_gamma_kappa} lists the projected stellar mass density and its uncertainty at this image position after fitting with MCMC sampling. We noticed that the projected stellar mass density for Image S3 is more than an order of magnitude higher than that of the other two images, as computed for the image positions predicted by the \texttt{GLAFIC} model ($\log_{10}(\Sigma_*/ [{\,M_\odot\, \rm kpc^{-2}}]) = 7.03^{+0.21}_{-0.33}$ and $7.35^{+0.17}_{-0.29}$ for Image S1 and S2, respectively), with Image S3 being relatively close to the center of the BCG. We find that our results are in excellent agreement with the radial stellar mass density profiles presented in \cite{2017ApJ...846..139M} for a sample of six \textit{Hubble} Frontier Field clusters spanning the redshift range $0.3 < z < 0.6$.


Having obtained the convergence ($\kappa$), shear ($\gamma$), and smooth matter fraction ($s = 1 - \kappa_{*}/\kappa$) (see Table~\ref{tab:glaficpred_gamma_kappa}), we applied the code developed by \cite{huber_chromatic_microlensing} to obtain chromatically microlensed SN Ia spectra. We first generated microlensing maps utilizing the \texttt{GERLUMPH} code \citep{vernardos_gerlumph_micro_maps} modified by \cite{chan+21}, which employs the inverse ray-shooting method \citep{kayser_inverse_ray_shooting, wambsganss_inverse_ray_shooting, vernardos_inverse_ray_shooting} to compute two-dimensional maps of the magnification factor, $\mu$. In our case, we calculated maps with side lengths of 20 Einstein radii ($R_{\rm E}$) in the source plane, where 
\begin{equation}
    R_{\rm E} = \sqrt{\frac{4G\langle M \rangle}{c^{2}} \cdot \frac{D_{\rm s}D_{\rm ds}}{D_{\rm d}}} ,
    \label{eq:einstein_radius}
\end{equation} and a resolution of 20,000 pixels per side. Here, $G$ is the gravitational constant, $c$ is the speed of light, $\langle M\rangle$ is the mean mass of the microlenses, and $D_{\rm d}$, $D_{\rm s}$, and $D_{\rm ds}$ are, respectively the angular diameter distances from the observer to the lens, the observer to the source, and the lens to the source.  

We then randomly placed an SN on the microlensing map and calculated its magnified spectrum by multiplying an unlensed base spectrum by the local magnification factor for all pixels the SN covers. The emitted specific intensity is obtained from \texttt{ARTIS} simulations \citep{kromer_artis_code} for four different SN~Ia explosion models. First is the W7 model \citep{nomoto_w7_model}, where a Chandrasekhar-mass carbon-oxygen white dwarf (CO WD) deflagrates. Second is the N100 model \citep{seitenzahl_n100_model}, which is a delayed detonation model of a Chandrasekhar-mass CO WD. Third is a sub-Chandrasekhar-mass explosion model from \cite{sim_sub_ch_model}, where a $1.06\,\,M_\odot$ CO WD detonates. The fourth model is a merger from \cite{pakmor_merger_model}, where two CO WDs with masses of $0.9\,M_\odot$ and $1.1\,M_\odot$ coalesce and detonate.

Lensing magnification is generally defined as the ratio between the observed (lensed) flux of a source and its intrinsic (unlensed) flux. Denoting the unlensed flux as $F_{\rm unlensed}$, the flux magnified solely by the smooth main lens mass distribution as $F_{\rm macro}$, and the flux affected by both the main lens and additional mass substructures (including stellar microlenses, dark matter subhalos in the lens plane, and any intervening structures along the line of sight between the observer and the source) as $F_{\rm full}$, the corresponding magnifications can be expressed as $\mu_{\rm macro} = F_{\rm macro}/F_{\rm unlensed}$ and 
$\mu_{\rm full} = F_{\rm full}/F_{\rm unlensed}$, respectively. In the previous section, we observed that the effect of dark matter substructure millilensing is relatively small. In this work, we thus assume that the total lensing effect arises only from the combination of the smooth mass component and stellar microlensing, and that the magnification contributions are approximately additive, despite this not being strictly valid in general. Under this assumption, the full magnification can be approximated as $\mu_{\rm full}\approx \mu_{\rm macro} + \Delta \mu_{\rm micro}$. Consequently, the ratio of the total lensed flux to the flux from the smooth lens alone can be written as \begin{equation}
\frac{F_{\rm full}}{F_{\rm macro}} \approx 1 + \frac{\Delta \mu_{\rm micro}}{\mu_{\rm macro}}.
\end{equation}

For each of the SN~Ia explosion models described previously, we generated a set of 1000 chromatically microlensed spectra, along with a single achromatic macrolensed spectrum. Using these spectra, we employed the \texttt{Astrolib PySynphot} package \citep{pysynphot2015} to compute synthetic photometry in the \textit{HST}-WFC3/IR F110W and F160W filters. The resulting synthetic apparent magnitudes are related to the corresponding magnification through the relation \citep{2006ApJ...653.1391D, 2024MNRAS.531.4349W, 2025PhRvD.112b3002S}
\begin{equation}
\Delta m={\rm mag}_{\rm full}-{\rm mag}_{\rm macro} \approx -2.5 \log_{10} \left( 1 + \frac{\Delta \mu_{\rm micro}}{\mu_{\rm macro}} \right).
\end{equation} Consequently, the additional magnification component introduced by stellar microlensing can be expressed as 
\begin{equation}
\Delta \mu_{\rm micro} \approx \mu_{\rm macro} \left( 10^{-\frac{\Delta m}{2.5}} -1 \right).
\end{equation}Table \ref{tab:microlensing_F160W} summarizes the impact of stellar microlensing on the absolute magnification of the last-to-arrive image of SN\,2022riv. In addition to the simulations using the Chabrier IMF, we performed a second set of simulations for each lens model adopting the \cite{Salpeter_IMF} IMF, which we approximate as having roughly twice the stellar mass of the Chabrier IMF \citep{Chabrie_2003_IMF}. These results are also listed in Table \ref{tab:microlensing_F160W}. The associated uncertainties are estimated from the 16th and 84th percentiles of the simulated magnification distributions. 

Interestingly, the last-to-arrive image of SN\,2022riv (Image S3) exhibits a substantially stronger impact from stellar microlensing, whereas we found that the other two images experience only a very small effect from microlensing, with $\Delta \mu_{\rm micro}$ values at or below 10\% in comparison to Image S3, as seen in the \texttt{GLAFIC} model. We find that microlensing introduces a change of SN\,2022riv magnification at a level of $\sim 20$\% -- 50\% across all lens models, assuming either the Chabrier or Salpeter IMF. 

\begin{figure*}
\centering
\includegraphics[clip, trim=0cm 0cm 0cm 0cm, width=1.0\textwidth]{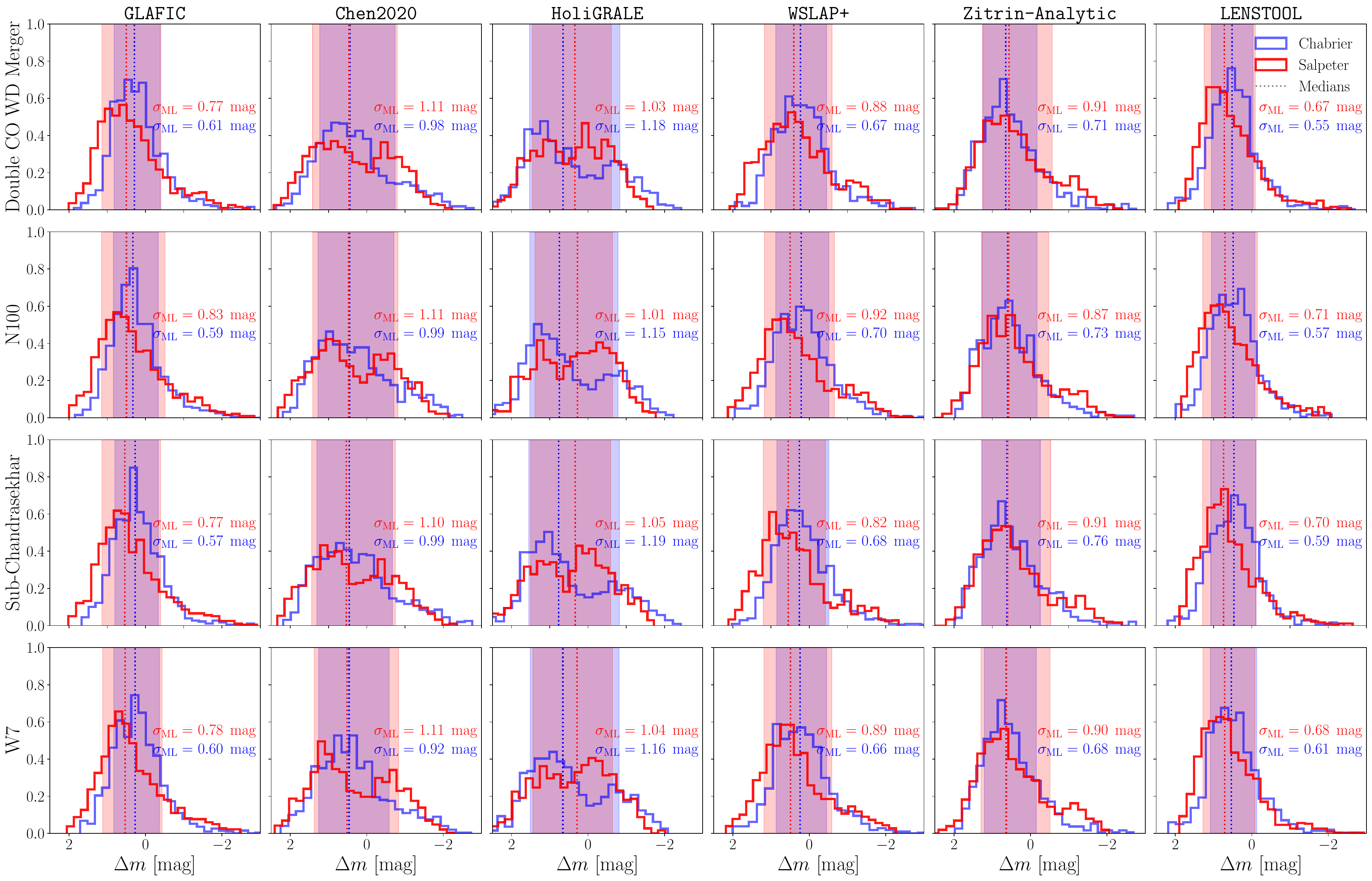}
\caption{The PDFs of the change in WFC3/IR F160W magnitude for each lens model and each SN Ia explosion model due to stellar microlensing. The blue (red) histograms represent the PDFs from microlensing simulations using the Chabrier (Salpeter) IMF. Dotted vertical lines indicate the median value for each case, while the shaded color bands represent the 16th and 84th percentiles. The scatter, $\sigma_{\rm ML}$, is shown on the right side of each panel. The $x$- and $y$-axis ranges are the same across all panels.
\label{delta_mag_plot_f160w}}
\end{figure*}

Overall, all the models exhibit demagnifications after accounting for stellar microlensing, consistent with the expectation that macrosaddle point images, such as Image S3, are more susceptible to microlensing-induced demagnification \citep{2002ApJ...580..685S, 2024MNRAS.531.4349W}. Figure~\ref{delta_mag_plot_f160w} presents the probability density functions (PDFs) of the magnitude change ($\Delta m$) of the F160W band due to microlensing. The scatter ($\sigma_{\rm ML}$) indicated on the right of each panel is estimated as half the difference between the 16th and 84th percentiles. These results are in excellent agreement with those reported by \cite{2018MNRAS.478.5081F}. The PDFs of some models, particularly those of \texttt{Chen2020} and \texttt{HoliGRALE}, show two distinct peaks. The leftmost peak of these bimodal distributions corresponds to cases where no microminima form at macrosaddles in the time-delay surface, and thus only microsaddles and micromaxima are present. The second peak, on the right side of the PDFs, corresponds to cases in which the SN lies in regions where the macrosaddle produces one or more microminima \citep{1992ApJ...386...30R, 2002ApJ...580..685S, 2003ApJ...583..575G, 2007MNRAS.377..977D, 2011MNRAS.411.1671S, 2024MNRAS.531.4349W}.

\begin{deluxetable*}{l|
>{\raggedleft\arraybackslash}p{1.7cm}|
>{\raggedleft\arraybackslash}p{1.7cm}|
>{\raggedleft\arraybackslash}p{1.7cm}|
>{\raggedleft\arraybackslash}p{1.7cm}|
>{\raggedleft\arraybackslash}p{1.7cm}|
>{\raggedleft\arraybackslash}p{1.7cm}|
>{\raggedleft\arraybackslash}p{1.7cm}|
>{\raggedleft\arraybackslash}p{1.7cm}
}
\setlength{\tabcolsep}{2.8pt}
\tablecaption{Effect of stellar microlensing on the magnification of SN\,2022riv for each lens model of the RX\,J2129 galaxy cluster, considering four different SN~Ia  explosion models as well as Chabrier and Salpeter IMFs. Synthetic photometry calculations were performed using the \textit{HST} WFC3/IR F160W band. Uncertainties are determined from the 16th and 84th percentiles.} \label{tab:microlensing_F160W}
\tablehead{
\multirow{4}{*}{Model} & \multicolumn{8}{c}{$\Delta \mu_{\rm micro}$}\\ [0.2cm]
\cline{2-9}\\[-0.4cm]
& \multicolumn{2}{c|}{Double CO WD Merger} & \multicolumn{2}{c|}{N100} & \multicolumn{2}{c|}{Sub-Chandrasekhar} & \multicolumn{2}{c}{W7}\\
\cline{2-9}\\[-0.45cm]
& Chabrier  & Salpeter  & Chabrier  & Salpeter  & Chabrier  & Salpeter  & Chabrier  & Salpeter }
\startdata
\renewcommand{\arraystretch}{1.5}
\texttt{GLAFIC} & $-1.1^{+3.1}_{-1.4}$  	&	 $-1.7^{+3.9}_{-1.3}$  	&	 $-1.2^{+2.9}_{-1.3}$  	&	 $-1.7^{+4.5}_{-1.3}$  	&	 $-1.0^{+2.8}_{-1.4}$  	&	 $-1.8^{+3.9}_{-1.2}$  	&	 $-1.0^{+2.9}_{-1.5}$  	&	 $-1.8^{+4.1}_{-1.2}$  	\\
\texttt{Chen2020} & $-2.4^{+9.4}_{-2.5}$  	&	 $-2.5^{+10.4}_{-2.8}$ 	&	 $-2.5^{+9.1}_{-2.5}$  	&	 $-2.6^{+10.6}_{-2.7}$ 	&	 $-2.5^{+8.7}_{-2.6}$  	&	 $-2.8^{+10.0}_{-2.5}$  	&	 $-2.5^{+7.7}_{-2.4}$  	&	 $-2.7^{+11.1}_{-2.5}$ 	\\
\texttt{HoliGRALE} &  $-6.6^{+23.7}_{-4.5}$ 	&	 $-4.0^{+15.1}_{-6.8}$ 	&	 $-7.3^{+23.0}_{-3.8}$ 	&	 $-3.3^{+15.1}_{-7.3}$ 	&	 $-7.5^{+24.3}_{-3.7}$ 	&	 $-3.9^{+14.6}_{-7.2}$ 	&	 $-6.7^{+23.0}_{-4.4}$ 	&	 $-3.4^{+15.3}_{-7.4}$ 	\\
\texttt{WSLAP+} &  $-1.0^{+3.7}_{-1.9}$  	&	 $-1.6^{+5.4}_{-1.8}$  	&	 $-0.9^{+4.1}_{-2.0}$  	&	 $-1.9^{+6.3}_{-1.5}$  	&	 $-1.1^{+4.2}_{-1.7}$  	&	 $-2.1^{+4.6}_{-1.4}$  	&	 $-1.0^{+3.7}_{-1.8}$  	&	 $-1.9^{+5.7}_{-1.6}$  	\\
\texttt{Zitrin-Analytic} & $-3.1_{-1.6}^{+4.3}$  	&	 $-2.8_{-1.9}^{+7.5}$  	&	 $-2.9_{-1.8}^{+4.1}$  	&	 $-2.8_{-1.9}^{+6.6}$  	&	 $-3.0_{-1.7}^{+4.8}$  	&	 $-2.9_{-1.8}^{+7.2}$  	&	 $-3.0_{-1.6}^{+4.0}$  	&	 $-3.0_{-1.7}^{+7.1}$  	\\
\texttt{LENSTOOL} & $-1.7_{-1.1}^{+1.8}$  	&	 $-2.1_{-0.9}^{+2.5}$  	&	 $-1.6_{-1.2}^{+1.9}$  	&	 $-2.1_{-1.0}^{+2.7}$  	&	 $-1.5_{-1.2}^{+2.0}$  	&	 $-2.2_{-0.9}^{+2.6}$  	&	 $-1.7_{-1.1}^{+2.3}$  	&	 $-2.1_{-0.9}^{+2.5}$  	\\
\enddata
\tablecomments{The values tabulated here are based on $\kappa$, $\gamma$, and $\kappa_*/\kappa$ values provided in Table~\ref{tab:glaficpred_gamma_kappa}.}
\end{deluxetable*}

\vspace{-0.5cm}

\subsection{Comparing Magnification to Lens Model Predictions}

In this section, we compare the lens model predictions of image magnifications ($\mu_{\rm model}$) with the absolute magnification derived from fitting the light curve of SN\,2022riv using the SALT3-NIR SN Ia spectral time-series model (hereafter $\mu_{\rm SALT3-NIR}$), and we examine the potential effects of stellar microlensing and dark matter substructure millilensing.

The lens model predictions for the magnifications account only for the structure on the scale of galaxy cluster members and the cluster itself. \cite{2006ApJ...653.1391D} demonstrated that localized gravitational perturbations from stars, beyond the smooth mass distribution of a lens galaxy, can produce observable microlensing signatures in the light curves of lensed SNe. Here, we assume that $\mu_{\rm SALT3-NIR}$ is approximately the sum of this smooth model, the magnification change due to millilensing ($\Delta \mu_{\rm milli}$), and the additional magnification from microlensing ($\Delta \mu_{\rm micro}$). For substructure millilensing, we adopt a subhalo projected mass fraction of $f_{\rm sub}=5\%$, and we consider all four different SN Ia explosion models discussed earlier, using the WFC3/IR F160W band for synthetic photometry. 
\begin{figure*}[h!]
\centering
\includegraphics[clip, trim=0cm 0cm 0cm 0cm, width=0.97\textwidth]{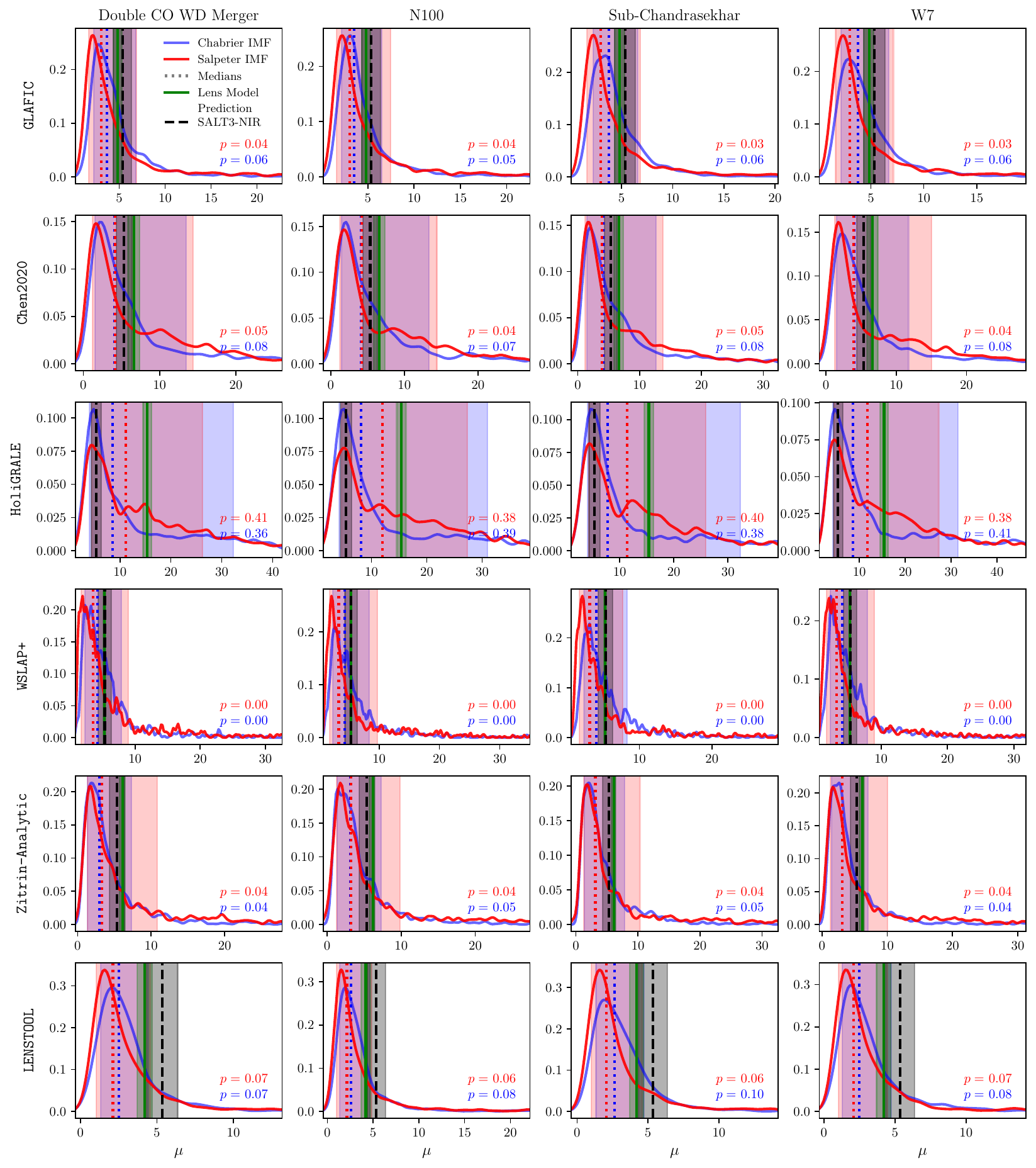}
\caption{The PDFs of the absolute magnifications for each lens model and each SN~Ia explosion model, after correcting the initial absolute magnification estimates 
from lens modeling ($\mu_{\rm model}$) for stellar microlensing and dark matter substructure millilensing. The blue (red) curves represent the PDFs derived assuming the Chabrier (Salpeter) IMF in the microlensing simulations. Dotted vertical lines indicate the median value for each case, while the shaded color bands represent the 16th and 84th percentiles. The vertical green lines correspond to the absolute magnifications predicted by lens models, with the green color bands indicating the 16th to 84th percentile range for these model predictions. Microlensing effects are estimated using the \textit{HST} WFC3/IR F160W band, with a substructure fraction of $ f_{\rm sub} = 5\% $ is considered for dark matter millilensing. The black-dashed vertical line indicates the absolute magnification from the SALT3-NIR light-curve fitting, with the shaded region representing the uncertainty. The $p$-values in the bottom left of each panel represent the probability that the corrected absolute magnification 
falls between $\mu_{\rm model}$ and the absolute magnification derived from fitting the light curve using the SALT3-NIR model. 
\label{magnification_plot_f160w}}
\end{figure*}

Figure~\ref{magnification_plot_f160w} shows the PDFs of the absolute magnifications for each lens model and SN~Ia explosion model, after correcting $\mu_{\rm model}$ for the effects of stellar microlensing and dark matter substructure millilensing. It is important to note that we ignore the parity of $\mu_{\rm model}$ and consider only its absolute value in this analysis. The blue (red) curves correspond to the PDFs derived assuming the Chabrier (Salpeter) IMF in the microlensing simulations. The dotted vertical lines indicate the median value for each case, while the shaded color bands represent the 16th and 84th percentiles. Vertical green lines mark the absolute magnifications predicted by the lens models, with the green color bands highlighting the range between the 16th and 84th percentiles of these predictions. Table~\ref{tab:total_mag_F160W_Chabrier} (Table~\ref{tab:total_mag_F160W_Salpeter}) provides the estimated absolute magnifications for each lens model and SN Ia explosion model, based on these PDFs, assuming the Chabrier (Salpeter) IMF. The $p$-values in the bottom-left corners of each panel in Figure~\ref{magnification_plot_f160w} represent the probability that the corrected absolute magnification falls between $\mu_{\rm model}$ and $\mu_{\rm SALT3-NIR}$ (black-dashed vertical line in Figure~\ref{magnification_plot_f160w}), and are given by 
$p=P(\mu \in [\min(\mu_{\rm model}, \mu_{\rm SALT3-NIR}), \, \max(\mu_{\rm model}, \mu_{\rm SALT3-NIR})])$. If there is significant statistical tension between $\mu_{\rm model}$ and $\mu_{\rm SALT3-NIR}$, a large $p$-value suggests that this tension is mainly due to stellar microlensing effects. In this case, the lens model predictions are in excellent agreement with the measurements when microlensing is taken into account. In contrast, a smaller $p$-value, indicating a very low probability that the corrected absolute magnification falls within the expected range, suggests that stellar microlensing has a minimal effect. This could either imply excellent statistical agreement between $\mu_{\rm model}$ and $\mu_{\rm SALT3-NIR}$, or, if there is significant statistical tension, the primary cause of the tension is not stellar microlensing effects.

Among all the galaxy cluster lens models, we find excellent agreement between the magnification prediction from the \texttt{WSLAP+} lens model --- which uses a free-form lensing reconstruction method that does not rely on assumptions about the dark matter distribution --- and our measurement from the SALT3-NIR light-curve fitting, with a statistical tension of $0.03\sigma$ ($p = 0.0$). This excellent agreement, especially without the inclusion of stellar microlensing effects, is surprising. If corroborated by additional magnified SNe, it could suggest that our understanding of the microlenses along the line of sight through the BCG may be incomplete. We also note that the inclusion of these effects in the model slightly increases the statistical tension between $\mu_{\rm model}$ and $\mu_{\rm SALT3-NIR}$ ($\sim0.3 \sigma$). Among the remaining five models, four show very good statistical agreement with the measured absolute magnification, with tensions ranging from $0.5\sigma$ to $1\sigma$ and $p$-values of $\sim0.1$ or smaller. This suggests that the lens model predictions remain robust, even when microlensing effects are not considered. In contrast, we observe a significant statistical tension of $7.6\sigma$ between the magnification predicted by the remaining \texttt{HoliGRALE} lens model --- which employs a hybrid mass reconstruction of the galaxy cluster using the lens inversion method \texttt{GRALE} --- and our measurement, accompanied by notably large $p\approx 0.4$, compared to the other lens model predictions. This $p > 0.05$ value suggests that the discrepancy can be primarily attributed to stellar microlensing effects, considering that the last-to-arrive image of the SN is closer to the BCG, where the stellar mass density is expected to be significantly higher. When stellar microlensing effects are incorporated, assuming either the Chabrier (Salpeter) IMF, the resulting statistical tension becomes significantly smaller, at around $0.7\sigma$ ($0.9\sigma$). As noted previously, all models show microlensing-induced demagnifications, consistent with macrosaddle point images such as S3 being particularly susceptible \citep{2002ApJ...580..685S, 2024MNRAS.531.4349W}. Even when stellar microlensing effects are included, they do not improve the agreement between $\mu_{\rm model}$ and $\mu_{\rm SALT3-NIR}$ for some lens models; however, the microlensing-corrected magnification predictions still remain in good statistical agreement with $\mu_{\rm SALT3-NIR}$ for all models.

\begin{deluxetable*}{l|
>{\raggedleft\arraybackslash}p{1.7cm}|
>{\raggedleft\arraybackslash}p{1.7cm}|
>{\raggedleft\arraybackslash}p{1.7cm}|
>{\raggedleft\arraybackslash}p{1.7cm}|
>{\raggedleft\arraybackslash}p{1.7cm}|
>{\raggedleft\arraybackslash}p{1.7cm}|
>{\raggedleft\arraybackslash}p{1.7cm}|
>{\raggedleft\arraybackslash}p{1.7cm}
}
\setlength{\tabcolsep}{2.8pt}
\tablecaption{The absolute magnifications for each lens model and each SN~Ia explosion model, after correcting $\mu_{\rm model}$ for stellar microlensing (see Table~\ref{tab:microlensing_F160W}) and dark matter substructure millilensing (see Table~\ref{tab:millilensing}), with $ f_{\rm sub} = 5\% $. Microlensing effects are estimated using the \textit{HST} WFC3/IR F160W band and the Chabrier IMF. The tension associated with each lens model and each Type Ia SN explosion model indicates the statistical tension between the absolute magnification derived from fitting the light curve of SN\,2022riv using the SALT3-NIR model and $\mu_{\rm model}$, after accounting for both millilensing and microlensing effects.} \label{tab:total_mag_F160W_Chabrier}
\tablehead{
\multirow{4}{*}{Model} & \multicolumn{8}{c}{SN Ia Explosion Model}\\ [0.2cm]
\cline{2-9}\\[-0.4cm]
& \multicolumn{2}{c|}{Double CO WD Merger} & \multicolumn{2}{c|}{N100} & \multicolumn{2}{c|}{Sub-Chandrasekhar} & \multicolumn{2}{c}{W7}\\
\cline{2-9}\\[-0.45cm]
& $\mu$  & $\rm Tension$  & $\mu$  & $\rm Tension$  & $\mu$  & $\rm Tension$  & $\mu$  & $\rm Tension$ }
\startdata
\texttt{GLAFIC} & $3.7_{-1.5}^{+3.1}$ & $0.5\sigma$ & $3.6_{-1.4}^{+3.0}$ & $0.6\sigma$ & $3.8_{-1.5}^{+2.8}$ & $0.5\sigma$ & $3.8_{-1.6}^{+2.9}$ & $0.5\sigma$ \\
\texttt{Chen2020} & $4.1_{-2.6}^{+9.4}$ & $0.1\sigma$ & $4.1_{-2.7}^{+9.3}$ & $0.1\sigma$ & $4.1_{-2.7}^{+8.5}$ & $0.1\sigma$ & $4.0_{-2.5}^{+7.7}$ & $0.2\sigma$ \\
\texttt{HoliGRALE} & $8.5_{-4.5}^{+23.8}$ & $0.7\sigma$ & $8.1_{-4.0}^{+22.9}$ & $0.7\sigma$ & $7.8_{-3.8}^{+24.4}$ & $0.6\sigma$ & $8.5_{-4.4}^{+23.0}$ & $0.7\sigma$\\
\texttt{WSLAP+} & $4.2_{-1.9}^{+3.7}$ & $0.3\sigma$ & $4.3_{-2.0}^{+4.1}$ & $0.3\sigma$ & $4.1_{-1.7}^{+4.2}$ & $0.3\sigma$ & $4.2_{-1.9}^{+3.8}$ & $0.3\sigma$ \\
\texttt{Zitrin-Analytic} & $3.0_{-1.7}^{+4.3}$ & $0.5\sigma$ & $3.2_{-1.8}^{+4.1}$ & $0.5\sigma$ & $3.1_{-1.8}^{+4.7}$ & $0.5\sigma$ & $3.1_{-1.6}^{+4.0}$ & $0.6\sigma$ \\
\texttt{LENSTOOL} & $2.5_{-1.2}^{+1.9}$ & $1.3\sigma$ & $2.6_{-1.3}^{+2.1}$ & $1.2\sigma$ & $2.7_{-1.3}^{+2.1}$ & $1.2\sigma$ & $2.5_{-1.2}^{+2.3}$ & $1.2\sigma$ \\
\enddata 
\end{deluxetable*}

\begin{deluxetable*}{l|
>{\raggedleft\arraybackslash}p{1.7cm}|
>{\raggedleft\arraybackslash}p{1.7cm}|
>{\raggedleft\arraybackslash}p{1.7cm}|
>{\raggedleft\arraybackslash}p{1.7cm}|
>{\raggedleft\arraybackslash}p{1.7cm}|
>{\raggedleft\arraybackslash}p{1.7cm}|
>{\raggedleft\arraybackslash}p{1.7cm}|
>{\raggedleft\arraybackslash}p{1.7cm}
}
\setlength{\tabcolsep}{2.8pt}
\tablecaption{Same as in Table~\ref{tab:total_mag_F160W_Chabrier}, but using the Salpeter IMF in the microlensing simulations.} \label{tab:total_mag_F160W_Salpeter}
\tablehead{
\multirow{4}{*}{Model} & \multicolumn{8}{c}{SN Ia Explosion Model}\\ [0.2cm]
\cline{2-9}\\[-0.4cm]
& \multicolumn{2}{c|}{Double CO WD Merger} & \multicolumn{2}{c|}{N100} & \multicolumn{2}{c|}{Sub-Chandrasekhar} & \multicolumn{2}{c}{W7}\\
\cline{2-9}\\[-0.45cm]
& $\mu$  & $\rm Tension$  & $\mu$  & $\rm Tension$  & $\mu$  & $\rm Tension$  & $\mu$  & $\rm Tension$ }
\startdata
\texttt{GLAFIC} & $3.1_{-1.4}^{+3.9}$ & $0.6\sigma$ & $3.1_{-1.4}^{+4.4}$ & $0.5\sigma$ & $3.0_{-1.4}^{+3.9}$ & $0.6\sigma$ & $3.0_{-1.4}^{+4.1}$ & $0.5\sigma$ \\
\texttt{Chen2020} & $4.2_{-3.0}^{+10.2}$ & $0.1\sigma$ & $4.2_{-2.9}^{+10.2}$ & $0.1\sigma$ & $3.9_{-2.7}^{+9.9}$ & $0.1\sigma$ & $4.0_{-2.7}^{+11.1}$ & $0.1\sigma$ \\
\texttt{HoliGRALE} & $11.1_{-6.6}^{+15.1}$ & $0.9\sigma$ & $12.0_{-7.4}^{+15.3}$ & $0.9\sigma$ & $11.3_{-7.1}^{+14.6}$ & $0.8\sigma$ & $11.8_{-7.3}^{+15.5}$ & $0.9\sigma$ \\
\texttt{WSLAP+} & $3.6_{-1.8}^{+5.4}$ & $0.3\sigma$ & $3.3_{-1.5}^{+6.4}$ & $0.3\sigma$ & $3.2_{-1.5}^{+4.6}$ & $0.5\sigma$ & $3.3_{-1.6}^{+5.6}$ & $0.4\sigma$ \\
\texttt{Zitrin-Analytic} & $3.3_{-2.0}^{+7.5}$ & $0.3\sigma$ & $3.2_{-1.9}^{+6.7}$ & $0.3\sigma$ & $3.1_{-1.9}^{+7.2}$ & $0.3\sigma$ & $3.1_{-1.8}^{+7.0}$ & $0.3\sigma$ \\
\texttt{LENSTOOL} & $2.1_{-1.1}^{+2.4}$ & $1.2\sigma$ & $2.2_{-1.1}^{+2.7}$ & $1.1\sigma$ & $2.1_{-1.1}^{+2.5}$ & $1.2\sigma$ & $2.1_{-1.1}^{+2.4}$ & $1.3\sigma$ \\
\enddata 
\end{deluxetable*}

\vspace{-1.4cm}

\section{Summary and Conclusions}
\label{sec:discussion}

We have presented the discovery and observations of SN\,2022riv, an SN~Ia strongly lensed by the RX\,J2129.7+0005 galaxy cluster and hosted by a lensed galaxy at $z = 1.522$. In this study, we classified SN\,2022riv as an SN Ia from both its spectrum measured using {\it JWST} NIRSpec and its light curve measured using {\it JWST} and {\it HST}. Using multiple SALT light-curve fitters, we estimated cosmology-independent measurements of the absolute magnification of the last-to-arrive image of the SN. The absolute magnification was estimated to be $5.35 \pm 1.01$ using the best-fitting SALT3-NIR light-curve model, and multiple SALT SN~Ia spectral time-series models yielded consistent magnifications. SN\,2022riv was observed relatively close to the BCG, leading to an estimated projected stellar mass density of $\log_{10}(\Sigma_*/ [{\,M_\odot\, \rm kpc^{-2}}]) = 8.729^{+0.014}_{-0.037}$ in that region. We note that this value is more than an order of magnitude higher than that of the predicted positions for the other two images, thus opening the door to probing microlensing effects on a lensed SN. Previous work by \cite{delaykellyrodneytreu23} on probing microlensing effects in the lensed SN\,Refsdal measured stellar mass densities between $\log_{10}(\Sigma_*/ [{\,M_\odot\, \rm kpc^{-2}}]) = 6.855^{+0.004}_{-0.004}$ and $7.254^{+0.004}_{-0.004}$ across five lensed images, which are  approximately 1--2 orders of magnitude smaller compared to the case of SN\,2022riv.

Six independent teams modeled the RX\,J2129 galaxy cluster lens, and they derived model predictions for the magnifications, time delays, and image positions of SN\,2022riv. We have used these lens model predictions to estimate the effect of dark matter substructure millilensing and stellar microlensing on the magnification predictions. We found that millilensing introduces a magnification effect on SN\,2022riv at a level of $\sim 0.1$--2\%,  relatively small compared to that estimated for SN\,Refsdal ($\sim 10$ times) and SN\,H0pe ($\sim 6$ times). In contrast, we find that stellar microlensing should introduce strong variation in the magnification, at a level of about 20--50\%, for either a Chabrier or Salpeter IMF. From a double-blind lens modeling analysis, we find that the statistical tension between the five lens model predictions for the magnification of the SN last-to-arrive image and the absolute magnification derived from fitting the light curve of SN\,2022riv with the SALT3-NIR model ranges from 0.03$\sigma$ to 1$\sigma$, corresponding to $p$-values of $\sim0.1$ or smaller. Here, $p$ represents the probability that the corrected absolute magnification (after microlensing and millilensing) lies between the model prediction and the measurement from light-curve fitting, thereby quantifying the statistical agreement between the measured magnification and the lens model predictions. This level of statistical agreement is unexpected, given the large effect anticipated from microlensing and the fact that microlensing is not included in the lens models. Among these five models, the free-form \texttt{WSLAP+} lens model shows the best statistical agreement ($0.03\sigma$ or $p=0.0$) without accounting for microlensing and millilensing, but incorporating these effects results in a increase in the tension ($\sim 0.3 \sigma$), though the model remains statistically consistent overall. 

We note that, perhaps intriguingly, several models yield agreement with the measured magnification that is significantly more precise than the microlensing simulations would predict ($p < 0.05$). Analysis of future SNe~Ia observed through a high mass density of stars in the BCG are needed to examine whether the agreement is simply an unlikely coincidence.

Finally, the hybrid \texttt{GRALE} lens reconstruction method shows a $\sim8\sigma$ tension with notably larger $p$-values of $\sim 0.4$, suggesting that this discrepancy can be explained when stellar microlensing effects are taken into account. Incorporating these effects alleviates the discrepancy, reducing the tension significantly to $0.7\sigma$ ($0.9\sigma$) when considering the Chabrier (Salpeter) IMF.

In conclusion, constructing lens models at the scale of galaxy clusters is inherently a complex undertaking, dependent on assumptions about how dark matter traces luminous matter and necessarily involving simplification. Cluster-lensed SNe therefore provide powerful observational tests for evaluating the performance and reliability of these model assumptions. In this work, we find that most lens models not only largely agree with each other but are also statistically consistent with the measured magnification prior to accounting for stellar microlensing, despite predictions that stellar microlenses along the sightline to SN\,2022riv should introduce $20-50\%$ modulations in the magnification. Such close agreement with the microlensing-free prediction is unlikely ($p<0.05$), making the precise agreement challenging to interpret and motivating the need for additional strongly lensed SNe observed in regions of high stellar mass density. A sample of such SNe will help determine whether this discrepancy is an unlikely coincidence, or instead may point to a need to revise our understanding of microlensing by the stellar population in the BCG and ICL, or the population itself.

This analysis considers only microlensing from the stellar mass distribution in the BCG and ICL. Nevertheless, dark compact objects could also act as microlenses, and strongly lensed SNe Ia provide an excellent opportunity to probe their presence \citep[see e.g.,][]{2023MNRAS.524.5762D, 2025OJAp....8E.166A}, particularly in systems with low foreground stellar mass density, where the expected stellar microlensing is optically thin, thus providing the strongest constraints on compact dark matter, since its microlensing signatures are more readily detectable. In contrast, SN\,2022riv is located in a region of exceptionally high foreground stellar mass density, and we expect microlensing along this line of sight to be dominated by the stellar component rather than by dark compact objects. Furthermore, we find that our microlensing predictions are not significantly affected by adopting either Chabrier or Salpeter IMFs for the stellar populations in the BCG and ICL, even though the Salpeter IMF yields roughly twice the stellar mass density of the Chabrier IMF; this insensitivity suggests that the presence of an additional, observationally constrained population of dark compact objects is likewise unlikely to significantly alter our results. Incorporating these microlensing effects into the time delays, a companion paper (Dalrymple et al.; in preparation) will provide constraints on the relative time delay of the earlier-arriving image, thus enabling a comparison with lens model predictions.


\vspace{0.6cm}


\section{Acknowledgments}

This research is based on observations made with the NASA/ESA/CSA {\it JWST} and NASA/ESA {\it HST}. The data were obtained from the Mikulski Archive for Space Telescopes (MAST) at the Space Telescope Science Institute (STScI), which is operated by the Association of Universities for Research in Astronomy, Inc., under NASA contracts NAS 5-03127 for {\it JWST} and NAS 5–26555 for \textit{HST}. The specific observations analyzed can be accessed via doi:\dataset[10.17909/gs33-s803]{https://doi.org/10.17909/gs33-s803}. These observations are associated with the \textit{JWST} program DDT-2767 and the \textit{HST} programs GO-16729, GO-16264, and GO/DDT-17253. The authors also acknowledge the VENUS (Cycle 4, \textit{JWST} GO-6882, PIs S. Fujimoto \& D. Coe) team for developing their observing program with a zero-exclusive-access period. We express our appreciation for Program Coordinators Tricia Royle and Blair Porterfield, as well as Instrument Scientists Armin Rest, Diane Karakala, and Patrick Ogle of  STScI for their help carrying out the {\it JWST} and {\it HST} observations. 

We would like to express our gratitude to the anonymous referee for providing insightful comments and suggestions on this manuscript for improvements. P.K. was supported by NSF grant AST-2408655, and {\it HST} programs GO-16729, GO-17253, and GO-17504, and \textit{JWST} program DD-2767 from STScI, which is operated by AURA, Inc., under NASA contract NAS 5-26555. J.P. was supported by \textit{HST} program GO-16264 from STScI. M.O. was supported by JSPS KAKENHI grants JP25H00662 and JP22K21349. A.Z. acknowledges support from Israel Science Foundation grant 864/23. J.H. was supported by research grants VIL16599 and VIL54489 from Villum Fonden. 
E.M. and S.H.S. thank the Max Planck Society for support through the Max Planck Fellowship of S.H.S. S.W.J. gratefully acknowledges support from a Guggenheim Fellowship. A.V.F. is grateful for the financial support of {\it HST} programs GO-16729 and GO-17504 from STScI, as well as many individual donors. M.J. was supported by the United Kingdom Research and Innovation (UKRI) Future Leaders Fellowship ``Using Cosmic Beasts to uncover the Nature of Dark Matter'' (grant number MR/X006069/1).


%

\facilities{{\it JWST} (NIRSpec spectroscopy and NIRCam imaging) and {\it HST} (WFC3/IR imaging)}


\software{\texttt{jwst} \citep{bushouse_2022_7229890}, \texttt{SNID} \citep{blondin07} \texttt{DrizzlePac} \url{https://github.com/spacetelescope/drizzlepac}, \texttt{photutils} \citep{larry_bradley_2024_13989456}, \texttt{StarDust2} \citep{satrdust_2014}, \texttt{snosmo} \citep{barbary_2024_14025775}, \texttt{SExtractor} \citep{1996A&AS..117..393B}, \texttt{pyHalo} \citep{Gilman_pyHalo}, \texttt{lenstronomy} \citep{Birrer:2018xgm, Birrer2021}, \texttt{Prospector} \citep{2021ApJS..254...22J}, \texttt{Astrolib PySynphot} \citep{pysynphot2015}, \texttt{PythonPhot3} \citep{2015ascl.soft01010J}, and \texttt{Astropy} \citep{astropy:2013, astropy:2018, astropy:2022}} 



\appendix
\section{Spectra of SN\,2022\MakeLowercase{riv} and Core-collapse SN Subclasses}\label{App: A}

\begin{figure*}[th!]
\centering
\includegraphics[clip, trim=0cm 0cm 0cm 0cm, width=0.9\textwidth]{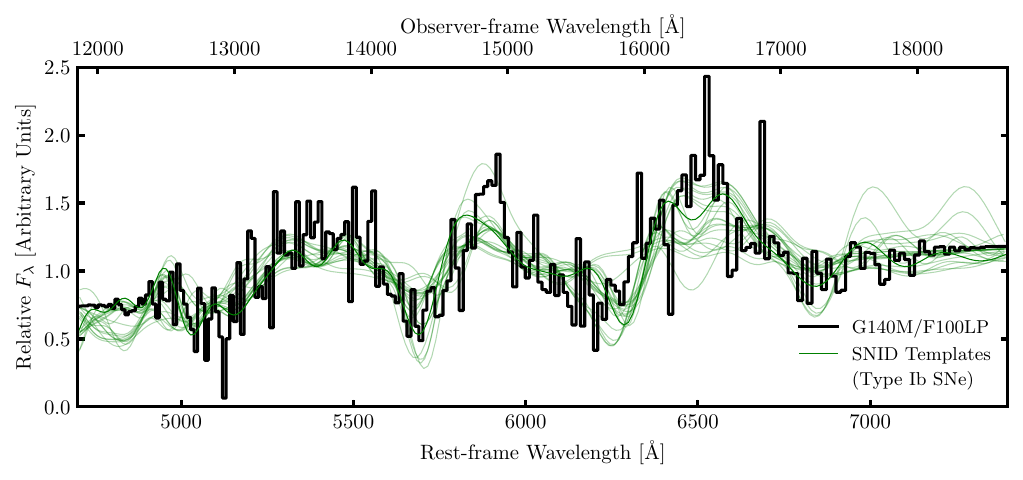}
\includegraphics[clip, trim=0cm 0cm 0cm 0cm, width=0.9\textwidth]{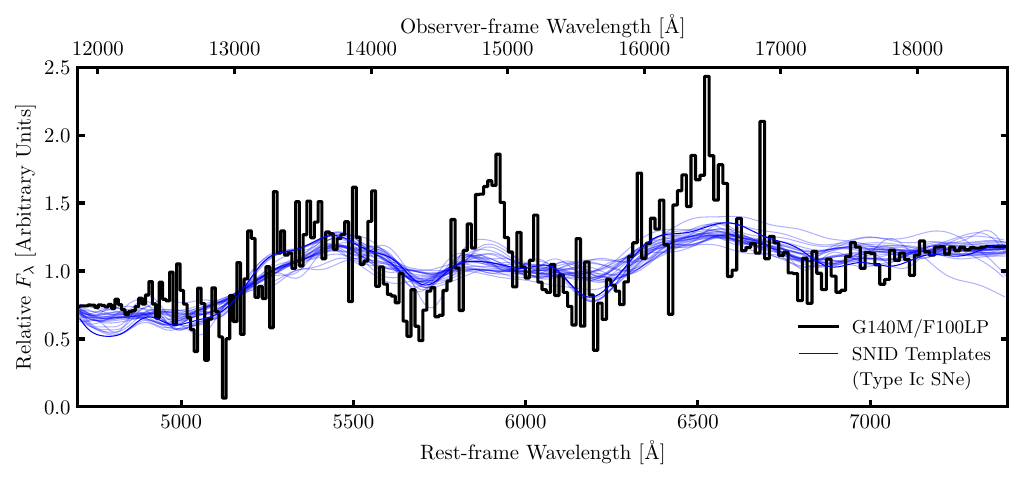}
\caption{Flattened G140M/100LP spectrum of SN\,2022riv used for the \texttt{SNID} analysis (black line). The 23 best-matching SN~Ib templates (green curves in the top panel) and 30 best-fitting SN~Ic templates (blue curves in the bottom panel) with $r \times lap >5$ from \texttt{SNID} are shown for comparison. No SN~II template matches were found.
\label{fig:100lp_Ib/c}}
\end{figure*}

\begin{figure*}[th!]
\centering
\includegraphics[clip, trim=0cm 0cm 0cm 0cm, width=0.9\textwidth]{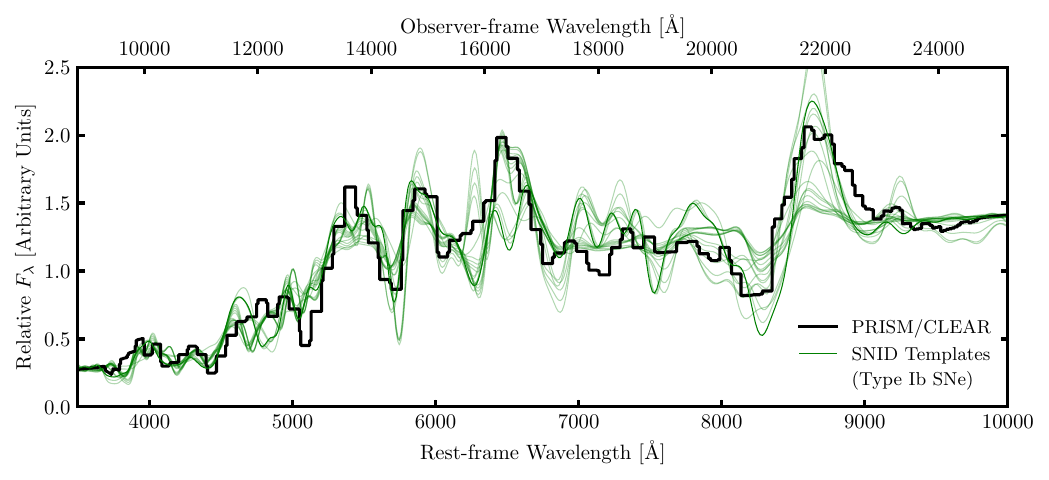}
\includegraphics[clip, trim=0cm 0cm 0cm 0cm, width=0.9\textwidth]{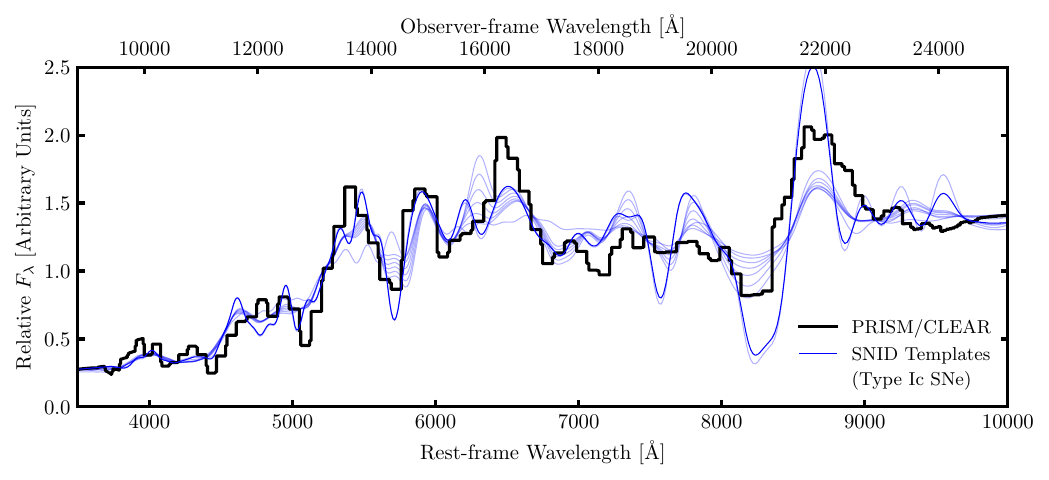}
\includegraphics[clip, trim=0cm 0cm 0cm 0cm, width=0.9\textwidth]{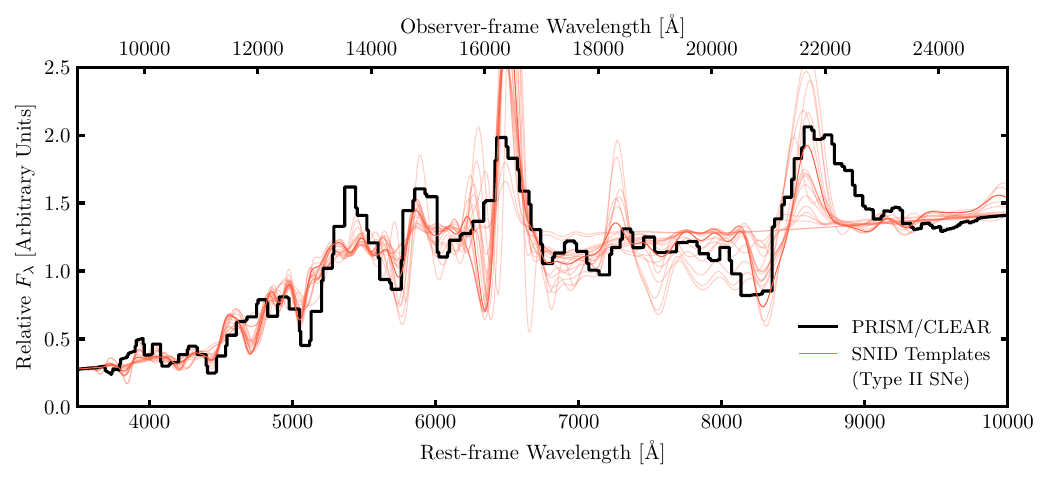}
\caption{Same as in Figure~\ref{fig:100lp_Ib/c}, but for the CLEAR/PRISM spectrum (black line). The 20 best-matching SN~Ib templates (green curves in the top panel), nine best-fitting SN~Ic templates (blue curves in the middle panel), and 19 best-fitting SN~II templates (orange curves in the bottom panel) with $r \times lap >5$ from \texttt{SNID} are shown for comparison.
\label{fig:PRISM_Ib/c_II}}
\end{figure*}

Figures~\ref{fig:100lp_Ib/c} and~\ref{fig:PRISM_Ib/c_II} in this appendix present the continuum-removed G140M/100LP and PRISM spectra of SN\,2022riv, respectively, overlaid with the best-fitting SN Ib, SN Ic, and SN II templates, each with an $r \times lap$ value greater than 5.

\section{Lensed Images} \label{lensedImages}

\begin{figure}[h!]
\centering
\includegraphics[clip, trim=0cm 0cm 0cm 0cm, width=0.45\textwidth]{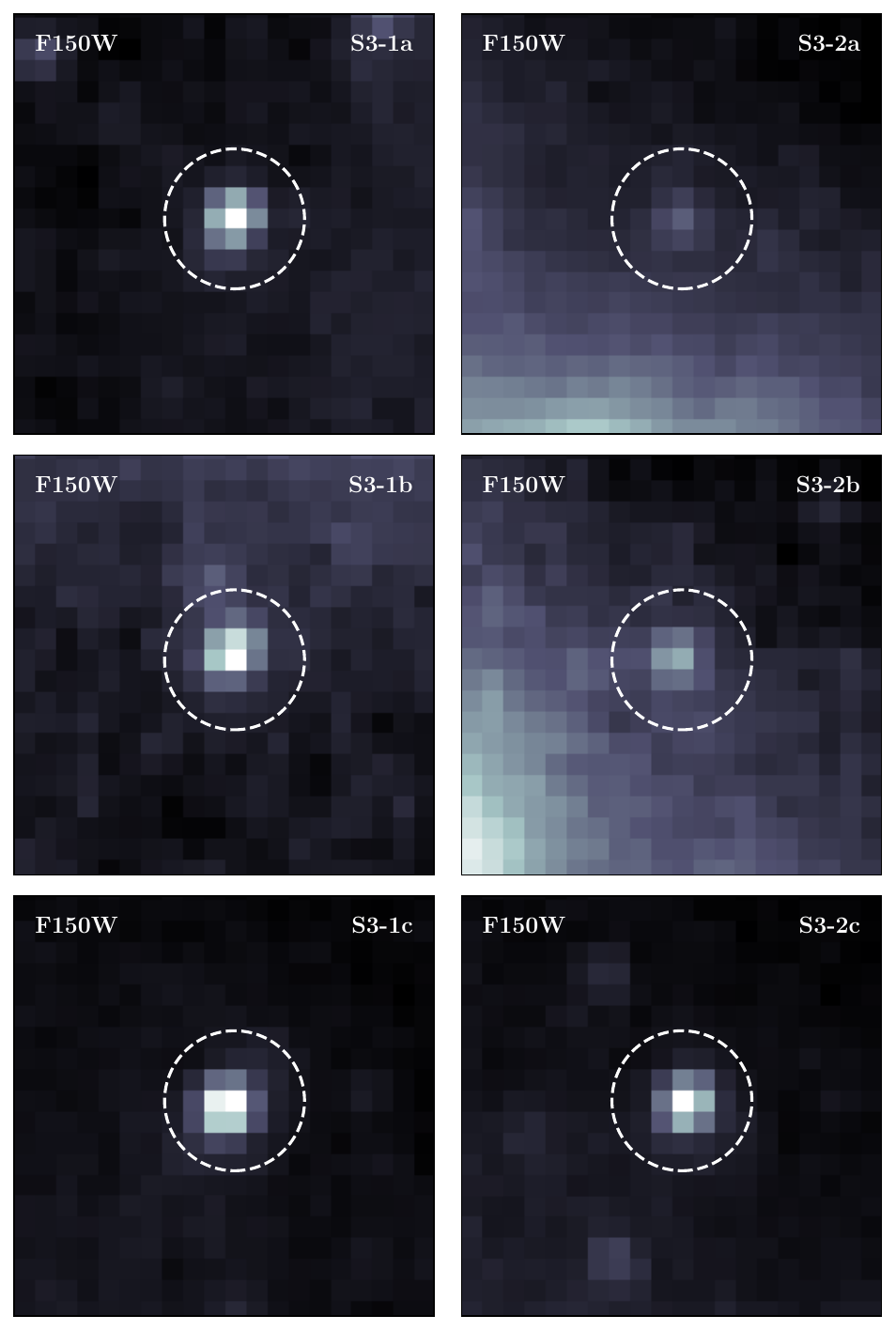}
\caption{Cutout images of the newly identified systems S3-(1--2)a, S3-(1--2)b, and S3-(1--2)c, all with spectroscopic redshifts of $z=1.5194$, as detected and localized using \textit{JWST} NIRCam observations.. 
\label{cutouts_11_S}}
\end{figure}

\begin{figure}
\centering
\includegraphics[clip, trim=0cm 0cm 0cm 0cm, width=0.47\textwidth]{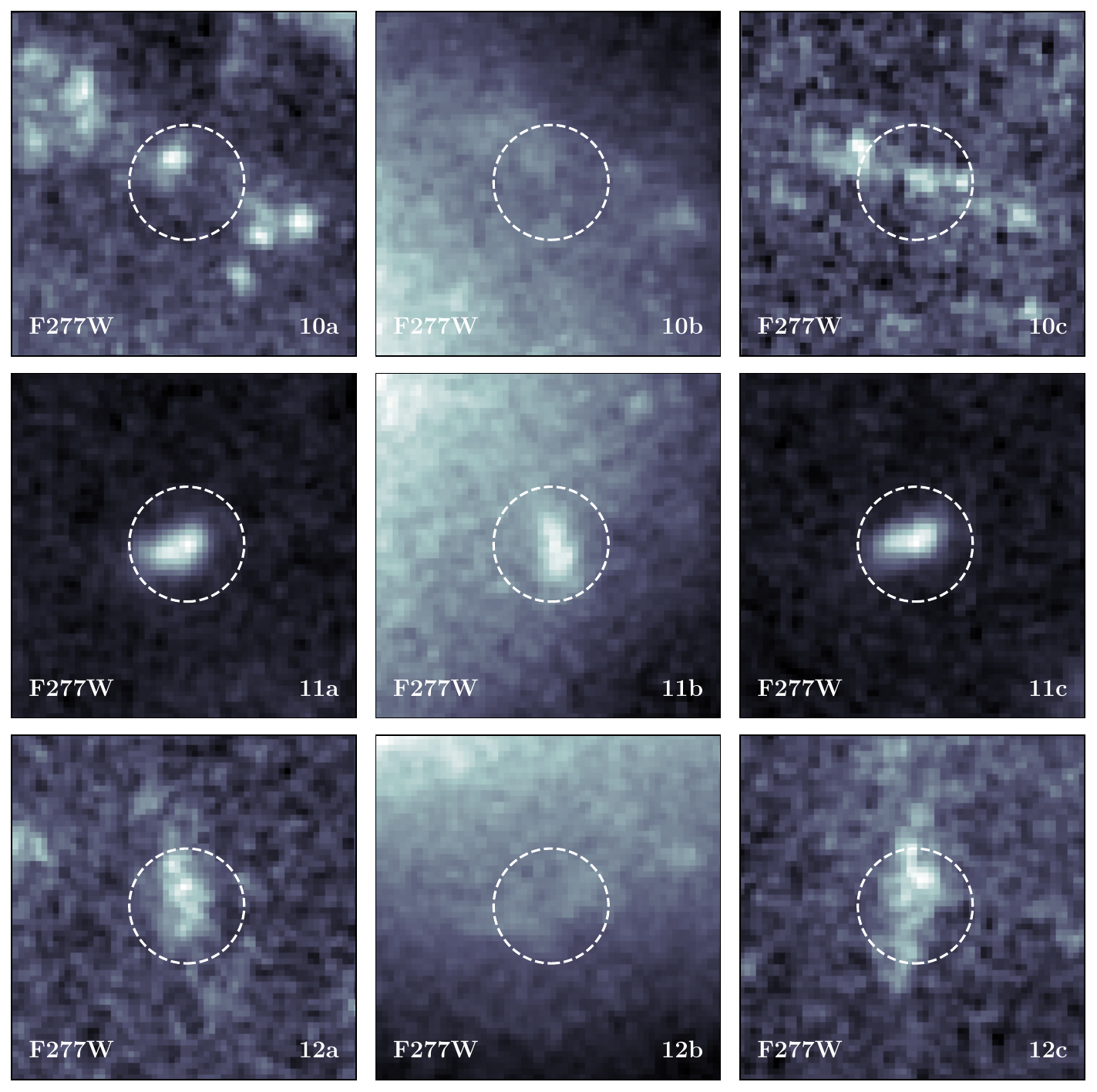}
\caption{Cutout images showing prospective multiple-image candidates in the RX\,J2129 galaxy cluster field. System 11 is well localized using \textit{JWST}/NIRCam observations. Systems 10 and 12 are not clearly identifiable owing to their extended and complex morphologies. Dashed circles indicate the inferred positions of these systems. 
\label{new_systems_invisible}}
\end{figure}

This appendix summarizes the lensed image systems used to generate the cluster lens models discussed in Section~\ref{lens_models}, as well as several prospective lensed images that were not included due to the lack of spectroscopic redshifts or their extended morphologies, which complicate secure identification in the NIRCam images. Table~\ref{Systems_RA_and_dec} lists the lens systems used to construct the galaxy cluster lens models, including three newly identified systems, whose cutouts are shown in Figure~\ref{cutouts_11_S}. We also report three additional prospective image systems that were not included as lens model constraints. These systems are listed in Table~\ref{fuzzy_systems}, and their cutouts are shown in Figure~\ref{new_systems_invisible}, with dashed circles indicating their inferred positions.

\begin{deluxetable}{lclc}[h!]
\setlength{\tabcolsep}{10pt}
\tablecaption{Catalog of multiple-image systems identified in the RX\,J2129 galaxy cluster field based on \textit{JWST}/NIRCam observations, including their corresponding spectroscopic redshifts.} \label{Systems_RA_and_dec}
\tablehead{
\colhead{ID} & \colhead{R.A. [deg.]} & \colhead{Decl. [deg.]} & \colhead{$z_{\rm spec}$}\\[-0.4cm]}
\startdata
1a & 322.4149230 & 0.0904156 & 0.6786\\
1b & 322.4151893 & 0.0889741 & 0.6786\\
1c & 322.4166530 & 0.0867530 & 0.6786\\
\hline
2a & 322.4146395 & 0.0923864 & 0.9160\\
2b & 322.4162941 & 0.0881023 & 0.9160\\
2c & 322.4165862 & 0.0877771 & 0.9160\\
\hline
3a & 322.4159519 & 0.0915111 & 1.5194\\
3b & 322.4173436 & 0.0906621 & --\\
3c & 322.4169561 & 0.0903403 & --\\
3d & 322.4185684 & 0.0849278 & 1.5194\\
\hline
4a & 322.4155685 & 0.0921535 & 1.5202\\
4b & 322.4174895 & 0.0901781 & --\\ 
4c & 322.4184261 & 0.0853785 & 1.5202\\
\hline
5a & 322.4179753 & 0.0932720 & --\\
5b & 322.4201821 & 0.0897722 & 1.5210\\
5c & 322.4204071 & 0.0883142 & 1.5210\\
\hline
6a & 322.4137586 & 0.0941984 & 3.0815\\
6b & 322.4167395 & 0.0877614 & 3.0815\\
6c & 322.4169923 & 0.0873924 & 3.0815\\
\hline
7a & 322.4137266 & 0.0920863 & 3.4270\\
7b & 322.4144323 & 0.0886347 & 3.4270\\
7c & 322.4175405 & 0.0838683 & 3.4270\\
\hline
G1 & 322.4202364 & 0.0936441 &  9.5100\\
G2 & 322.4215532 & 0.0916940  & 9.5100\\
G3 & 322.4224106 & 0.0889363 &  9.5100\\
\hline
S3-1a & 322.4163448 & 0.0912444 & $1.5194^*$\\
S3-2a & 322.4169115 & 0.0905028 & $1.5194^*$\\
\hline
S3-1b & 322.4161198 & 0.0911278 & $1.5194^*$\\
S3-2b & 322.4167364 & 0.0904193 & $1.5194^*$\\
\hline
S3-1c & 322.4163198 & 0.0907945 & $1.5194^*$\\
S3-2c & 322.4165032 & 0.0905862 & $1.5194^*$\\
\hline
SN\,2022riv & 322.4175552 & 0.0900571 & 1.5220\\
\enddata 
\tablecomments{For systems 1–7, we revised the positions originally reported by \cite{caminharosatigrillo19} and adopted the spectroscopic redshifts obtained from MUSE observations. System G corresponds to the gravitationally lensed high-redshift galaxy identified by \cite{williamskellychen23}. \\
$^*$Systems S3-(1--2)a, S3-(1--2)b, and S3-(1--2)c are newly identified and are assumed to belong to System 3; consequently, they are assigned the same spectroscopic redshift as System 3.
\label{tab:images}}
\end{deluxetable}

\begin{deluxetable}{lcl}
\setlength{\tabcolsep}{22pt}
\tablecaption{Prospective multiple-image system candidates in the RX\,J2129 galaxy cluster field.} \label{fuzzy_systems}
\tablehead{
\colhead{ID} & \colhead{R.A. [deg.]} & \colhead{Decl. [deg.]} \\[-0.4cm]}
\startdata
10a & 322.4140779 & $0.0912750^*$\\ 
10b & 322.4144767 & $0.0895131^*$\\  
10c & 322.4179046 & $0.0833583^*$\\
\hline
11a & 322.4138783 & 0.0922445\\ 
11b & 322.4148032 & 0.0884279\\
11c & 322.4172449 & 0.0849278\\
\hline
12a & 322.4138783 & $0.0930039^*$\\ 
12b & 322.4154533 & $0.0880447^*$\\  
12c & 322.4170658 & $0.0859786^*$\\ 
\enddata 
\tablecomments{ $^*$The inferred positions for systems that are either not clearly identifiable due to their extended and complex morphology. \label{tab:candidates}}
\end{deluxetable}

\section{Predicted Stellar Microlensing Effects on Absolute Magnification Using the \textit{HST} WFC3/IR F110W Band}

\begin{figure*}[h!]
\centering
\includegraphics[clip, trim=0cm 0cm 0cm 0cm, width=1\textwidth]{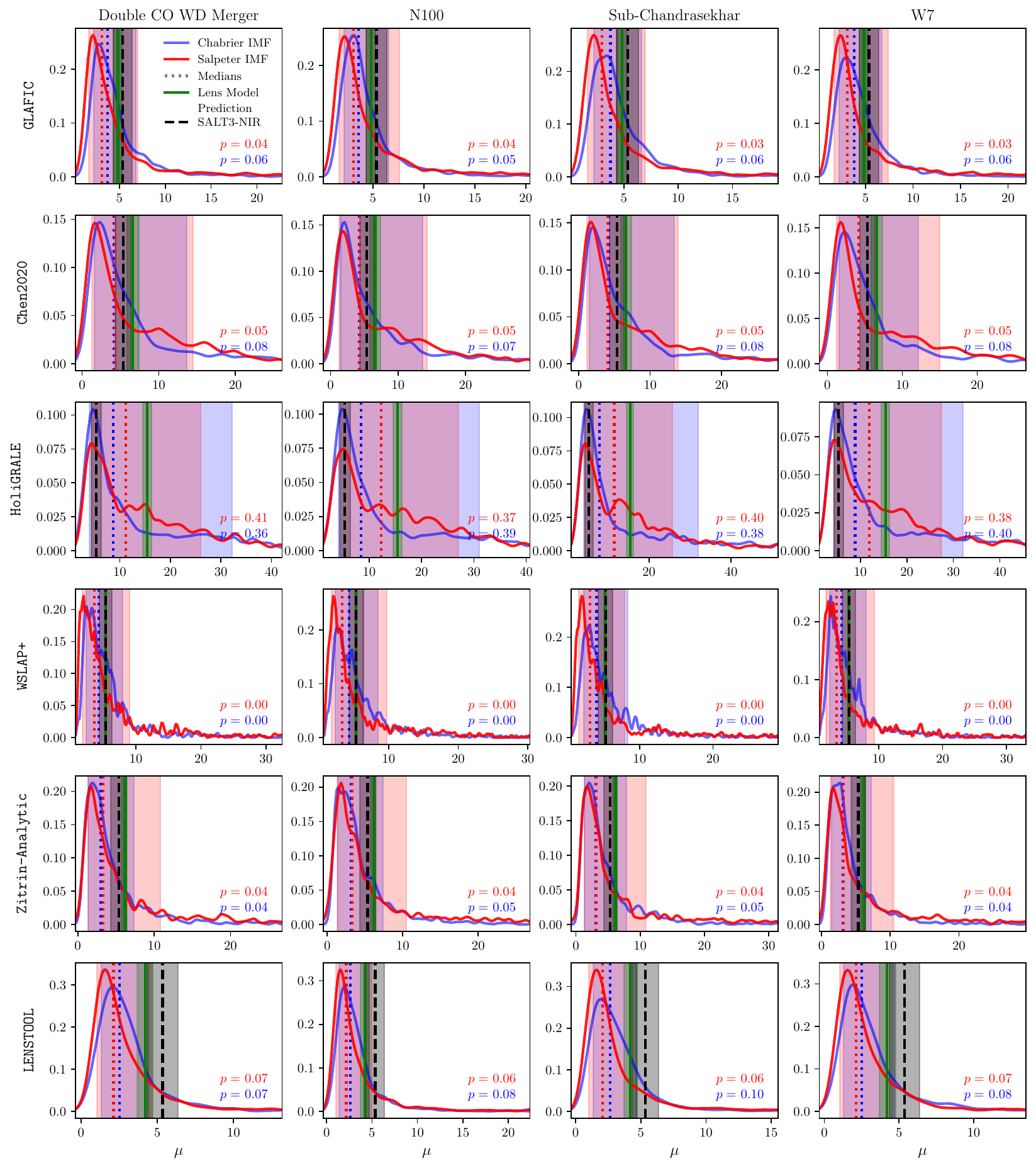}
\caption{The same as in Fig.~\ref{magnification_plot_f160w}, but with synthetic photometric calculations performed using the \textit{HST} WFC3/IR F110W band.
\label{magnification_plot_f110w}}
\end{figure*}

This appendix presents the effects of stellar microlensing constrained using the \textit{HST} WFC3/IR F110W band in the synthetic photometry calculations, considering four different SN~Ia explosion models. Table~\ref{tab:microlensing_F110W}, analogous to Table~\ref{tab:microlensing_F160W}, summarizes the change in magnification due to microlensing. Tables~\ref{tab:total_mag_F110W_Chabrier} and ~\ref{tab:total_mag_F110W_Salpeter} present the absolute magnifications for each lens model and SN~Ia explosion model after correcting $\mu_{\rm model}$ for stellar microlensing (see Table~\ref{tab:microlensing_F110W}) and dark matter substructure millilensing (see Table~\ref{tab:millilensing}), assuming Chabrier (Table~\ref{tab:total_mag_F110W_Chabrier}) and Salpeter (Table~\ref{tab:total_mag_F110W_Salpeter}) IMFs. The PDFs of the corrected absolute magnifications are shown in Figure~\ref{magnification_plot_f110w}. These results are consistent with those obtained using the WFC3/IR F160W band.

\begin{deluxetable*}{l|
>{\raggedleft\arraybackslash}p{1.7cm}
>{\raggedleft\arraybackslash}p{1.7cm}|
>{\raggedleft\arraybackslash}p{1.7cm}
>{\raggedleft\arraybackslash}p{1.7cm}|
>{\raggedleft\arraybackslash}p{1.7cm}
>{\raggedleft\arraybackslash}p{1.7cm}|
>{\raggedleft\arraybackslash}p{1.7cm}
>{\raggedleft\arraybackslash}p{1.7cm}
}[h]
\setlength{\tabcolsep}{2.8pt}
\tablecaption{Same as in Table~\ref{tab:microlensing_F160W}, but using the \textit{HST} WFC3/IR F110W band for synthetic photometry.} \label{tab:microlensing_F110W}
\tablehead{
\multirow{4}{*}{Model} & \multicolumn{8}{c}{$\Delta \mu_{\rm micro}$}\\ [0.2cm]
\cline{2-9}\\[-0.4cm]
& \multicolumn{2}{c}{Double CO WD Merger} & \multicolumn{2}{c}{N100} & \multicolumn{2}{c}{Sub-Chandrasekhar} & \multicolumn{2}{c}{W7}\\
\cline{2-9}\\[-0.45cm]
& Chabrier & Salpeter & Chabrier & Salpeter & Chabrier & Salpeter & Chabrier & Salpeter}
\startdata
\texttt{GLAFIC}	&	 $-1.1^{+3.1}_{-1.4}$  	&	 $-1.7^{+4.0}_{-1.3}$  	&	 $-1.2^{+2.9}_{-1.3}$  	&	 $-1.7^{+4.6}_{-1.4}$  	&	 $-1.0^{+2.8}_{-1.4}$  	&	 $-1.8^{+4.0}_{-1.2}$  	&	 $-1.0^{+3.0}_{-1.5}$  	&	 $-1.8^{+4.5}_{-1.2}$  	\\
\texttt{Chen2020} & $-2.4^{+9.5}_{-2.5}$ & $-2.4^{+10.3}_{-2.9}$ & $-2.3^{+9.3}_{-2.7}$ & $-2.4^{+10.1}_{-2.8}$ & $-2.5^{+9.6}_{-2.6}$ & $-2.7^{+10.0}_{-2.6}$ & $-2.4^{+7.8}_{-2.5}$ & $-2.4^{+10.8}_{-2.8}$ \\
\texttt{HoliGRALE}	&	 $-6.3^{+23.4}_{-4.8}$ 	&	 $-4.0^{+14.7}_{-6.9}$ 	&	 $-6.7^{+22.3}_{-4.3}$ 	&	 $-3.0^{+14.8}_{-7.6}$ 	&	 $-7.3^{+24.1}_{-3.9}$ 	&	 $-3.6^{+13.9}_{-7.3}$ 	&	 $-6.4^{+23.1}_{-4.7}$ 	&	 $-3.4^{+15.4}_{-7.4}$ 	\\
\texttt{WSLAP+}	&	 $-1.0^{+3.7}_{-1.9}$  	&	 $-1.6^{+5.4}_{-1.8}$  	&	 $-0.9^{+4.2}_{-2.0}$  	&	 $-1.9^{+6.4}_{-1.5}$  	&	 $-1.1^{+4.2}_{-1.7}$  	&	 $-2.0^{+4.7}_{-1.5}$  	&	 $-1.0^{+3.8}_{-1.8}$  	&	 $-1.9^{+5.9}_{-1.6}$  	\\
\texttt{Zitrin-Analytic}	&	 $-3.1_{-1.6}^{+4.2}$  	&	 $-2.7_{-2.0}^{+7.5}$  	&	 $-2.9_{-1.8}^{+4.2}$  	&	 $-2.7_{-2.0}^{+7.2}$  	&	 $-3.0_{-1.7}^{+4.7}$  	&	 $-2.9_{-1.9}^{+7.8}$  	&	 $-3.0_{-1.6}^{+4.1}$  	&	 $-2.9_{-1.8}^{+7.3}$  	\\
\texttt{LENSTOOL}	&	 $-1.6_{-1.1}^{+1.8}$  	&	 $-2.1_{-0.9}^{+2.5}$  	&	 $-1.6_{-1.2}^{+2.0}$  	&	 $-2.1_{-1.0}^{+2.8}$  	&	 $-1.5_{-1.2}^{+2.0}$  	&	 $-2.2_{-0.9}^{+2.6}$  	&	 $-1.7_{-1.1}^{+2.3}$  	&	 $-2.1_{-0.9}^{+2.5}$  	\\
\enddata 
\end{deluxetable*}

\begin{deluxetable*}{l|
>{\raggedleft\arraybackslash}p{1.7cm}|
>{\raggedleft\arraybackslash}p{1.7cm}|
>{\raggedleft\arraybackslash}p{1.7cm}|
>{\raggedleft\arraybackslash}p{1.7cm}|
>{\raggedleft\arraybackslash}p{1.7cm}|
>{\raggedleft\arraybackslash}p{1.7cm}|
>{\raggedleft\arraybackslash}p{1.7cm}|
>{\raggedleft\arraybackslash}p{1.7cm}
}
\setlength{\tabcolsep}{2.8pt}
\tablecaption{Same as in Table~\ref{tab:total_mag_F160W_Chabrier}, but using the \textit{HST} WFC3/IR F110W band for synthetic photometry. } \label{tab:total_mag_F110W_Chabrier}
\tablehead{
\multirow{4}{*}{Model} & \multicolumn{8}{c}{SN Ia Explosion Model}\\ [0.2cm]
\cline{2-9}\\[-0.4cm]
& \multicolumn{2}{c|}{Double CO WD Merger} & \multicolumn{2}{c|}{N100} & \multicolumn{2}{c|}{Sub-Chandrasekhar} & \multicolumn{2}{c}{W7}\\
\cline{2-9}\\[-0.45cm]
& $\mu$  & $\rm Tension$  & $\mu$  & $\rm Tension$  & $\mu$  & $\rm Tension$  & $\mu$  & $\rm Tension$ }
\startdata
\texttt{GLAFIC} & $3.7_{-1.5}^{+3.1}$ & $0.5\sigma$ & $3.6_{-1.4}^{+3.0}$ & $0.6\sigma$ & $3.8_{-1.5}^{+2.9}$ & $0.5\sigma$ & $3.8_{-1.6}^{+3.0}$ & $0.5\sigma$ \\
\texttt{Chen2020} & $4.2_{-2.6}^{+9.5}$ & $0.1\sigma$ & $4.3_{-2.8}^{+9.4}$ & $0.1\sigma$ & $4.2_{-2.7}^{+9.2}$ & $0.1\sigma$ & $4.1_{-2.6}^{+8.0}$ & $0.2\sigma$ \\
\texttt{HoliGRALE} & $8.7_{-4.7}^{+23.5}$ & $0.7\sigma$ & $8.4_{-4.3}^{+22.6}$ & $0.7\sigma$ & $8.0_{-3.9}^{+24.0}$ & $0.6\sigma$ & $8.9_{-4.7}^{+23.2}$ & $0.7\sigma$ \\
\texttt{WSLAP+} & $4.3_{-1.9}^{+3.7}$ & $0.3\sigma$ & $4.3_{-2.0}^{+4.2}$ & $0.2\sigma$ & $4.1_{-1.7}^{+4.3}$ & $0.3\sigma$ & $4.2_{-1.9}^{+3.8}$ & $0.3\sigma$ \\
\texttt{Zitrin-Analytic} & $3.0_{-1.7}^{+4.3}$ & $0.5\sigma$ & $3.2_{-1.8}^{+4.2}$ & $0.5\sigma$ & $3.1_{-1.8}^{+4.8}$ & $0.5\sigma$ & $3.1_{-1.6}^{+4.1}$ & $0.5\sigma$ \\
\texttt{LENSTOOL} & $2.5_{-1.2}^{+1.9}$ & $1.2\sigma$ & $2.6_{-1.3}^{+2.1}$ & $1.2\sigma$ & $2.7_{-1.3}^{+2.1}$ & $1.2\sigma$ & $2.5_{-1.2}^{+2.3}$ & $1.1\sigma$ \\
\enddata 
\end{deluxetable*}

\begin{deluxetable*}{l|
>{\raggedleft\arraybackslash}p{1.7cm}|
>{\raggedleft\arraybackslash}p{1.7cm}|
>{\raggedleft\arraybackslash}p{1.7cm}|
>{\raggedleft\arraybackslash}p{1.7cm}|
>{\raggedleft\arraybackslash}p{1.7cm}|
>{\raggedleft\arraybackslash}p{1.7cm}|
>{\raggedleft\arraybackslash}p{1.7cm}|
>{\raggedleft\arraybackslash}p{1.7cm}
}
\setlength{\tabcolsep}{2.8pt}
\tablecaption{Same as in Table~\ref{tab:total_mag_F160W_Chabrier}, but using the Salpeter IMF in the microlensing simulations and the \textit{HST} WFC3/IR F110W band for synthetic photometry.} \label{tab:total_mag_F110W_Salpeter}
\tablehead{
\multirow{4}{*}{Model} & \multicolumn{8}{c}{SN Ia Explosion Model}\\ [0.2cm]
\cline{2-9}\\[-0.4cm]
& \multicolumn{2}{c|}{Double CO WD Merger} & \multicolumn{2}{c|}{N100} & \multicolumn{2}{c|}{Sub-Chandrasekhar} & \multicolumn{2}{c}{W7}\\
\cline{2-9}\\[-0.45cm]
& $\mu$  & $\rm Tension$  & $\mu$  & $\rm Tension$  & $\mu$  & $\rm Tension$  & $\mu$  & $\rm Tension$ }
\startdata
\texttt{GLAFIC} & $3.1_{-1.4}^{+3.9}$ & $0.6\sigma$ & $3.1_{-1.5}^{+4.5}$ & $0.5\sigma$ & $3.0_{-1.4}^{+4.0}$ & $0.6\sigma$ & $3.1_{-1.4}^{+4.3}$ & $0.5\sigma$ \\
\texttt{Chen2020} & $4.3_{-3.0}^{+10.2}$ & $0.1\sigma$ & $4.3_{-3.1}^{+9.9}$ & $0.1\sigma$ & $4.0_{-2.8}^{+9.9}$ & $0.1\sigma$ & $4.2_{-2.9}^{+10.8}$ & $0.1\sigma$ \\
\texttt{HoliGRALE} & $11.2_{-6.7}^{+14.8}$ & $0.9\sigma$ & $12.3_{-7.7}^{+14.7}$ & $0.9\sigma$ & $11.6_{-7.2}^{+14.2}$ & $0.8\sigma$ & $11.9_{-7.4}^{+15.5}$ & $0.9\sigma$ \\
\texttt{WSLAP+} & $3.6_{-1.9}^{+5.4}$ & $0.3\sigma$ & $3.3_{-1.6}^{+6.4}$ & $0.3\sigma$ & $3.2_{-1.5}^{+4.8}$ & $0.4\sigma$ & $3.4_{-1.6}^{+5.9}$ & $0.3\sigma$ \\
\texttt{Zitrin-Analytic} & $3.4_{-2.0}^{+7.5}$ & $0.3\sigma$ & $3.3_{-2.0}^{+7.2}$ & $0.3\sigma$ & $3.2_{-1.9}^{+7.7}$ & $0.3\sigma$ & $3.2_{-1.9}^{+7.3}$ & $0.3\sigma$ \\
\texttt{LENSTOOL} & $2.1_{-1.1}^{+2.5}$ & $1.2\sigma$ & $2.2_{-1.1}^{+2.8}$ & $1.1\sigma$ & $2.1_{-1.1}^{+2.5}$ & $1.2\sigma$ & $2.1_{-1.1}^{+2.5}$ & $1.2\sigma$\\
\enddata 
\end{deluxetable*}

\bibliography{ms, sample631}{}
\bibliographystyle{aasjournal}




\end{document}